\documentclass[12pt]{article}
\usepackage{graphicx,rotating,epsfig,amsmath,amssymb,color,multirow,pstricks,pst-plot}
\newlength{\digitwidth} 

\newcommand{\HRule}{\rule{0.4\linewidth}{0.3mm}}
\newcommand{\mlab}[1]%
    {\mbox{}\marginpar{\raggedright\hspace{0pt}\footnotesize #1}}
\newcommand{\elab}{$E_{\mathrm{lab}}$}
\newcommand{\pt}{$p_{\mathrm{T}}$}
\newcommand{\xf}{$x_{\mathrm F}$}
\newcommand{\dd}{{\mathrm d}}
\newcommand{\ccbar}{${\rm c}\bar{\rm c}$}
\newcommand{\bbbar}{${\rm b} \bar{\rm b}$}
\newcommand{\qqbar}{$q\bar{q}$}
\newcommand{\BBbar}{${\rm B}\overline{\rm B}$}
\newcommand{\DDbar}{${\rm D}\overline{\rm D}$}
\newcommand{\ppbar}{${\rm p}\overline{\rm p}$}
\newcommand{\D}{{\rm D}}
\newcommand{\B}{{\rm B}}
\newcommand{\jpsi}{J/$\psi$}
\newcommand{\psip}{$\psi^\prime$}
\newcommand{\sqrts}{$\sqrt{s}$}

\newcommand{\degree}{$^\circ$}

\begin{document}

\begingroup
\thispagestyle{empty}
\baselineskip=14pt
\parskip 0pt plus 5pt

\begin{center}
{\large EUROPEAN LABORATORY FOR PARTICLE PHYSICS}
\end{center}

\bigskip
\begin{flushright}
CERN--PH--EP\,/\,2006--013--rev\\
September 8, 2006
\end{flushright}

\bigskip
\begin{center}
{\Large\bf
Heavy-Flavour Hadro-Production\\
from Fixed-Target to Collider Energies}

\bigskip\bigskip

C.~Louren\c{c}o\\
CERN-PH, Geneva, Switzerland\\[0.5cm]

H.K.~W\"ohri\,$^*$\\
CFTP-IST, Lisbon, Portugal and\\
CERN-PH, Geneva, Switzerland

\bigskip\bigskip\bigskip
\textbf{Abstract}

\end{center}

\begingroup
\leftskip=0.4cm
\rightskip=0.4cm
\parindent=0.pt

We review the hadro-production data presently available on open charm and
beauty absolute production cross-sections, collected by experiments at
CERN, DESY and Fermilab.  The published charm production cross-section
values are updated, in particular for the ``time evolution'' of the
branching ratios.  These measurements are compared to LO pQCD
calculations, as a function of the collision energy, using recent
parameterisations of the parton distribution functions.  We then
estimate, including nuclear effects on the parton densities, the charm
and beauty production cross-sections relevant for measurements at SPS
and RHIC energies, in proton-proton, proton-nucleus and
nucleus-nucleus collisions.  The calculations are also compared with
measurements of single D and B kinematical distributions, and \DDbar\
pair correlations.  We finish with two brief comments, concerning the
importance of beauty production as a feed-down source of \jpsi\
production, and open charm measurements performed using
leptonic decays.

\bigskip


\endgroup
\bigskip\bigskip

\begin{center}
\emph{To appear in Physics Reports}
\end{center}

\vfill
\HRule

$^*$~Now at INFN-Cagliari, Italy.

\endgroup

\newpage
\thispagestyle{empty}
\tableofcontents

\cleardoublepage
\pagenumbering{arabic}
\setcounter{page}{1}

\section{Introduction}

In the context of the study of heavy ion collisions, heavy flavour
production is becoming increasingly more interesting, as the energies
available for particle production increase from fixed target (SPS) to
collider (RHIC, LHC) experiments.  Charm is the heaviest flavour which
can be produced in nucleus-nucleus collisions at SPS energies, where
experiments with high intensity beams and a dimuon trigger made
detailed studies of the production and suppression of charmonium
states (J/$\psi$, $\psi^{\prime}$).  Besides being the natural
reference for charmonia studies (same initial state), open charm
significantly contributes to the yield of dimuons measured in the mass
range between the $\phi$ and the J/$\psi$ resonances, through the
simultaneous semi-muonic decays of a pair of D mesons.  Enhanced
production of continuum dileptons in this mass range, as seen by
several experiments at the CERN SPS, could be a signal of thermal
dimuon production from a Quark-Gluon Plasma phase, emphasising the
importance of understanding the charm ``background''.

At the higher energies of RHIC and LHC, also beauty production comes
into play, not only as a direct probe of the properties of the very
early phase of the collision system but also as a source of J/$\psi$
mesons.  Indeed, the very important study of J/$\psi$ production (and
suppression or enhancement) at collider energies requires a good
understanding of the fraction of J/$\psi$'s coming from the feed-down
of B decays.  If this feed-down source is not well understood, no
proper physics interpretation of the J/$\psi$ data will be possible.

Therefore, it is very important to have a good evaluation of the yield
of charm and beauty production as a function of the collision energy.
This is necessary for the understanding of the SPS data and to make
reliable estimates of the yields expected at the collider experiments.
To ensure their usefulness, these evaluations should be made using a
well known and easily available calculation procedure, such as the one
provided by the event generator Pythia.  However, it is essential to
ensure that the generator properly describes existing data, collected
over the last years, essentially in fixed target experiments with
proton and pion beams.

Since the calculations critically depend on the distribution functions
of the quarks and gluons present in the interacting hadrons, we must
see how the results vary if we use different sets of parton
distribution functions (PDFs).  Finally, since we want to use these
evaluations in the context of nuclear collisions, it is also important
to correctly estimate the effects of the nuclear modifications of the
PDFs on the production of heavy flavours. However, the degree of
nuclear shadowing at low values of $x$ is not well known for the
valence and sea quark distributions, and there is no \emph{direct}
information at all on the nuclear effects on the gluon distribution
function, at any value of $x$.  This makes the predictions of heavy
flavour production particularly uncertain for nuclear collisions,
especially at the higher energies (lower $x$ values) available at the
colliders, where gluon fusion dominates the production cross-sections.

In the following section we will discuss some basic concepts relevant
to the topic of heavy flavour hadro-production, with emphasis on the
importance of the parton distribution functions, including nuclear
effects.  In Section~\ref{sec:exp} we introduce the experiments which
provided the data included in this report.  The data are critically
reviewed and compiled in Section~\ref{sec:dataCharm} for open charm
production and in Section~\ref{sec:dataBeauty} for the beauty case.
These data are then compared, in Section~\ref{sec:calculations}, to
the LO pQCD calculations provided by the event generator Pythia. In
this section we also estimate the cross-sections for experiments made
at SPS and RHIC energies, and we compare the calculations with
existing data on single meson kinematical distributions and on pair
correlations.  We finish with brief comments on the relevance of the
beauty feed-down to the studies of \jpsi\ suppression at RHIC and
higher energies, and on some (indirect) measurements of charm production
yields using leptonic decays.

We should clarify that several issues related to heavy-flavour
production are \emph{not} addressed in this review.  In particular, we
only addressed hadro-production data; see, for instance,
Ref.~\cite{Mangano1997} (and references therein) for information
related to photo-production or $e^+e^-$ experiments.  We have also not
covered in here any effects related to ``heavy-ion physics'', such as
charm flow, gluon saturation, recombination, dead-cone effect, etc;
see Refs.~\cite{yellow, HP04}, for instance, for details on such
issues.  And it is clear that we only addressed \emph{open} heavy
flavour production; see, for instance, Refs.~\cite{HP04, QWG} for
details on quarkonia production, in elementary and nuclear collisions.

\section{Heavy flavour production in pQCD}
\label{sec:hvinpqcd}

In this section we will briefly review some basic issues related to
the physics of heavy-flavour production in hadronic collisions.  For a
more detailed introduction, see e.g.\ Ref.~\cite{Mangano1997}.

\begin{figure}[htb]
\centering
\begin{tabular}{cccc}
\resizebox{0.25\textwidth}{!}{%
\includegraphics*{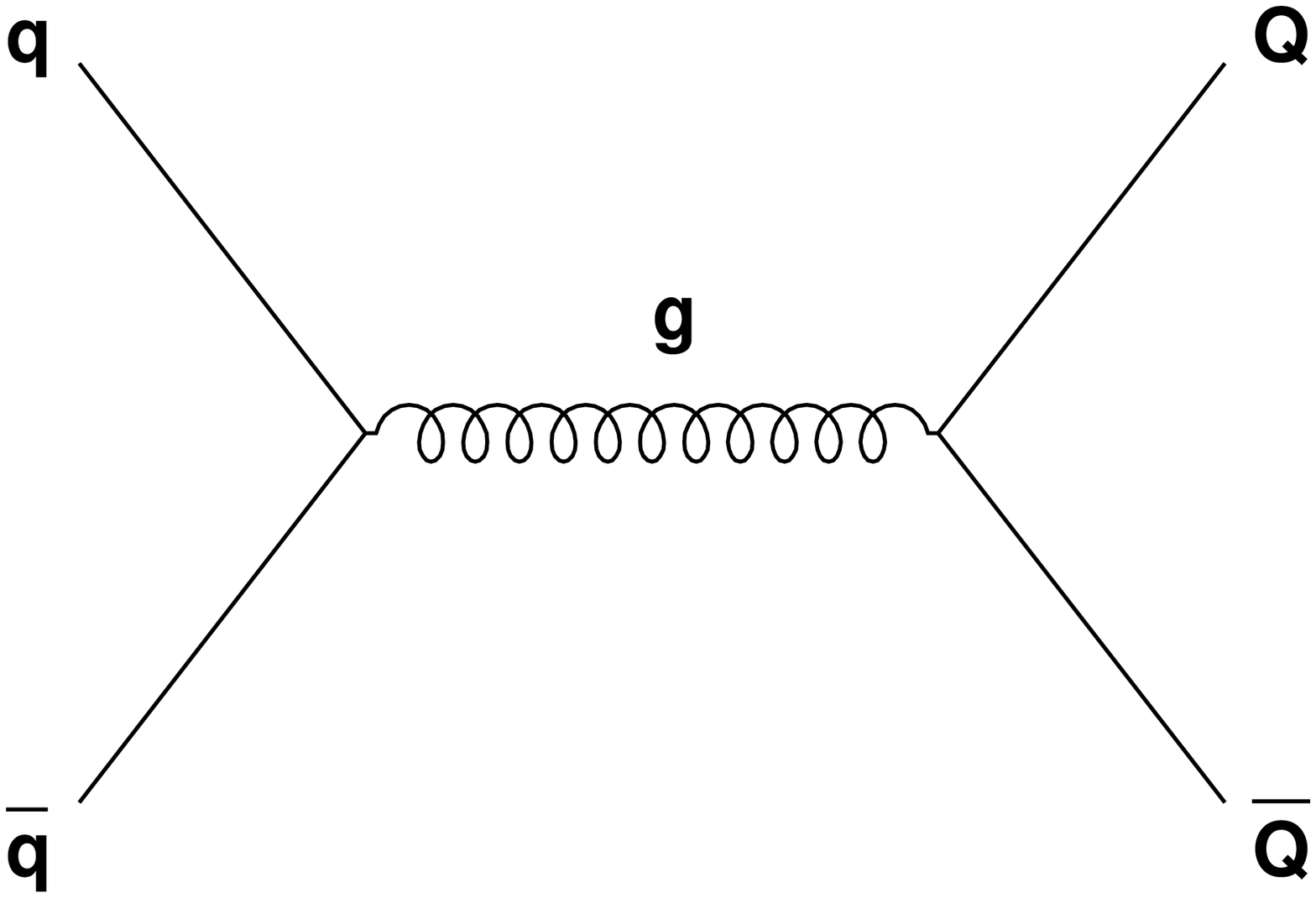}}&
\resizebox{0.25\textwidth}{!}{%
\includegraphics*{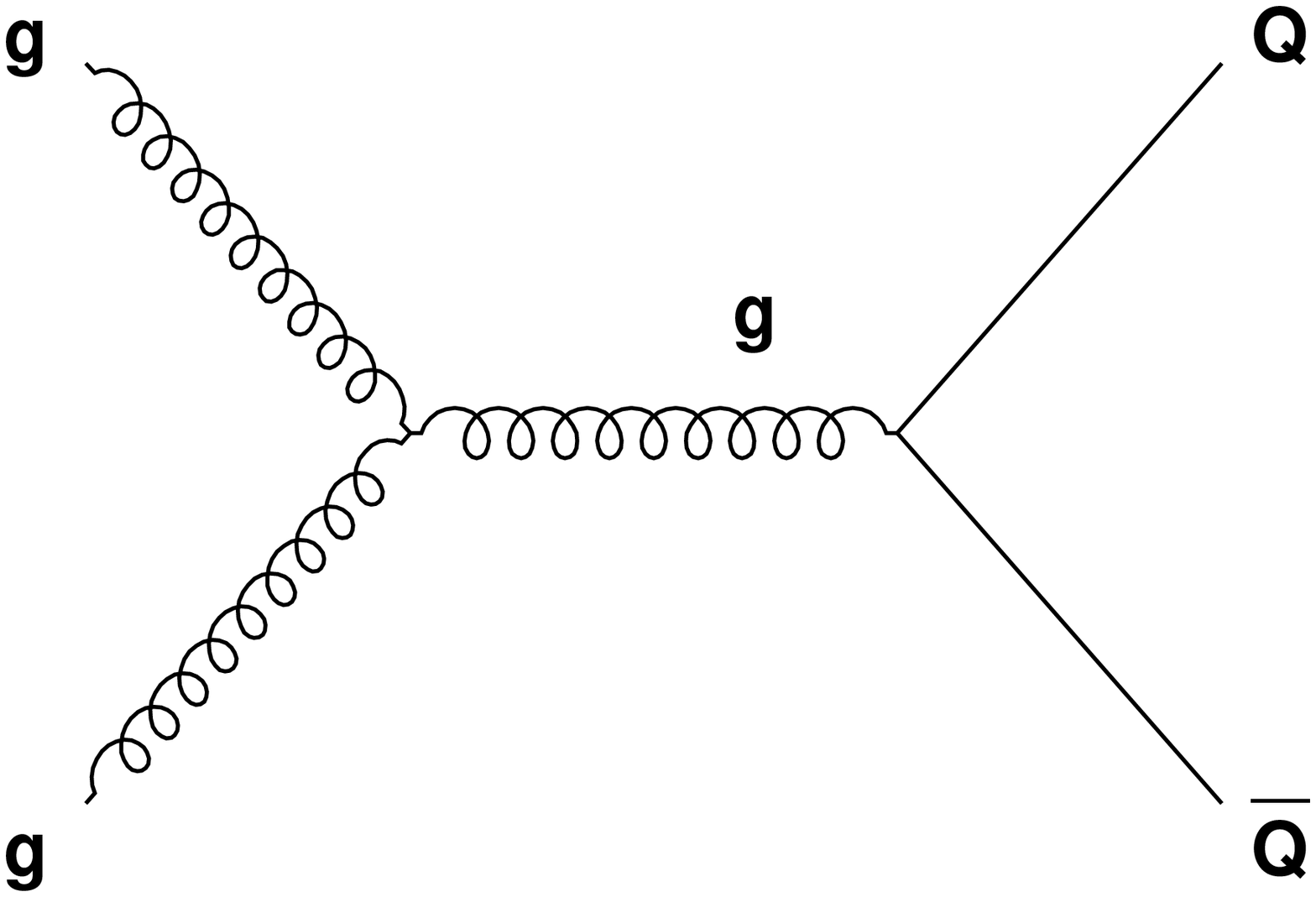}}&
\resizebox{0.16\textwidth}{!}{%
\includegraphics*{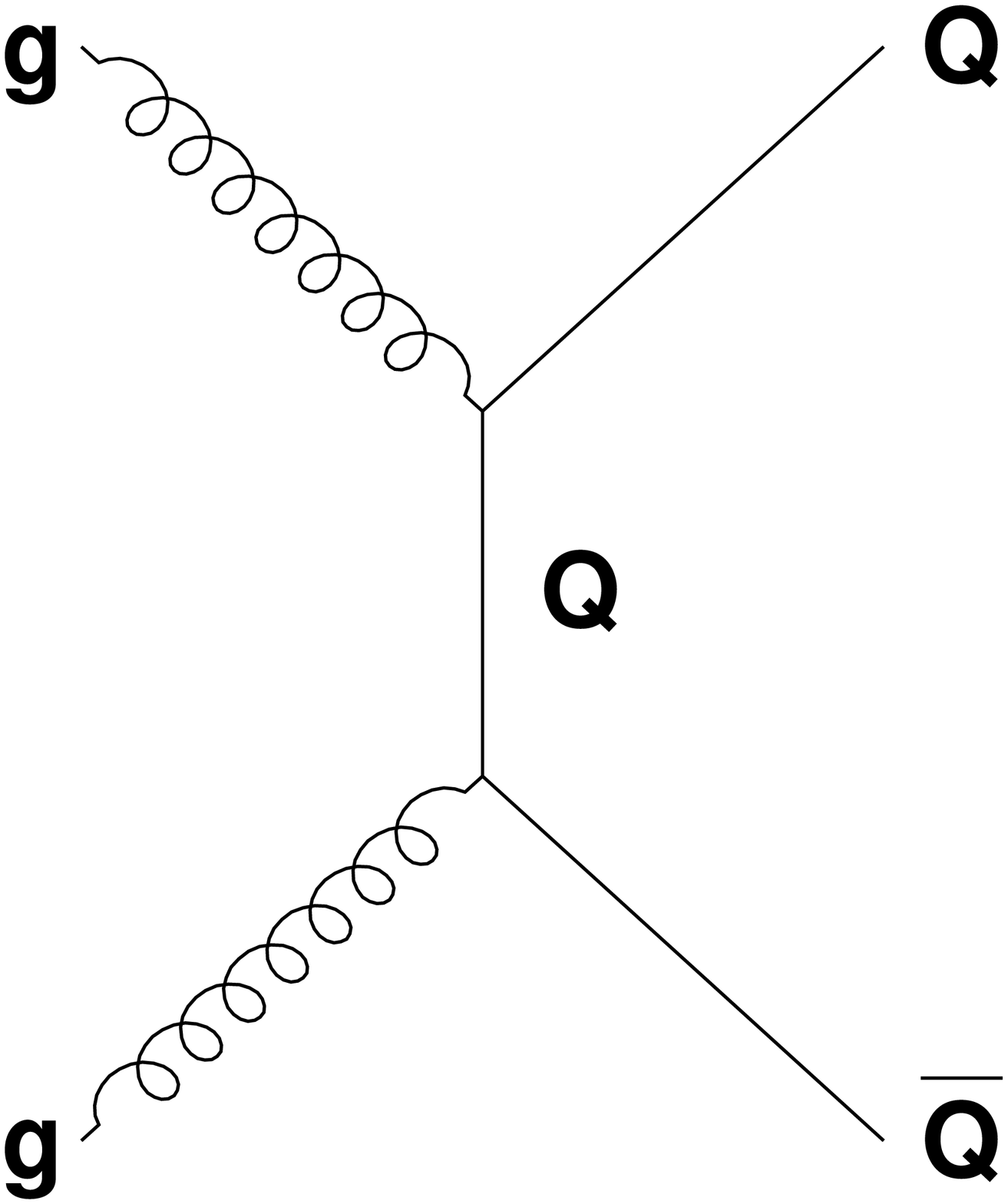}}&
\resizebox{0.23\textwidth}{!}{%
\includegraphics*{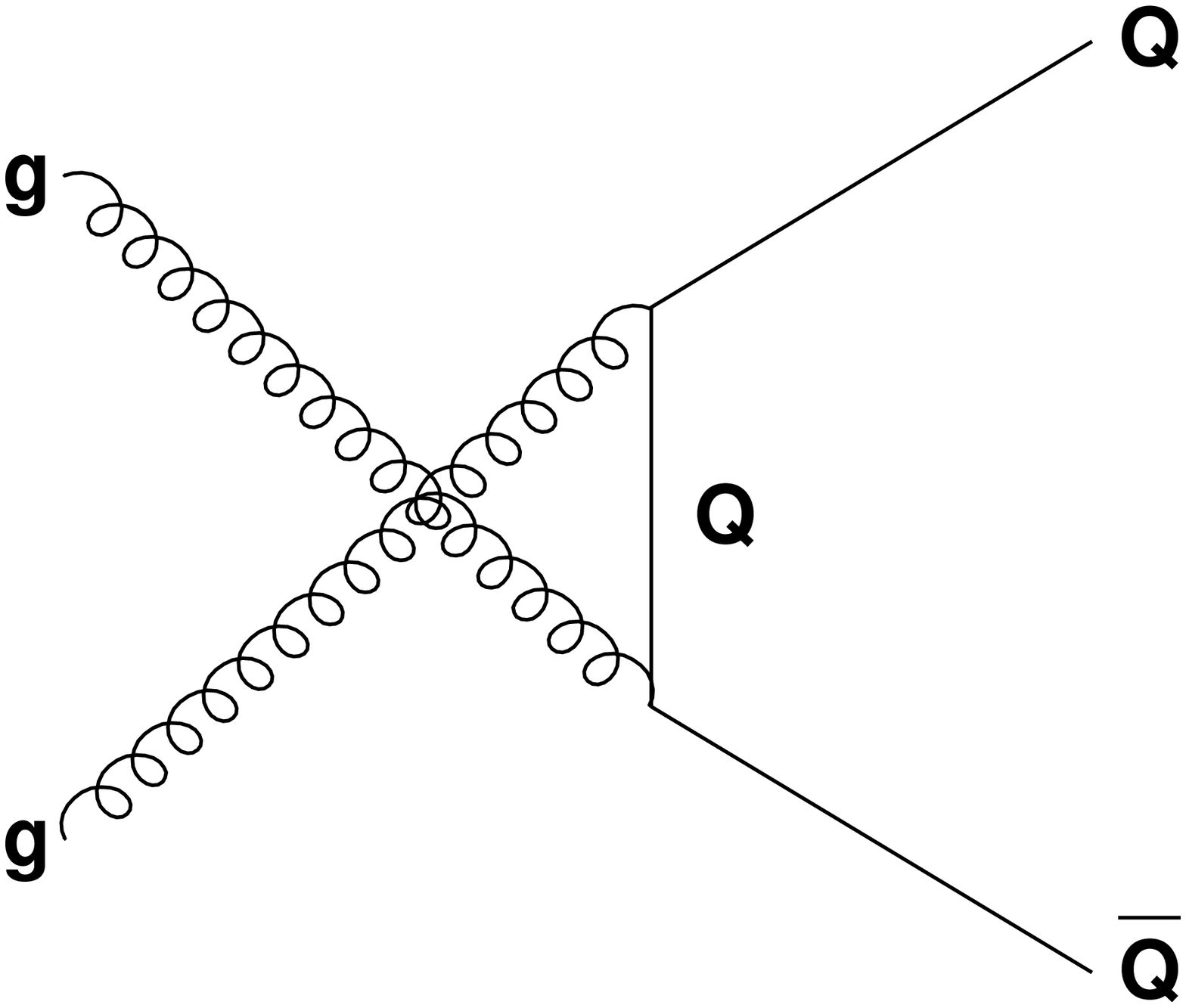}}\\
\end{tabular}
\caption{Heavy flavour production mechanisms at leading order.}
\label{fig:prod}
\end{figure}

At leading order (LO), the only processes which can lead to heavy
flavour production are quark-antiquark annihilation and gluon fusion,
as illustrated in Fig.~\ref{fig:prod}.  In general, pQCD processes can
be factorised into three different parts: the non-perturbative initial
conditions, describing the state \emph{before} the collision takes
place, the hard process itself, perturbatively calculable, and the
subsequent step of hadronisation (also non-perturbative).

The first part is mainly determined by the interacting quarks' and
gluons' fractional momenta, $x_{1,2}=p_{\mathrm{parton}} / p_{1,2}$,
where 1 and 2 stand for the projectile and target nucleons. It also
depends on the squared energy-momentum transfer between the two
partons, $Q^2$, and on the intrinsic transverse momenta,
$k_\mathrm{T}$, which the partons carry inside the projectile (proton,
pion or nucleus). The distributions of the fractional momenta of the
various quarks inside protons and pions were studied in deep inelastic
scattering (DIS) experiments, among others, and were parameterised by
several groups. They will be discussed in the next section.  The
``strength'' of the process is given by the partonic cross-section,
$\hat{\sigma}_{ij}$, which depends, in particular, on the available
energy.  At a given energy, in LO calculations, the partonic
cross-section is determined by the mass of the heavy quark, $m_{\rm
Q}$, and by the strong coupling constant, $\alpha_{\rm S}$, evaluated
at the scale $Q^2$,
\begin{equation}
  \hat{\sigma}_{ij}^{\rm LO}(\hat{s}, m_{\rm Q}^2, Q^2) =
  \frac{\alpha_{\rm S}^2(Q^2)}{m_{\rm Q}^2}\cdot
  f_{ij}^{0,0}(\frac{m_{\rm Q}^2}{\hat{s}}) \quad .
\end{equation}
$f_{ij}^{0,0}$ is a dimensionless scaling function which determines
the energy dependence of the charm or beauty production cross-section,
and which depends on the ratio $m_{\rm Q}^2/\hat{s}$, where $\hat{s}$
is the squared partonic centre of mass energy, $\hat{s} =
x_1x_2s$~\cite{RVogt}.  The indices represent the interacting partons
($ij = q\bar{q}$ or $gg$).

The cross-section to produce a heavy quark pair in a proton-proton
collision, $\sigma^{\rm pp}_{Q\overline{Q}}$, is then obtained by convoluting
the perturbatively calculated partonic cross-section with the
(non-perturbative) parton distribution functions, $f^{\rm p}$, of the interacting
hadrons,
\begin{equation} \label{eq:cross}
\sigma^{\rm pp}_{Q\overline{Q}} = \sum_{i,j} \int{\dd x_1\cdot \dd x_2\cdot
  f_i^{\rm p}(x_1,Q^2) \cdot f_j^{\rm p}(x_2,Q^2)\cdot
  \hat{\sigma}_{ij}(\hat{s})} \quad .
\end{equation}
If the protons are inside nuclei, their partons have modified
distribution functions. In Eq.~(\ref{eq:cross}) the parton
distribution functions should then be represented by
$f^A(x,Q^2)$ instead of $f^{\rm p}(x,Q^2)$, where \emph{A} represents the
mass number of the nucleus.  The implications of this effect on the
total cross-sections will be studied in Sections~\ref{sec:nucl}
and~\ref{sec:nuclData}.

In the last step, the hadronisation, the heavy quark pair fragments
into hadrons, including the neutral and charged D or B mesons.  We
will discuss this step in Section~\ref{sec:frag}.

\subsection{Parton distribution functions}
\label{sec:pdf}

\begin{figure}[ht!]
\centering
\resizebox{.48\textwidth}{!}{%
\includegraphics*{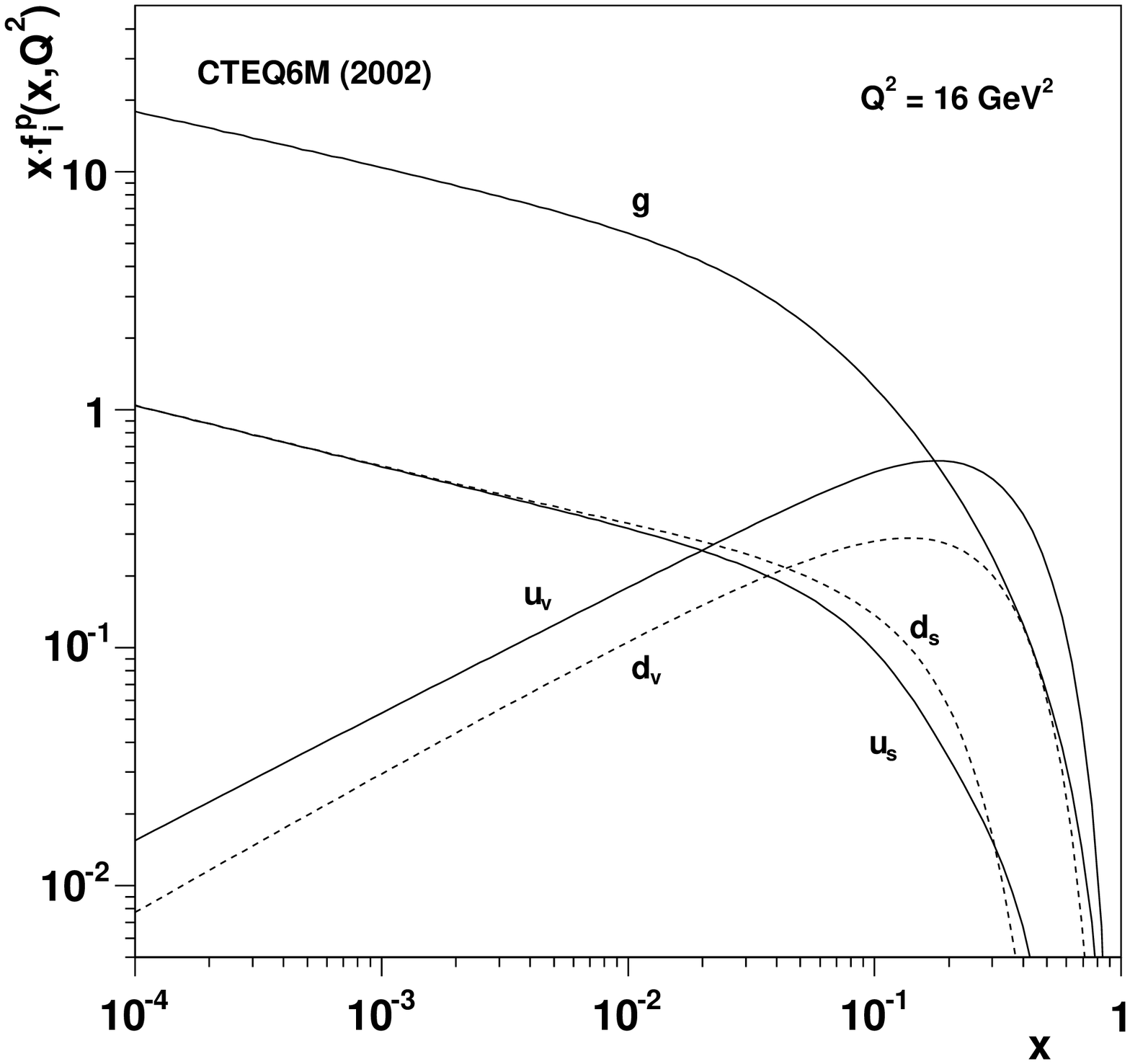}}
\resizebox{.48\textwidth}{!}{%
\includegraphics*{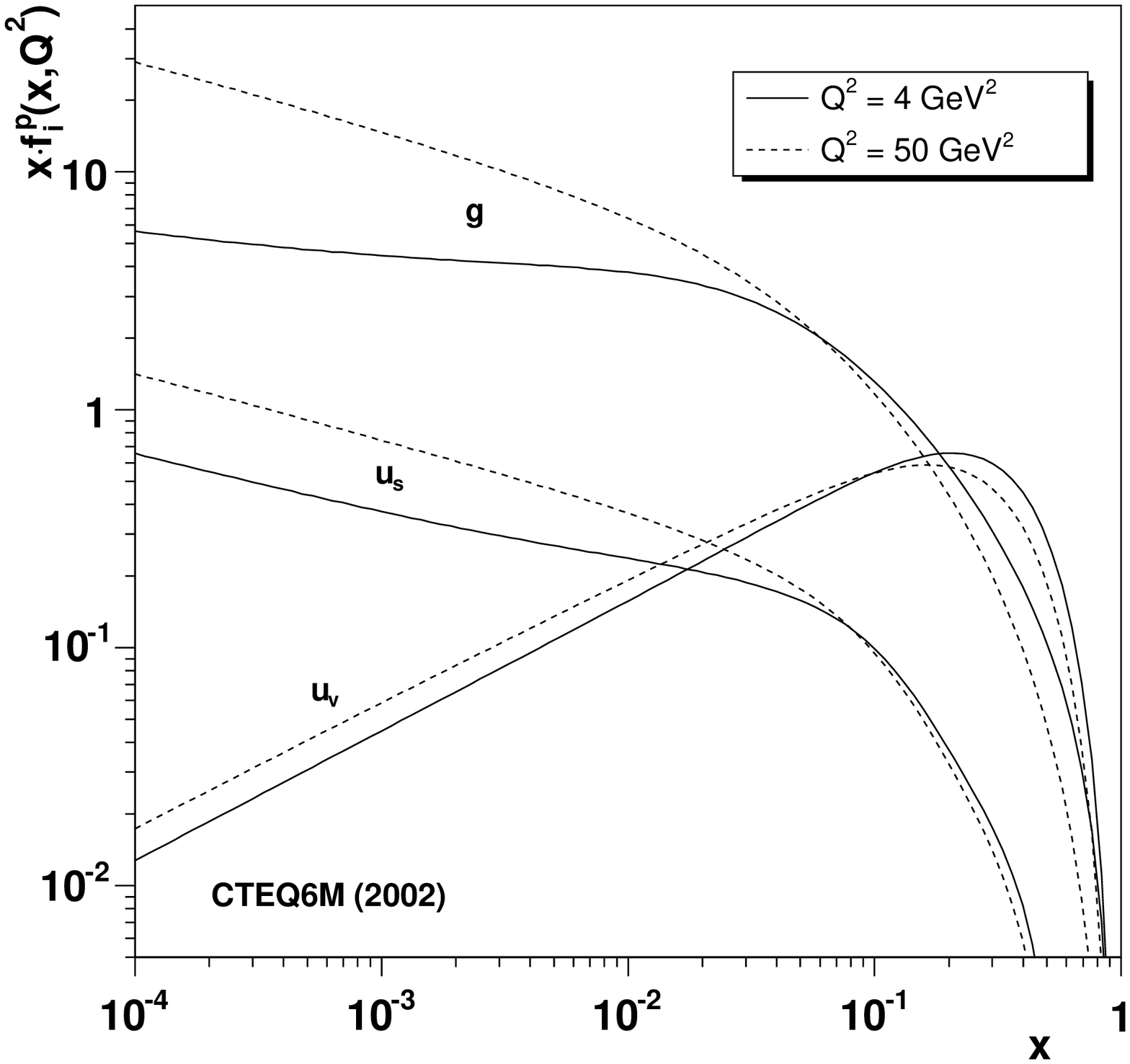}}
\caption{Valence quark ($\rm u_v$, $\rm d_v$), sea quark ($\rm u_s$,
  $\rm d_s$) and gluon (g) distributions inside a proton, evaluated at
  $Q^2$ values relevant for charm and beauty production.}
\label{fig:pdfs}
\vglue2mm
\centering
\resizebox{0.48\textwidth}{!}{%
\includegraphics*{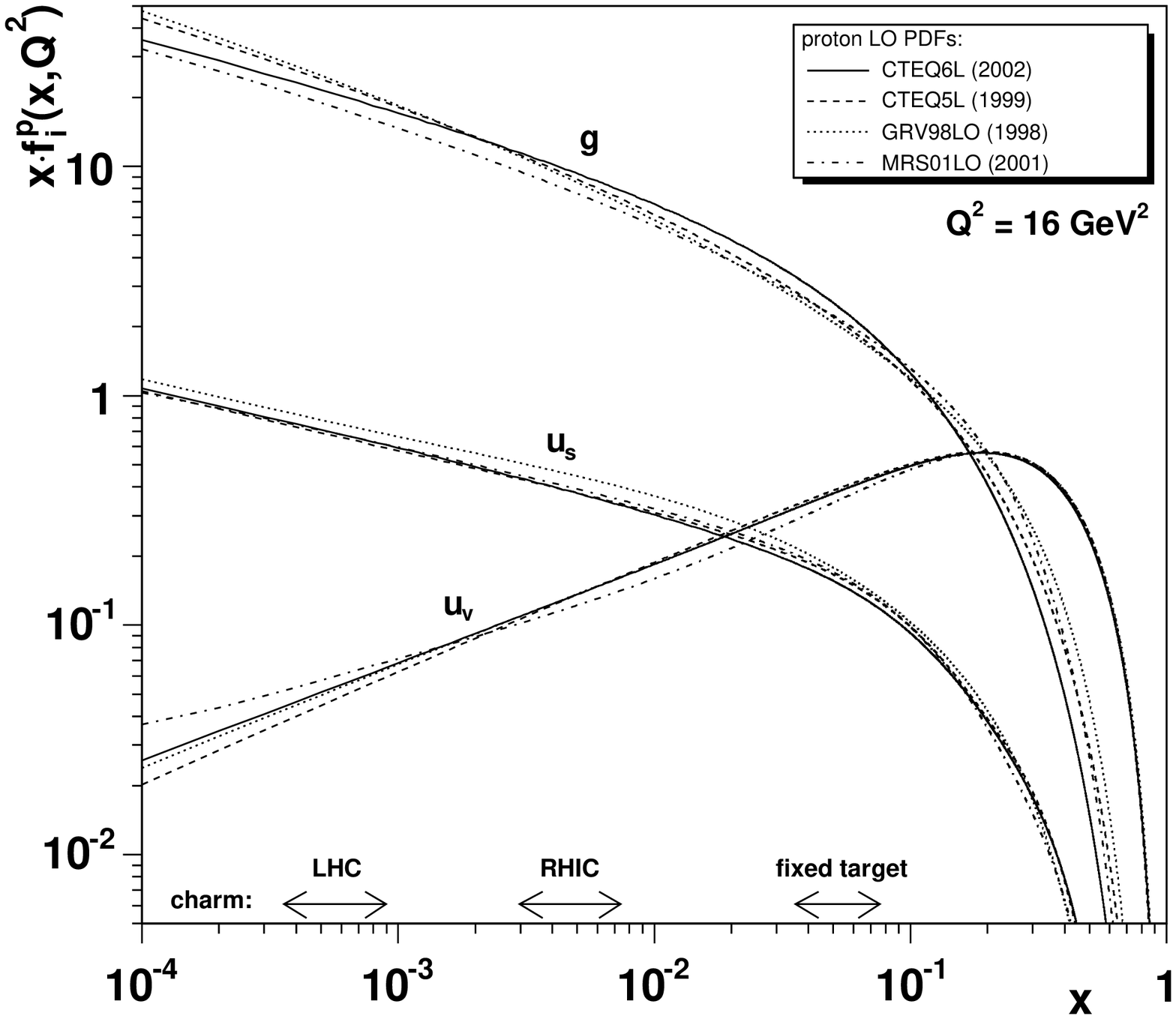}}
\resizebox{0.48\textwidth}{!}{%
\includegraphics*{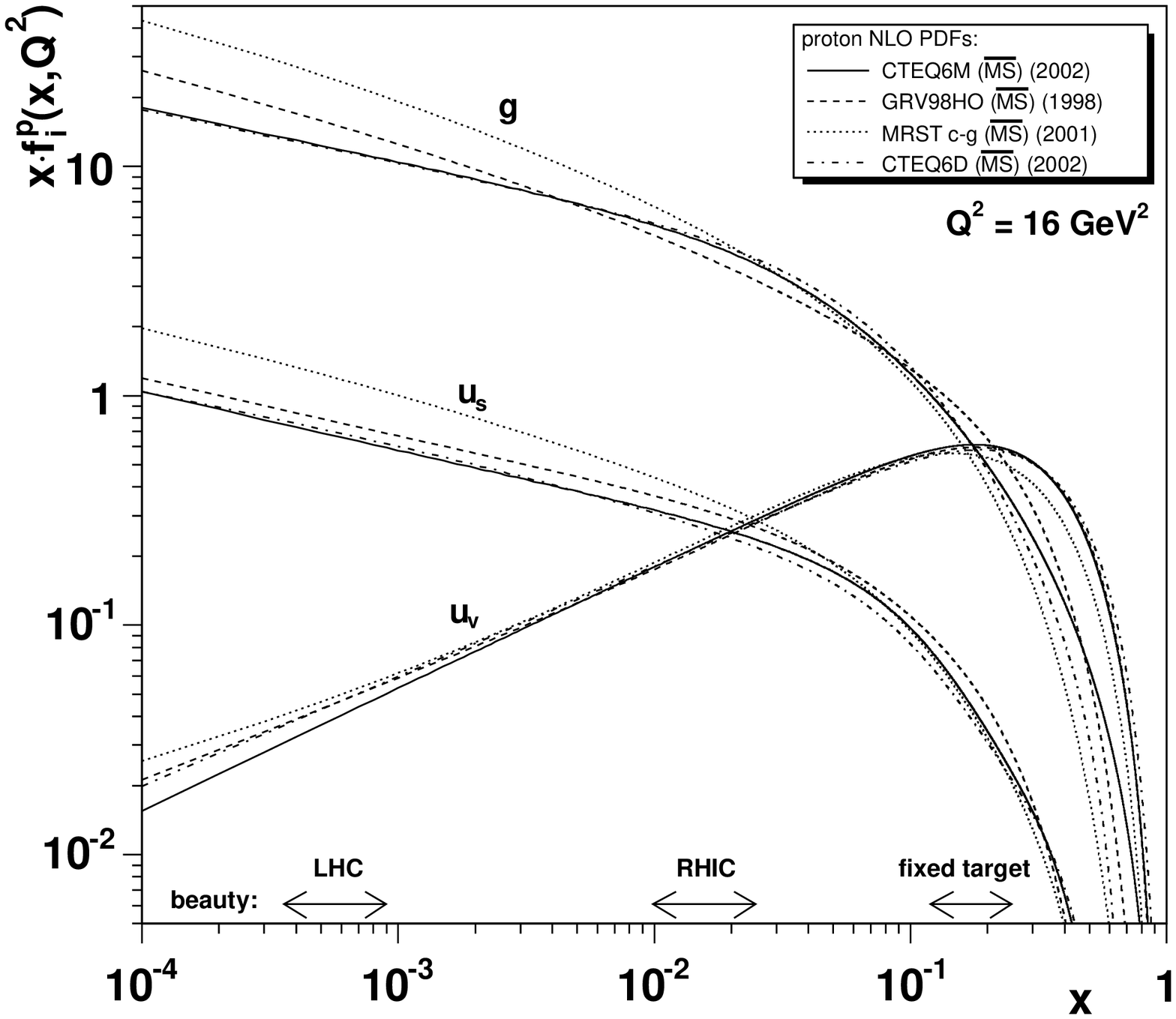}}
\caption{Various sets of proton LO (left) and NLO (right) PDFs.}
\label{fig:CompPdfs}
\end{figure}

Figure~\ref{fig:pdfs} shows the fractional momentum distributions, in
protons, of the valence quarks, sea quarks and gluons.  The valence
quarks usually have relatively high momentum fractions while the sea
quarks and the gluons are mostly found at low $x$ values.
As first observed by NA51~\cite{NA51} and then studied in more detail
by E866~\cite{E866}, the $\bar{\rm u}$ and $\bar{\rm d}$ distributions are not
identical; at $x\sim0.2$ there are almost twice as many $\bar{\rm d}$ than
$\bar{\rm u}$ in the proton sea.  The right panel of Fig.~\ref{fig:pdfs}
shows how the probabilities of finding certain partons at given
momentum fractions change when we probe them in reactions of different
$Q^2$ values.  The higher the energy-momentum transfer, $Q^2$, the
higher the probability that low-$x$ partons participate in the
particle production process.

Later in this report we will calculate the \sqrts\ dependence of
\ccbar\ and \bbbar\ production, using different PDF sets.  In view of
this work, we have upgraded the PDFLIB package~\cite{PDFLIB} to
include the recent CTEQ6~\cite{cteq6} and MRST 2001~\cite{mrst}
parameterisations.

Figure~\ref{fig:CompPdfs} shows different sets of LO and NLO proton
PDFs.  While the four LO sets show a similar behaviour, the NLO
parameterisations show a more significant spread, in particular for
the gluons.  The CTEQ6M parameterisation differs from other sets
mainly in the following two aspects: an extended $\chi^2$ function is
used to fit the data points, including correlated systematic errors,
and new measurements are included, with improved precision and
expanded $(x,Q^2)$ ranges. Among the new data sets, the inclusive jet
cross-section measurements of the D\O\ experiment at Fermilab, giving
access to the \emph{x} range $0.01 < x < 0.5$, are particularly
important, since they have a big impact on the CTEQ6M gluon
distribution functions.  In Ref.~\cite{cteq6} the CTEQ Collaboration
gives a detailed description of their new method, which mainly uses
the $\overline{\rm MS}$ scheme, and outlines the differences with respect
to the MRST~2001 sets.

At the bottom of Fig.~\ref{fig:CompPdfs} we roughly indicate the ranges probed
by fixed target (SPS, FNAL, DESY) and collider (RHIC, LHC) experiments, for
charm and beauty production.  They were evaluated using the expression
$x = M / \sqrt{s} \cdot \exp(y^*)$, with masses of 1.5 and 5~GeV/$c^2$ for
charm and beauty, respectively.  We set $y^* = 0$, where $y^*$ is the
rapidity in the centre of mass frame, and varied the energy within the
ranges of the available experimental measurements.

\begin{figure}[htb]
\centering
\resizebox{0.33\textwidth}{!}{%
\includegraphics*{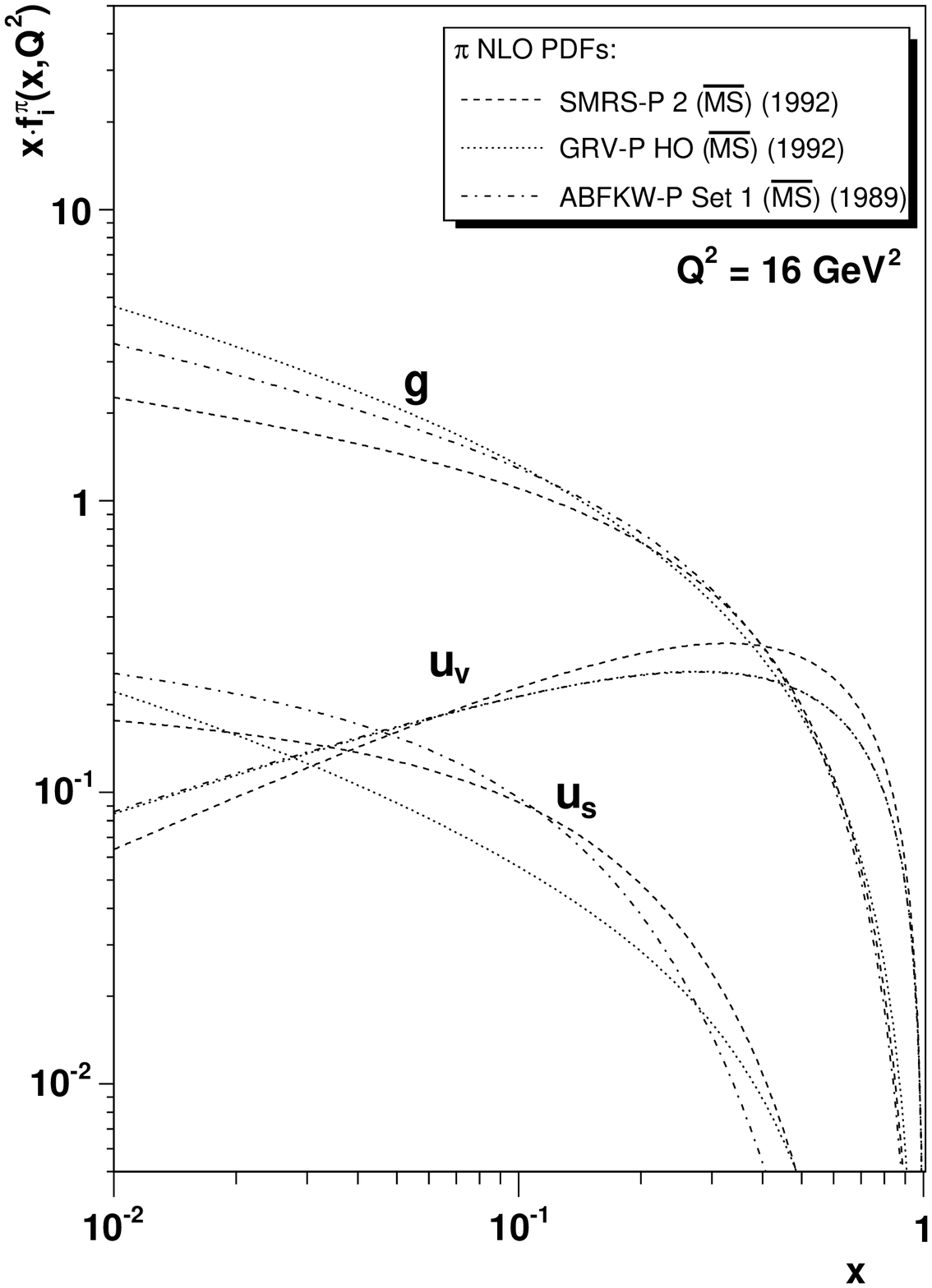}}
\resizebox{0.33\textwidth}{!}{%
\includegraphics*{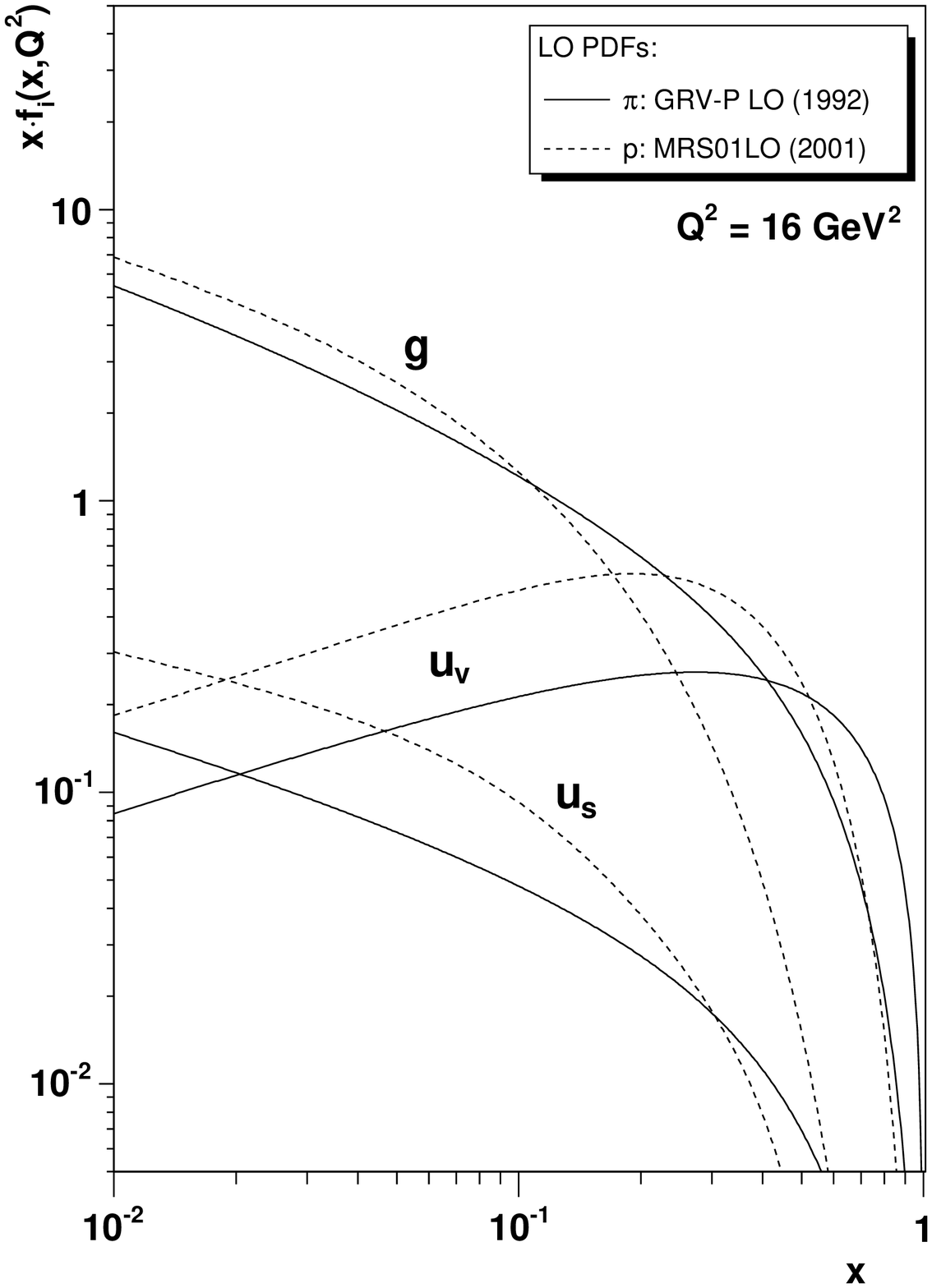}}
\caption{Three different sets of NLO pion PDFS (left) and comparison
  of pion and proton LO PDFs (right).}
\label{fig:pionPDFs}
\end{figure}

In this report we will consider measurements of open charm and beauty
production cross-sections from data obtained with proton and pion
beams.  In the left panel of Fig.~\ref{fig:pionPDFs} we show three
\emph{pion} PDFs, calculated at NLO.  It should be noted that all
available pion PDF sets are more than 10 years old.  These three pion
PDFs are significantly different from each other.  On the right panel
of this figure we compare LO parton distribution functions in a pion
and in a proton.  We can see that the valence quark distributions are
peaked at $x \sim 0.2$ in protons and $x \sim 0.45$ in pions, where
the gluons are much harder.

\begin{figure}[ht!]
\centering
\resizebox{\textwidth}{!}{
\includegraphics*{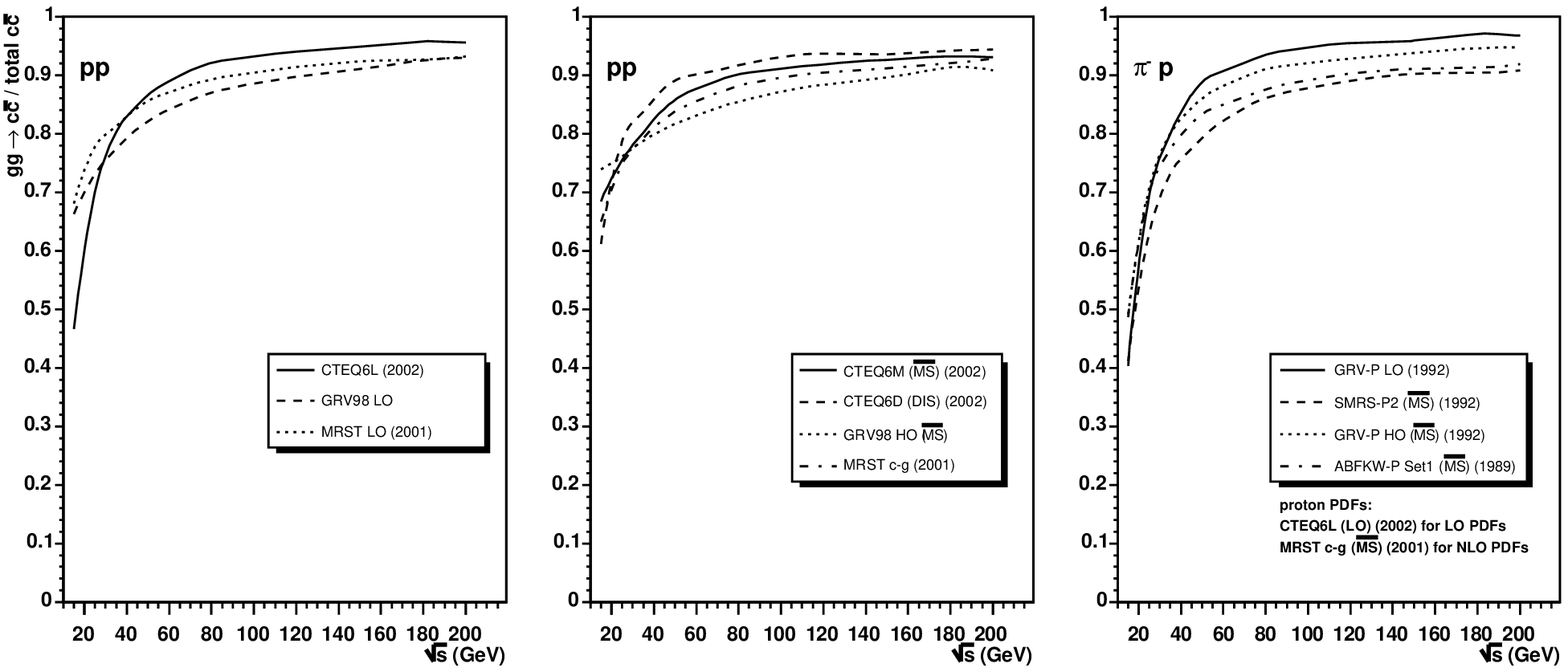}}
\caption{Relative contribution of gluon fusion to the total \ccbar\
production cross-section, as a function of $\sqrt{s}$, in pp (left and
middle) and $\pi^-$p (right) collisions.}
\label{fig:ccFrac}
\centering
\resizebox{\textwidth}{!}{
\includegraphics*{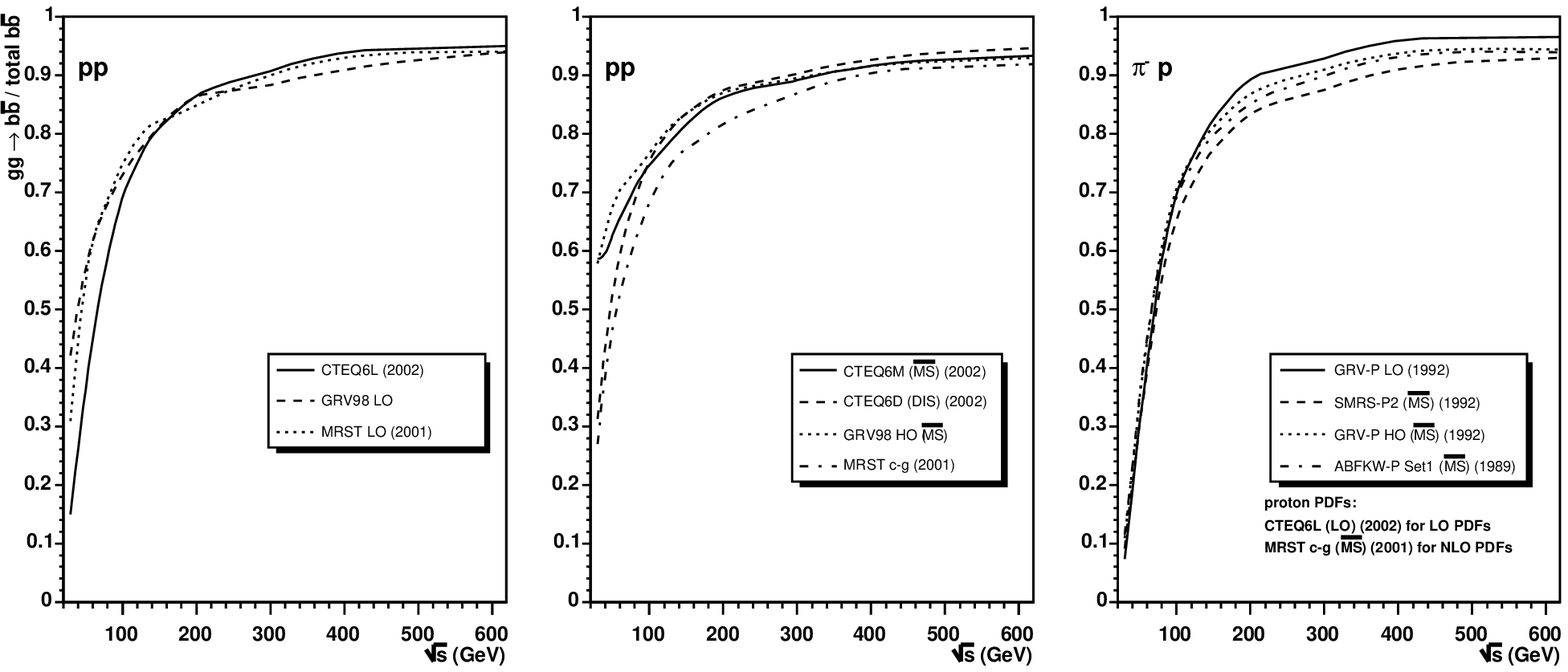}}
\caption{Same as previous figure but for \bbbar\ production, and up to
   higher energies.}
\label{fig:bbFrac}
\end{figure}

It is clear that the use of different PDF sets will change the
calculated total production cross-section of heavy flavour production
(see Eq.~\ref{eq:cross}).  It also influences the relative importance
of gluon fusion and \qqbar\ annihilation.  Figure~\ref{fig:ccFrac}
shows the relative contribution of gluon fusion to the total \ccbar\
production cross-section, as a function of $\sqrt{s}$, as calculated
by the Monte Carlo event generator Pythia~\cite{Pythia6}, for pp and
$\pi^-$p collisions.  The remainder of the total
cross-section is due to $q\bar{q}$ annihilation, since at LO there are
only these two processes.  More details on the calculations will be
given later.  Note that for the $\pi^-$p collisions we used the CTEQ6L
(2002) and MRST~c-g (2001) sets to describe the parton distributions
inside the target \emph{proton}, but other sets of proton PDFs give
similar results.  Most measurements of the \ccbar\ cross-section were
made in the range $200<E_{\mathrm{lab}}<920$~GeV, or
$\sqrt{s}=20$--40~GeV, where the contribution from gluon fusion is
around 80\,\% in pp and around 70\,\% in $\pi^-$p collisions.

In Fig.~\ref{fig:bbFrac} we show the \sqrts\ dependence of the
relative contribution of gluon fusion to the total \bbbar\
cross-section.  The higher values obtained with the CTEQ6M and
GRV98~HO PDFs, at the lowest energies, result from the harder gluon
distributions of these sets.  Comparing Figs.~\ref{fig:ccFrac}
and~\ref{fig:bbFrac} we see that in the energy range of the fixed
target experiments the production of the two heavy flavours is
dominated by different mechanisms: while \qqbar\ annihilation is
responsible for only $\sim$\,20\,\% of the total \ccbar\ production
cross-section, it is the dominant process in \bbbar\ production.  At
higher energies, both charm and beauty production are dominated by
gluon fusion.

\subsection{Nuclear effects in p-A and A-A collisions}
\label{sec:nucl}

The parton distribution functions in the proton, $f^{\rm p}(x, Q^2)$, are
essentially extracted from the structure functions ($F_1$, $F_2$ and
$F_3$) measured in deep inelastic scattering experiments. These
experiments are performed with various nuclear targets, and indicate
that the distributions of partons inside bound protons are different
from those in hydrogen. These nuclear effects are expressed as the
ratio of the PDFs observed in a nucleus with respect to those of a
``free'' proton,
\begin{equation}
R_i^A(x,Q^2) = \frac{f_i^A(x,Q^2)}{f_i^{\rm p}(x,Q^2)}\quad,
\end{equation}
with $i$ representing the valence quarks, the sea quarks, or the
gluons.

\begin{figure}[h!]
\centering
\resizebox{0.5\textwidth}{!}{%
\includegraphics*{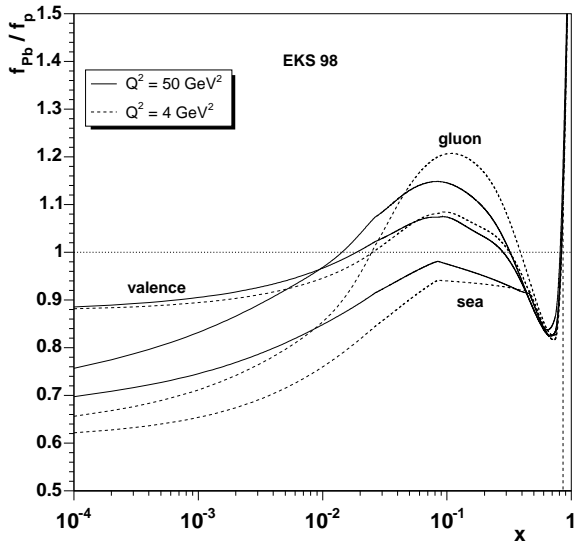}}
\caption{Nuclear modifications of the PDFs, for the Pb nucleus,
  according to the EKS~98~\cite{eks} weight functions.}
\label{fig:nwf}
\end{figure}

Figure~\ref{fig:nwf} shows this ``nuclear weight function'', for the
Pb nucleus, as a function of $x$, according to the EKS~98~\cite{eks}
parameterisation.  The curves are shown for two values of $Q^2$.  A
detailed discussion of the dependence of the nuclear effects on $Q^2$,
among many other related issues, can be found in Refs.~\cite{eks,ekv}.
The interested reader will find in Ref.~\cite{yellow-report-npdfs} a
recent and detailed review of nuclear parton distribution functions.

In different regions of $x$, the nuclear effects are traditionally
referred to by the following expressions:
\begin{itemize}
\parskip=0.pt \parsep=0.pt \itemsep=0.pt
\item \emph{shadowing}: at low \emph{x}, where $R^A(x,Q^2) < 1$
\item \emph{anti-shadowing}: at medium \emph{x}, where $R^A(x,Q^2) >
1$
\item \emph{EMC effect}: at relatively high \emph{x}, where
  $R^A(x,Q^2) < 1$
\item \emph{Fermi motion}: at the highest \emph{x}, where
$R^A(x,Q^2) > 1$
\end{itemize}

\begin{figure}[ht]
\centering
\resizebox{0.5\textwidth}{!}{%
\includegraphics*{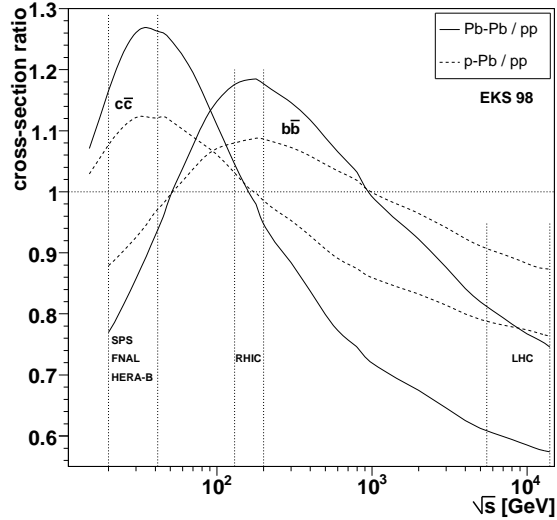}}
\caption{Changes induced on the \ccbar\ and \bbbar\ cross-sections by
  the nuclear modifications of the PDFs, at mid-rapidity.}
\label{fig:nucl-ccbb}
\end{figure}

The impact of these nuclear effects on heavy flavour production can be
seen in Fig.~\ref{fig:nucl-ccbb}, which shows the ratio between the
heavy flavour (\ccbar\ and \bbbar) production cross-sections
calculated for pp collisions taking into account that those protons
are inside Pb nuclei (p-Pb, Pb-Pb) and the same cross-sections
calculated for pp collisions in the vacuum.  According to the EKS~98
parameterisation, and in what concerns charm production, the
experiments carried out at the SPS and FNAL energies are in the
anti-shadowing regime.  Therefore, inside a heavy nucleus the parton
distributions are expected to be harder than in a ``free'' proton,
leading to higher production cross-sections in p-A and A-A collisions
with respect to the linear scaling from pp collisions.  For instance,
at $\sqrt{s}\sim$\,30--40~GeV, the \ccbar\ production cross-sections
in p-Pb collisions should be 10\,\% higher than in the absence of
gluon anti-shadowing.  The RHIC experiments, at mid-rapidity, are just
in the $x$ range where the nuclear effects on charm production change
from the anti-shadowing to the shadowing region.  Therefore,
measurements in the central detectors should not be very sensitive to
nuclear effects on the parton distribution functions.  However, this
is no longer the case for the detectors placed away from mid-rapidity.
At $\sqrt{s}=200$~GeV, and for charm production, we have the same gluon
anti-shadowing ($\sim$\,15\,\%) at $y=-2.0$ as we have at the SPS
($\sqrt{s}=27.4$~GeV) at mid-rapidity.  In the case of d-Au
collisions, there is a significant difference between the expected
nuclear effects on the PDFs in the ``North'' and ``South'' muon arms
of Phenix, for instance.  In the ``forward hemisphere'' (with respect
to the d beam), at $y=+2.0$, there is $\sim$\,20\,\% \emph{shadowing}
effects on charm production, instead of the $\sim$\,15\,\%
\emph{anti-shadowing} expected on the ``backward'' side.  It is crucial,
hence, to keep the PHENIX charm and charmonia d-Au analyses
independent for each of the three covered rapidity ranges.
In what concerns beauty, the nuclear effects
are expected to influence the production cross-sections in opposite
ways when going from measurements done in the energy range of
fixed-target experiments (EMC region) to those done at RHIC
(anti-shadowing region), with the nuclear cross-sections changing from
suppressed to enhanced, with respect to the linear extrapolation from
pp collisions.  At the LHC energies, we will certainly be in the
shadowing region, both for charm and for beauty.  Charm production in
the \mbox{Pb-Pb} collision system, for instance, is expected to be
suppressed by around 40\,\% with respect to a linear extrapolation of
nucleon-nucleon collisions.

It should be noted, however, that there are no measurements today
which can constrain the nuclear \emph{gluon} distribution function.
The nuclear gluon densities provided by the EKS98 parameterisation are
only indirectly constrained, through the scale evolution of $F_2^{\rm
A}(x, Q^2)$ and through momentum conservation.  Accurate measurements
of open charm production in proton-nucleus collisions, using
several nuclear targets and over a broad range of energies, 
would be crucial to significantly reduce the
present uncertainties on the nuclear gluon densities~\cite{ekv}.

\subsection{Fragmentation}
\label{sec:frag}

In the hadronisation step, the outgoing heavy quarks fragment into
hadrons.  The energy carried by the formed hadron with respect to the
quark's energy, $z = E_{H}/E_{Q}$, is distributed according to the
fragmentation function, $D_{Q}^{H}(z)$, measured in $e^+e^-$ reactions
and \emph{assumed} to be the same in hadronic collisions.
We should note that this definition of $z$ is not unique; in
theoretical studies it is more common to use the lightcone fraction,
where $E$ is replaced by $E+p_{\rm L}$, with the longitudinal momentum
defined by the quark's direction.
Light quark (u,d,s) fragmentation is usually parameterised as
$D_{q}^{h}(z) \propto z^{-1}(1-z)^n$ while heavy quarks only
experience a relatively small deceleration when combining with a
(slow) light quark.
In 1983, Peterson \emph{et al.}~\cite{peterson} proposed the following
heavy quark fragmentation function
\begin{equation}
D_Q^H(z) \propto \frac{1}{z[1-(1/z)-\epsilon_Q/(1-z)]^2} \quad,
\end{equation}
which peaks at $z\approx 1 - \sqrt{\epsilon_Q}$.
In principle, $\epsilon_Q$ is fixed by the light and heavy quark
masses, $\epsilon_Q=(m_q/m_Q)^2$, but in practice it is a free
parameter, usually taken to be 0.06 for charm and 0.006 for beauty.

Alternatively, the Lund string fragmentation scheme~\cite{lund} leads
to the expression
\begin{equation}
D_Q^H(z) \propto \frac{(1-z)^a}{z^{1+b\, m_Q^2}} \cdot \exp({-b\, m_{\rm
T}^2/z}) \quad,
\end{equation}
which is sensitive to the \pt\ of the produced hadron, through the
transverse mass.  It implies a harder fragmentation function for
heavier quarks, through the explicit mass dependence. The term
$z^{-b\, m_Q^2}$ was introduced by Bowler~\cite{bowler} in 1981, to
improve the agreement with the $e^+e^-$ B meson data available at that
time.

\begin{figure}[htb]
\centering
\resizebox{0.65\textwidth}{!}{%
\includegraphics*{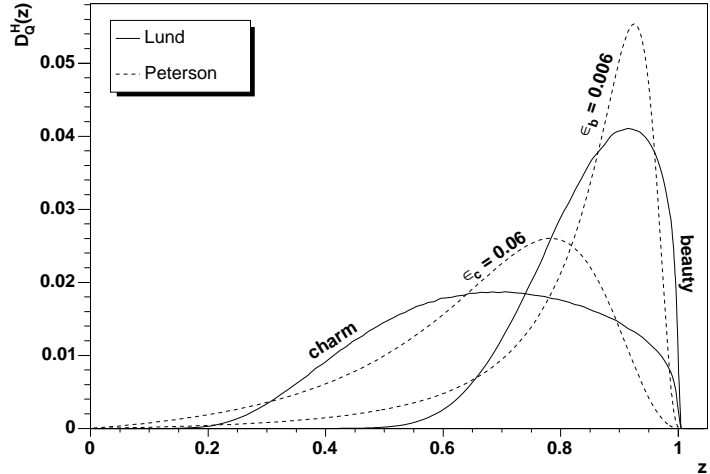}}
\caption{Heavy flavour fragmentation functions according to the
  Peterson~\cite{peterson} and Lund string~\cite{lund} expressions.
  All curves are normalised to unity.}
\label{fig:fragFF}
\end{figure}

Figure~\ref{fig:fragFF} shows both fragmentation functions, for charm
and beauty hadrons.  The Lund model curve was drawn with $a=0.3$,
$b=0.58$~GeV$^{-2}$, $m_{\rm c}=1.5$~GeV/$c^2$ and $m_{\rm b}=4.8$~GeV/$c^2$
(values used by the Pythia event generator~\cite{Pythia6}), and with
an exponential \pt\ spectrum, $\dd N/\dd p_{\rm T} \propto \exp(-5
\cdot p_{\rm T})$.

\section{Heavy flavour experiments}
\label{sec:exp}

Most of the experiments considered in this report were designed to
study the properties of charm and/or beauty hadrons.  To select events
with charm or beauty particles, the experiments used high-resolution
detectors in the target region, to observe primary and secondary
vertices, signaling the decay of the heavy flavoured hadrons.  Three
types of vertex detectors have been commonly used: bubble chambers,
emulsions and silicon tracking telescopes.  The spacial resolution of
the silicon detectors is not as good as that of the other systems, but
they can be operated at much higher interaction rates, a crucial
feature when looking for rare processes.  Most of the experiments had
particle tracking devices and a muon spectrometer.  In addition, the
experiments which measured the charm (or beauty) hadrons in hadronic
decay channels also had electromagnetic and/or hadronic calorimeters,
and particle identification detectors, such as \v{C}erenkov counters, to
distinguish pions, kaons and protons.  Furthermore, many of them, in
particular those which studied beauty production, implemented triggers
to enrich their collected event sample with charm and/or
beauty events.  With only one exception (E789), the fixed target
experiments could detect the charm or beauty hadrons in the full
forward hemisphere.  Some experiments had \emph{active} targets, where
the vertex detector itself was used as target.

The heavy-flavour hadro-production experiments we consider in this
report are summarised in Table~\ref{tab:overview}, and described in
some detail in the next pages, roughly in chronological order.

\begin{table}[h]
\centering
\begin{tabular}{|c|l|}\hline
Flavour & \hfil Experiment\hfil \\ \hline
charm & NA11, NA16, NA27, E743, NA32, WA75, WA82, E769, E791, E706 \\
beauty & NA10, WA78, UA1, E706/E672, E771 \\
both & E653, E789, WA92, CDF, HERA-B \\ \hline
\end{tabular}
\caption{Heavy-flavour hadro-production experiments considered in this
  report.}
\label{tab:overview}
\end{table}

\subsection{NA16, NA27 and E743: the LEBC Experiments}

The purpose of NA16 was the ``study of hadronic production and
properties of new particles with a lifetime $10^{-13} < \tau <
10^{-10}$~s, using LEBC-EHS''. It used the high-resolution hydrogen
bubble chamber ``LEBC'' (Lexan bubble chamber) and a prototype version
of the European Hybrid Spectrometer, ``EHS''~\cite{na16-nim}.  LEBC
was a rapid cycling liquid hydrogen bubble chamber, with a fiducial
volume of $12\times5\times2.5$~cm$^3$ photographed by two cameras.  It
served both as a liquid-hydrogen target and as a high-resolution
vertex detector.  The direct observation of the production and decay
vertices is one of the key features of this apparatus.  The decay
products were analysed downstream in the EHS spectrometer, which could
detect photons but had very limited particle identification
capabilities.  The acceptance for D mesons covered the positive \xf\
range and was independent of the observed decay mode. The data samples
were collected in the late seventies, at the CERN SPS, with 360~GeV
proton and $\pi^-$ beams~\cite{na16}.

NA27 was built ``to measure accurately the lifetime of the $\D^0$,
$\D^{\pm}$, $\rm F^{\pm}$, $\Lambda_c$ charm particles and to study
their hadronic production and decay properties''
(note that the $\rm F^{\pm}$ is now named $\D_{\rm s}^{\pm}$). 
It used the final version of the EHS spectrometer~\cite{na27-nim},
composed of three parts and extending over more than 40 metres.
Immediately downstream of LEBC were placed two wire chambers,
complemented by two small drift chambers, for track reconstruction and
triggering purposes.  The trigger simply required more than two hits
in each of the wire chambers.  Each of the two other parts had a
magnet and three large drift chambers, leading to a momentum
measurement with a relative resolution better than 1\,\% up to
250~GeV/$c$.  Electron and photon detection were provided by two
lead-glass electromagnetic calorimeters.  A hadron calorimeter was
also available.  Charged particle tracking and identification were
essentially performed by a 40~m$^3$ drift chamber, ISIS, through up to
320~$\dd E/\dd x$ measurements, complemented by two \v{C}erenkov detector
systems and a transition radiation detector.  In the 400~GeV proton
run, a total of 98 neutral and 119 charged D mesons were
found~\cite{na27p}.  Previously, NA27 had a $\pi^-$ run at 360~GeV,
with lower statistics~\cite{na27pi}. The data samples were taken in
the early eighties, at the CERN SPS.

Some of the CERN ISR experiments studied charm production indirectly,
by triggering on single electrons. The deduced cross-sections were
between ten and hundred times higher than those observed in the LEBC
experiments.  Since the 800~GeV proton beam of Fermilab provided
collisions with an energy half way between the SPS and the ISR data,
an experiment was proposed, E743, to investigate the discrepancy
between the previous measurements. The bubble chamber LEBC was
transported to Fermilab and complemented with a multi-particle
magnetic spectrometer, MPS, that had \v{C}erenkov and transition radiation
detectors for particle identification, besides proportional wire
chambers for tracking.  The interaction trigger was provided by two
proportional wire chambers placed just downstream of LEBC.  The
experiment collected data in 1985.  Like NA16 and NA27, also E743
identified the charm mesons in topological decays, by observing the
charge of the decaying particle and a given number of charged final
state particles~\cite{e743}.  The new measurement (10 neutral and 46
charged D mesons) agreed with the results of the previous SPS LEBC
experiments.

\subsection{NA10}

NA10 was designed to perform a ``high resolution study of the
inclusive production of massive muon pairs by intense pion beams'' and
took data in the early eighties, at the SPS~\cite{na10-nim}.  The muon
spectrometer, separated from the target region by a 5~m long carbon
muon filter, was composed of eight multi-wire proportional chambers
with three tracking planes each, and four trigger hodoscopes made of
plastic scintillator slabs, separated in two telescopes by an air core
toroidal magnet.  The last trigger hodoscope was protected by a 1.2~m
iron wall, placed after the tracking chambers, to ensure a clean
dimuon trigger without deteriorating the reconstruction of the muon
trajectories.  The highly selective dimuon trigger, optimised for
masses above 3~GeV/$c^2$, allowed to run at a beam intensity of
$\sim$\,$1.5 \times 10^9$ pions/burst.  The study of \BBbar\
production in $\pi^-$-W interactions was based on the selection of
events with three high \pt\ muons in the final state, coming from the
semi-muonic decays of both B mesons and from the semi-muonic decay of
one of the D mesons.  Beauty production cross-sections were
given~\cite{na10} for incident beam energies of 140, 194 and 286~GeV,
the largest statistics ($\sim$\,14 signal events) corresponding to the
highest energy.

\subsection{WA78, WA75}

WA78 was proposed to ``search for the hadroproduction of \BBbar\
pairs'' and took place in the early eighties, at the
SPS~\cite{wa78-nim}.  It followed a similar strategy as NA10, 
looking at three muons in the final state or at
like-sign muon pairs. In addition to the muon spectrometer, consisting
of drift and multi-wire proportional chambers surrounding a 1.5~T
superconducting 
dipole 
magnet, WA78 had extensive calorimetry.  Because of
the large mass difference between beauty and charm mesons, muons
produced in the B decay have larger \pt\ and are accompanied by more
energetic neutrinos than those produced by charm decays. The trigger
and event selection procedures were, therefore, designed to select
events with at least two high-\pt\ muons and large missing energy. The
final event samples, collected with a 320~GeV $\pi^-$ beam incident on
an U target, contained both tri-muon events and like-sign muon
pairs~\cite{wa78}.

The same muon spectrometer had previously been used by
WA75~\cite{wa75}, which collected a few hundred charm events (and
\emph{one} \BBbar\ event).  The online event selection required at
least one high-\pt\ muon, to enhance the fraction of events with
semi-leptonic decays of heavy flavour particles.  The primary and
secondary vertices were located in the emulsion target, within the
volume indicated by the tracks reconstructed with the silicon
microstrip planes.  A total of 339 events were observed with the
identified muon among the tracks of the decay vertices, presumably due
to charmed particle semi-leptonic decays.

\subsection{NA11 and NA32: the ACCMOR Experiments}

The ACCMOR Collaboration started their charm physics program in 1980,
as NA11, taking $\pi^-$-Be data at 120, 175 and
200~GeV~\cite{na11-stage1}, at the SPS.  Their large acceptance
forward magnetic spectrometer included two magnets, four sets of drift
chambers (a total of 48 planes), a complex system of multi-wire
proportional chambers, five \v{C}erenkov counters and Pb-scintillator
electromagnetic calorimeters.  The data samples were collected with a
single electron trigger.  In a second stage, a vertex telescope of
high resolution silicon microstrip detectors was added to the setup,
helping to reduce the background levels.  The statistics of the charm
event sample, however, remained small~\cite{na11-1,na11-2}.

In 1984, the NA32 experiment took over, to ``investigate charm
production in hadron interactions using high-resolution silicon
detectors''.  The ACCMOR spectrometer~\cite{na32-detector} was
complemented with a finely segmented active silicon target, made of 14
planes of 20~$\mu$m pitch silicon microstrip detectors, 280~$\mu$m
thick, preceeded by a silicon beam telescope and followed by two
silicon multiplicity counters.  Forward going particles were tracked
in 7 planes of 20~$\mu$m pitch silicon microstrip detectors, before
entering the ACCMOR spectrometer.  The data samples were taken with an
interaction trigger, defined using signals from the active target.
Most of the statistics came from runs with a 200~GeV hadron
beam~\cite{na32-88} consisting of $\pi^-$ and K$^-$, separately
identified by means of two threshold \v{C}erenkov beam counters.  Much
fewer events were also collected with a 200~GeV proton beam, allowing
the experiment to publish cross-sections for three different beam
particles.

The second stage of the NA32 experiment was performed in 1985/86 with
a 2.5~mm copper target, placed in vacuum, an improved vertex detector,
including CCDs of $22\times22~\mu$m$^2$ pixels, and a two-level
trigger to select $\rm \Lambda_c$ and $\D_{\rm s}$
decays~\cite{na32-91}.  In these runs, the 230~GeV hadron beam
contained 96\,\% $\pi^-$ and 4\,\% K$^-$ mesons.

\subsection{E653}

The E653 experiment~\cite{e653-nim} was designed to ``study charm
and beauty using hadronic production in a hybrid emulsion
spectrometer''.  It was the first experiment designed to measure both
charm and beauty hadrons, and took data in 1985 (with protons) and
1987 (with pions), at Fermilab.  It used a 1.47~cm long emulsion
target with $\rm A=26.6$ as the average nuclear mass number.  This
allowed to measure the primary vertex and at least one decay vertex
still within the emulsion volume. The trigger, which was optimised to
select \emph{semi-muonic} decays of charm particles, required an
interaction in the target and a high \pt\ muon candidate.  Tracking
started with a 18-plane silicon microstrip vertex detector and
continued with a magnetic spectrometer composed of a dipole magnet and
55~drift chamber planes.  After a 5~m long steel absorber was the muon
spectrometer, made of 12~drift chamber planes on each side of an iron
toroidal magnet.  The primary vertex was located visually, and the
muon trajectory was compared to the tracks in the emulsion in order to
find a good match.

Cross-sections for neutral and charged D meson production were
obtained with 800~GeV protons~\cite{e653p} and 600~GeV
pions~\cite{e653pi}, essentially on the basis of semi-muonic decays,
with a small contribution from purely hadronic decay channels.  For
the beauty study~\cite{e653B}, the 600~GeV pion data sample was
carefully re-analysed to look for additional vertices.  As before, the
emulsion analysis procedure selected events with muonic secondary
vertices of large muon transverse momentum, before proceeding with the
search for other decay vertices. 9~events were found where both the B
and the $\rm \overline{\B}$ decay vertices could be identified and the
whole decay chain reconstructed.

\subsection{E769, E791}

The open geometry TPL (``tagged photon laboratory'') spectrometer, at
Fermilab, after being used by E516 and E691 to measure charm
photo-production, was ``inherited'' by the E769 experiment to study
``pion and kaon production of charm and charm-strange states'', in
1987/88.  The E769 detector included an 11-plane silicon microstrip
vertex detector, 2 analysing magnets, 35 drift chambers, 2 multi-wire
proportional chambers, and 2 segmented threshold \v{C}erenkov counters to
identify kaons, pions and protons.  It also included electromagnetic
and hadronic calorimeters, a wall of scintillation counters for muon
identification, and a high-rate data acquisition system.  E769 used a
mixed 250~GeV secondary beam of both charges.  A much smaller data
sample was also collected with a 210~GeV negative beam.  The
composition of the negative beam was 93\,\% $\pi^-$, 5\,\% K$^-$ and
1.5\,\% $\bar{\rm p}$, while the positive beam was 61\,\% $\pi^+$,
4.4\,\% K$^+$ and 34\,\% p. The beam-particle identification was
provided by a differential \v{C}erenkov counter complemented, in the case
of the positive beam, by a transition radiation detector.  Eight
proportional wire chambers and two silicon microstrip planes were used
to track the beam.  This allowed the E769 Collaboration to study D
meson production in pion, kaon and proton induced
collisions~\cite{e769}.  The mixed hadron beam collided on a multifoil
target, consisting of 250~$\mu$m thick Be, Al and Cu foils, as well as
100~$\mu$m thick W foils, interspaced by 1.6~mm. In total, 26~target
foils were used, giving a total target thickness of 2\,\% of an
interaction length.  Having different target materials simultaneously
in the beam allowed the experiment to measure the nuclear dependence
of open charm production in pion induced collisions.

E791 was approved to study ``hadroproduction of heavy flavours at
the tagged photon laboratory''.  With respect to the E769 detector, the
E791 experiment increased the number of silicon microstrip planes in
the vertex telescope, added a second scintillator wall in the muon
identifier, and implemented a faster read-out system.  The 2\,\%
interaction length target was made of one 0.52~mm thick platinum and
four 1.56~mm thick diamond disks, interspaced by 1.53~cm.  This rather
large spacing ensured that the decay of a charm hadron would occur
\emph{between} the target foils.  It collected 88\,990~neutral D
mesons in 1991/92, with a pure $\pi^-$ beam of 500~GeV~\cite{e791}.

\vspace{-3mm}
\subsection{UA1}

The UA1 detector was built to find the intermediate vector bosons in
\ppbar\ collisions, using the SPS in a collider mode.  It was
basically composed of a cylindrical drift chamber and an
electromagnetic calorimeter immersed in a dipole magnetic field,
surrounded by a hadron calorimeter and a 8-layer muon detector.
``End-cap'' electromagnetic and hadronic calorimeters were installed in
the forward directions, giving the detector an excellent hermetic
coverage.  The beauty production cross-section measurement~\cite{ua1}
was performed in the 1988 and 1989 runs, at $\sqrt{s} = 630$~GeV, when
the central electromagnetic and forward calorimeters were removed in
preparation for the installation of new detectors.  The muon detection
system was improved by the addition of iron shielding in the forward
region. Muon trigger processors selected tracks in the muon chambers
pointing back to the interaction region. At high luminosity, the muon
trigger rate in the forward region was further reduced by requiring a
jet of transverse energy greater than 10~GeV in coincidence with the
muon trigger.  The search for beauty hadrons was performed in four
independent decay channels: ${\rm b}\to \mu X$, ${\rm b}\to {\rm J}/\psi (\to \mu^+
\mu^-) X$, ${\rm b}\to {\rm c} (\to \mu X)\, \mu X$ and ${\rm b} \to
\mu X ; \bar{\rm b} \to
\mu X$.  Cross-sections were measured for each of these processes,
each channel covering different ranges in \pt, from which B hadron and
b quark cross-sections were inferred. The combined cross-section was
then extrapolated to full phase space.

\vspace{-3mm}
\subsection{E672 and E706}

E706 was designed to perform ``a comprehensive study of direct
photon production in hadron induced collisions'', at Fermilab.  The
detector complemented a large acceptance liquid argon calorimeter,
containing a finely segmented electromagnetic section and a hadronic
section, with a charged particle tracking system composed of silicon
microstrip detectors, a large aperture dipole magnet, proportional
wire chambers and straw tube drift chambers.  The experiment collected
events triggered by high transverse momentum showers detected in the
electromagnetic calorimeter.  This requirement enhanced the fraction
of selected events containing charm by nearly an order of magnitude,
compared to a minimum bias trigger.  The measurements~\cite{e706},
restricted to charged D mesons, were performed in 1990, using a
negative 515~GeV beam, primarily composed of pions with a small
admixture of kaons ($<5\,\%$), not separated.  Two $780~\mu$m thick
copper targets were followed by two beryllium cylinders, 3.71 and
1.12~cm long.

Downstream of the E706 apparatus, 20~m away from the target and
protected by a steel wall to absorb most of the hadrons, was placed
the E672 muon spectrometer, aimed at ``studying hadronic final
states produced in association with high-mass dimuons''.  The E672 muon
spectrometer was composed of six proportional wire chambers, a
toroidal magnet and two scintillator hodoscopes, besides iron and
concrete shielding, to provide a clean dimuon trigger.  In 1990, the
E706 and E672 Collaborations joined efforts to study beauty production
in $\pi^-$-Be collisions at 515~GeV~\cite{e672}, using \jpsi\ mesons
coming from secondary vertices to tag the beauty candidates.  The
trigger selected dimuons in the proximity of the \jpsi\ mass; they had
a mass resolution of 68~MeV and an average vertex resolution of 14 and
350~$\mu$m, in the transverse and longitudinal coordinates,
respectively.

\subsection{E789}

E789 was proposed to ``measure the production and decay into
two-body modes of b-quark mesons and baryons'', and took data in
1990/91.  E789 upgraded the spectrometer previously used by the E605
and E772 experiments by adding a vertex telescope, made of 16 planes
of 50~$\mu$m pitch microstrip silicon detectors and placed between 37
and 94~cm downstream of the target, to identify the decays of neutral
D mesons.  Data samples were taken with either beryllium or gold
targets.  A vertex processor selected (on-line) track pairs consistent
with decay vertices at least 1.02~mm downstream of the target and
impact parameters of at least 51~$\mu$m relative to the target centre.
The spectrometer featured two large magnets.  Particles were
identified by electromagnetic and hadronic calorimeters, scintillation
hodoscopes, proportional-tube muon detectors and a ring-imaging
\v{C}erenkov counter.  E789 collected a large statistics data
sample~\cite{e789}, much bigger than all other charm measurements made
with proton beams, but only measured neutral D mesons and their
acceptance was limited to the $0<x_{\rm F}<0.08$ window.  It also
measured the nuclear dependence of neutral D meson production,
comparing data taken on beryllium and gold targets.

E789 was the first experiment to measure the beauty production
cross-section in proton-nucleus interactions~\cite{e789B}, using a
gold target of $50 \times 0.2$~mm$^2$ area and 3~mm thick.  The target
was placed in vacuum, to ensure that interactions in air would not be
confused with $b$-hadron decays.  Within its acceptance, the highly
energetic ($p_{\rm lab} \sim 150$~GeV) B hadrons had their production
and decay vertices separated by an average distance of 1.3~cm.  E789
triggered on events with a dimuon coming from the target region, to
look for beauty hadrons through their decay into a \jpsi\ meson.  The
spectrometer had excellent dimuon mass resolution: 16~MeV at the
\jpsi\ mass, dominated by multiple scattering in the target.  The
longitudinal vertex resolution was 700~$\mu$m.  \jpsi's from beauty
decays were requested to have their origin more than 7~mm downstream
of the target centre. The impact parameter of each muon, defined as
the vertical distance between the muon track and the target centre,
had to be larger than 150~$\mu$m.  $19\pm 5$ events survived these
rather strict selection cuts.

\subsection{E771}

The E771 Collaboration upgraded the Fermilab High Intensity Lab
spectrometer, previously used by E537 and E705, with silicon
microstrip detectors, pad chambers and resistive plate counters, to
``study charm and beauty states as detected by decays into
muons''~\cite{e771-nim}.  The beauty production
cross-sections~\cite{e771} were measured from data collected during
one month in 1991, using the 800~GeV proton beam.  The target
consisted of twelve 2~mm thick Si foils interspaced by 4~mm, giving a
total effective length of 5.2\,\%~$\lambda_{\rm int}$.  A silicon
microvertex detector was positioned downstream of the target for the
measurement of primary and secondary vertices.  Multiwire proportional
chambers and drift chambers were used, together with a dipole analysis
magnet, to determine charged particle trajectories and momenta. The
spectrometer finished with a muon detector made of three planes of
resistive plate counters, embedded in steel and concrete shielding.
E771 used a dimuon trigger to select two possible decay modes: a
\jpsi\ coming from the decay of a B meson, or a muon pair from the
simultaneous semi-muonic decays of two beauty hadrons.  Muons from the
semi-muonic decay of B mesons could be accepted if within the
$-0.25<x_{\mathrm F}<0.50$ window.

\subsection{WA92, WA82}

The Beatrice Collaboration, WA92, used the Omega Spectrometer, at the
CERN SPS, to ``measure beauty particle lifetimes and
hadroproduction cross-sections'', but also published results on charm
production~\cite{wa92-nim}.  It took data in 1992 and 1993 with a
350~GeV $\pi^-$ beam (including a 1.2\,\% K$^-$ contamination)
incident on Cu and W targets.  Charged particle tracking started with
a series of high granularity (10 to $50~\mu$m pitch)
silicon-microstrip detector planes, organised in a Beam Hodoscope, a
Decay Detector and a Vertex Detector.  Charged particles were then
tracked in multi-wire proportional chambers placed inside the
superconducting Omega dipole magnet, with a bending power of 7.2~Tm,
and in drift chambers placed downstream of the magnet.  The setup was
completed with an electromagnetic calorimeter, followed by a muon
identifier made of resistive plate counters protected by hadron
absorbers.  A multi-component trigger was developed to identify events
with beauty decays, which also slightly enriched the fraction of
collected charm events.  Due to the high statistics collected with the
Cu and W targets, WA92 also measured the nuclear dependence of neutral
and charged D meson production~\cite{wa92}.  The beauty production
cross-section measurement was derived from the $\pi^-$-Cu data
samples~\cite{wa92B}.

Part of these detectors had already been used by the WA82 experiment,
where the Omega Spectrometer was complemented by 23 silicon-microstrip
detector planes, with a pitch ranging from 10 to 50~$\mu$m, to
precisely reconstruct tracks and (secondary) vertices~\cite{wa82-nim}.
The fast online treatment of the silicon-microstrip data was used to
select events with at least one track missing the primary vertex.  To
study the nuclear dependence of charm production~\cite{wa82}, the
experiment took data (with a 340~GeV $\pi^-$ beam) using a 2~mm thick
target made of two materials (either Si/W or Cu/W), placed side by
side, transversely with respect to the beam axis.  The beam
illuminated simultaneously both target materials, reducing the
systematic uncertainties.

\subsection{CDF}

CDF is a ``general purpose'' experiment, at Fermilab, which studied
\ppbar\ collisions at $\sqrt{s} = 1.8$~TeV between 1992 and 1995
(``Run I''), and at 1.96~TeV from year 2001 onwards (``Run II'').
Charged track trajectories are reconstructed in 
a drift 
chamber and matched to strip clusters in the silicon vertex detectors.
These devices are immersed in a magnetic field of 1.4~T, generated by
a superconducting solenoid.  The central muon system (outside a hadron
calorimeter) consists of eight layers of drift chambers, four before
and four after a 60~cm thick steel absorber, and detects muons with
$p_{\mathrm T} >1.4$~GeV/$c$ in the range $|y|<1$.
The measurements of D and B meson production are based on rather large
data samples, but the results from Run~I are restricted to a
relatively high $p_{\mathrm T}$ window, $p_{\mathrm T} >
5.5$--6.0~GeV/$c$, at mid-rapidity, $|y|<1$.  In Run~I, CDF studied
charged B meson production~\cite{cdf} using the B$^+ \to {\rm
J}/\psi\, {\rm K}$ decay channel.  Charged B candidates were selected
by combining the \jpsi\ mesons with each charged particle track of
$p_{\mathrm T}>1.25$~GeV/$c$ (kaons from B meson decays have a harder
\pt\ spectrum than most other particles).  The dimuon and kaon tracks
were then constrained to come from a common vertex.  This study was
done using events triggered on two opposite-sign muons in the mass range
of the \jpsi, corresponding to an integrated luminosity of ${\cal
L}_{\rm int} = 98\pm4$~pb$^{-1}$.

Run~II provided much better data for heavy flavour studies,
largely because of the added ability to trigger on secondary vertices.
Results
on charm production~\cite{cdfCharm} were obtained from data collected
in early 2002.  The D mesons were selected by requiring two oppositely
charged tracks with $p_{\rm T}>2$~GeV/$c$ and $p_{\rm T,1}+p_{\rm
T,2}>5.5$~GeV/$c$, with a distance of closest approach to the beam
axis between $120~\mu$m and 1~mm.  Results on beauty production have
also been published~\cite{cdfII}, based on data collected in 2002,
with ${\cal L}_{\rm int} = 39.7\pm 2.3$~pb$^{-1}$.  Beauty hadrons
were measured in the inclusive B$\to {\rm J}/\psi\, X$ channel, using
the transverse distance between the \jpsi\ origin and the \ppbar\
collision vertex, within $|y|<0.6$ and down to $p_{\rm T} = 0$, a
rather impressive experimental achievement: close to $p_{\rm T} = 0$
the signal drops fast but the background does not, the acceptance
changes rapidly, etc.

\subsection{HERA-B}
\label{heraB-exp}

The HERA-B fixed-target experiment, at DESY, was designed to identify,
within a large geometrical coverage, the decays of B and \jpsi\ mesons
produced on target wires by the halo of the 920~GeV HERA proton beam.
Most of the events were collected with C and W targets, with only a
small fraction (less than 10\,\%) taken with Ti.  Primary and
secondary vertices were reconstructed by the vertex detector system, made
of double-sided silicon microstrip detectors integrated in the HERA
proton ring.  The main tracking system was placed upstream of a 2.3~T
dipole magnet.  The inner region, near the beam pipe, used microstrip
gas chambers, while the outer tracker was made of honeycomb drift
cells.  Particle identification was performed by a ring imaging
\v{C}erenkov detector, a muon spectrometer composed of four tracking
stations, and an electromagnetic calorimeter.  In the data taking
period of 2002/2003, $164 \cdot 10^{6}$ events were collected with a
dilepton \jpsi\ trigger.  Decays of \jpsi, $\psi^{\prime}$, $\chi_{c}$
and $\Upsilon$ mesons were reconstructed, in the $\mu^{+}\mu^{-}$ and
in the $e^{+}e^{-}$ decay channels.

A \bbbar\ production cross-section measurement was
published~\cite{heraB-beauty} from data collected in year 2000, with
dimuon and dielectron triggers, through the study of $b\to {\rm
J}/\psi \to l^+l^-$ decays in the window $-0.25 < x_{\mathrm F}^{\rm
J/\psi} < 0.15$, on the basis of $1.9^{+2.2}_{-1.5}$ dimuon events and
$8.6^{+3.9}_{-3.2}$ dielectron events.  Improved results from the
2002/2003 run were recently published~\cite{heraB-beauty2}, following
the same analysis, yielding $46.2^{+8.6}_{-7.9}$ and
$36.9^{+8.5}_{-7.8}$ events in the muonic and electronic channels,
respectively.  Results on open charm production are expected to be
published soon.  Preliminary values have already been made
public~\cite{heraB-charm}.

\section{Data on open charm production}
\label{sec:dataCharm}

In this section we collect and discuss the experimentally measured
production cross-sections for the charged and neutral D mesons,
$\sigma(\D^+)+\sigma(\D^-)$ and
$\sigma(\D^0)+\sigma(\overline{\D^0})$.  Within the last 30 years
various experiments, using different kinds of detectors, have
collected data on open charm production.  In the late seventies, four
experiments at the ISR pp collider, at CERN, reported results on charm
production, mostly triggering on single electrons, assumed to come
from the semi-electronic decay of a D$^-$ or a $\overline{\D^0}$. We
have not included these measurements in our study.  In some cases,
only upper limits or ranges were given for the cross-sections.
Moreover, due to lack of statistics, sometimes data collected at
$\sqrt{s} = 52$ and 62~GeV were merged to obtain a common result on
the production cross-section.  Besides, the published values differ
significantly between the different experiments, as discussed in
Ref.~\cite{tavernier}.

\begin{table}[ht]
\begin{center}
\begin{tabular}{|l|lc|l|l|r|r|}\hline
\hfil Experim.\hfil &  \multicolumn{2}{|l|}{Beam \hfill $E_{\mathrm{lab}}$} &
\hfil Target \hfil
& \hfil Phase space window \hfil & \multicolumn{2}{|c|}{Events}\\ \cline{6-7}
     &  & [GeV]     &        & \hfil ($p$, \pt\ in GeV/$c$) \hfil &
\rule{0pt}{0.5cm} \hfil D$^0$ \hfil & \hfil D$^+$ \hfil \\ \hline
NA16   & p & 360 & p              & \xf$>$--0.1 & 5 & 10 \\
NA27   & p & 400 & p              & \xf$>$--0.1 & 98 & 119\\
E743   & p & 800 & p              & \xf$>$--0.1 & 10 & 46 \\
E653   & p & 800 & emulsion & \xf$>$--0.2$,\,p^\mu$$>$$8,\,p^\mu_{\rm T}$$>$0.2
& 108 & 18\\E789   & p & 800 & Be,\,Au & 0$<$$x_{\rm
F}$$<$$0.08,\,p_{\rm T}$$<$$1.1$ & $>$4000 & ---\\
E769   & p & 250 & Be,\,Al,\,Cu,\,W  & \xf$>$$-0.1$ & 136 & 159 \\
HERA-B & p & 920 & C,\,Ti,\,W & $-0.1$$<$$x_{\rm F}$$<$$+0.05$ & 189 & 98\\\hline
NA11   & $\pi^-$ & 200 & Be & \xf$>$0.0 & 29 & 21\\
NA16   & $\pi^-$ & 360 & p  & \xf$>$$-0.1$ & 4 & 9 \\
NA27   & $\pi^-$ & 360 & p  & \xf$>$$0.0$ & 49 & 14\\
NA32   & $\pi^-$ & 200 & Si & \xf$>$$0.0$ & 75 & 39\\
NA32   & $\pi^-$ & 230 & Cu & \xf$>$$0.05$ & 543 & 249\\
E653   & $\pi^-$ & 600 & emulsion & \xf$>$$0.0,\,p^\mu$$>$8 & 325 & 351\\
E769   & $\pi^-$ & 210 & Be,\,Al,\,Cu,\,W & \xf$>$--0.1 & 62 & 73\\
E769   & $\pi^-$ & 250 & Be,\,Al,\,Cu,\,W & \xf$>$--0.1 & 353 & 414\\
WA92   & $\pi^-$ & 350 & Cu,\,W & \xf$>$0.0 & 3873 & 3299 \\
E791   & $\pi^-$ & 500 & C,\,Pt & \xf$>$--0.1 & 88990 & ---\\
E706   & $\pi^-$ & 515 & Be,\,Cu & \xf$>$--0.2,\,1$<$$p_{\rm T}$$<$8 &--- & 110 \\
E769   & $\pi^+$ & 250 & Be,\,Al,\,Cu,\,W & \xf$>$$-0.1$ & 144 & 169\\ \hline
\end{tabular}
\end{center}
\vglue-4mm
\caption{Experiments measuring the production cross-sections of
  neutral and charged D mesons. The average mass number of the
  emulsion target of E653 is $A=26.6$.}
\label{tab:ccdata}
\end{table}

There are other early experiments~\cite{earlyExp} which studied
open charm production but which we will not consider here, such as
NA18, NA25, E515 and E595.  They could not separate the
different charm hadrons, only giving ``associated charm production
cross-sections''.  Within their rather large uncertainties, their
values are consistent with the measurements we have considered.  We
have also ignored the result obtained at $\sqrt{s}=630$~GeV with a
modified UA2 detector~\cite{UA2}, given its huge uncertainty.

In Table~\ref{tab:ccdata} we summarise the data used in the present
study, obtained with proton and pion beams, at energies ranging from
$E_{\mathrm{lab}}=200$ to 920~GeV.  Very significant statistical
samples have been collected by WA92 and E791 ($\sim$\,7000 and
$\sim$\,90\,000 events, respectively) with pion beams, while the fewer
proton beam experiments collected much less data. The $\sim$\,300
events of E769 (adding neutral and charged D mesons and the statistics
of four different nuclear targets) constitute the highest statistics
proton event sample, among the fixed target experiments covering a
reasonably large phase space window.  

At the much higher energies of the Tevatron p$\overline{\rm p}$
collider, $\sqrt{s} = 1.96$~TeV, CDF collected (within the $|y| < 1$
window) 36\,804 D$^0$ mesons, of $p_{\rm T} > 5.5$~GeV/$c$, and
28\,361 D$^+$ mesons, of $p_{\rm T} > 6.0$~GeV/$c$.

\begin{table}[ht!]
\begin{center}
\begin{tabular}{|l|c|lr|c|}\hline
\hfil Decay channel\hfil & Experiment & \multicolumn{2}{|c|}{BR used} & BR
              \small(PDG04) \\ 
&     &  \multicolumn{2}{|c|}{[\%]}  &  [\%] \\ \hline
$\D^0\rightarrow {\rm K}^-\pi^+$ & NA11(86) & $5.1\pm0.6$ & \cite{na11-1} &
$3.80\pm0.09$\\                           & NA32(88), E653(92) &
$4.2\pm0.4\pm0.4$ & \cite{markIII-88} &  \\
\rule{0pt}{0.4cm} & NA32(91) & $3.77^{+0.37}_{-0.32}$ & \small PDG88 & \\
& E789(94) & $3.65\pm0.21$ & \small PDG92 & \\
& E769(96) & $4.01\pm0.14$ & \small PDG94 &\\
& WA92(97) & $3.83\pm0.12$ & \small PDG96 &\\
& E791(99) & $3.85\pm0.09$ & \small PDG98 &\\
& HERA-B(05), CDF(03)&$3.80\pm0.09$ & \small PDG02 & \\ \hline
$\D^0\rightarrow$ & NA11(86) &$11.5\pm 1.1$ & \cite{na11-1} &
$7.46\pm0.31$\\
${\rm K}^-\pi^+\pi^+\pi^-$      &NA32(88), E653(92) & $9.1\pm0.8\pm0.8$ 
& \cite{markIII-88} &\\
\rule{0pt}{0.4cm}        & NA32(91) & $7.9^{+1.0}_{-0.9}$ & \small PDG88 & \\
& WA92(97) & $7.5\pm0.4$ & \small PDG96 & \\
& E791(99) & $7.6\pm0.4$ & \small PDG98 & \\ \hline
$\D^0\rightarrow \mu^+X^-$ &E653(91) &$0.96\,(7.7\pm1.2)$ & \small PDG90 &
$6.5\pm0.8$\\ \hline
$\D^0\rightarrow {\rm K}^-\mu^+\nu_\mu$ & E653(92) & $2.95\pm0.30$
&\cite{e653pi} & $3.19\pm0.17$\\ \hline
\rule{0pt}{0.4cm}$\D^0\rightarrow$ & E653(92) & $e:1.98\pm0.26$&\cite{e653pi} & $e:2.15\pm0.35$ \\ 
$\overline{{\rm K}}^*(892)^-\mu^+\nu_\mu$ & & & & \\ \hline
$\D^+\rightarrow {\rm K}^-\pi^+\pi^+$ & NA11(86) & $11.3\pm1.5$ &
\cite{na11-1}& $9.2\pm0.6$ \\
& NA32(88), E653(92) &$9.1\pm1.3\pm0.4$ & \cite{markIII-88} & \\
\rule{0pt}{0.4cm}& NA32(91) & $7.8^{+1.1}_{-0.8}$ & \small
PDG88 &\\
& E769(96) & $9.1\pm0.6$ & \small PDG94 &\\
& WA92(97), E706(97) & $9.1\pm0.6$ & \small PDG96 &\\
& HERA-B(05), CDF(03) & $9.1\pm0.6$ & \small PDG02 & \\ \hline
$\D^+\rightarrow$ & NA11(87) & $3.9\pm0.8\pm0.7$ &\cite{na11-2} & $5.5\pm0.7$\\ 
$\overline{{\rm K}}^*(892)^0e^+\nu_e$ & & & & \\ \hline
$\D^+\rightarrow $ & E653(91) & $0.96\,(3.8\pm0.7)$ & \small PDG90 &
$3.7\pm0.3$\\
$\overline{{\rm K}}^*(892)^0\mu^+\nu_\mu$ & E653(92) & $4.99\pm0.48$ &
\cite{e653pi} &\\ \hline
\end{tabular}
\end{center}
\vglue-4mm
\caption{Comparison between the latest branching ratios~\cite{pdg04}
 for the various D meson decay channels and the ones used in the
original publications.  The number in parentheses indicates the
year of publication.  Since the branching ratio of the decay channel
$\D^0\rightarrow\overline{{\rm K}}^*(892)^-\mu^+\nu_\mu$ is not known,
the electronic value was used by E653, assuming lepton universality.}
\label{tab:ccBR}
\end{table}

\begin{figure}[h!]
\centering
\resizebox{0.43\textwidth}{!}{%
\includegraphics*{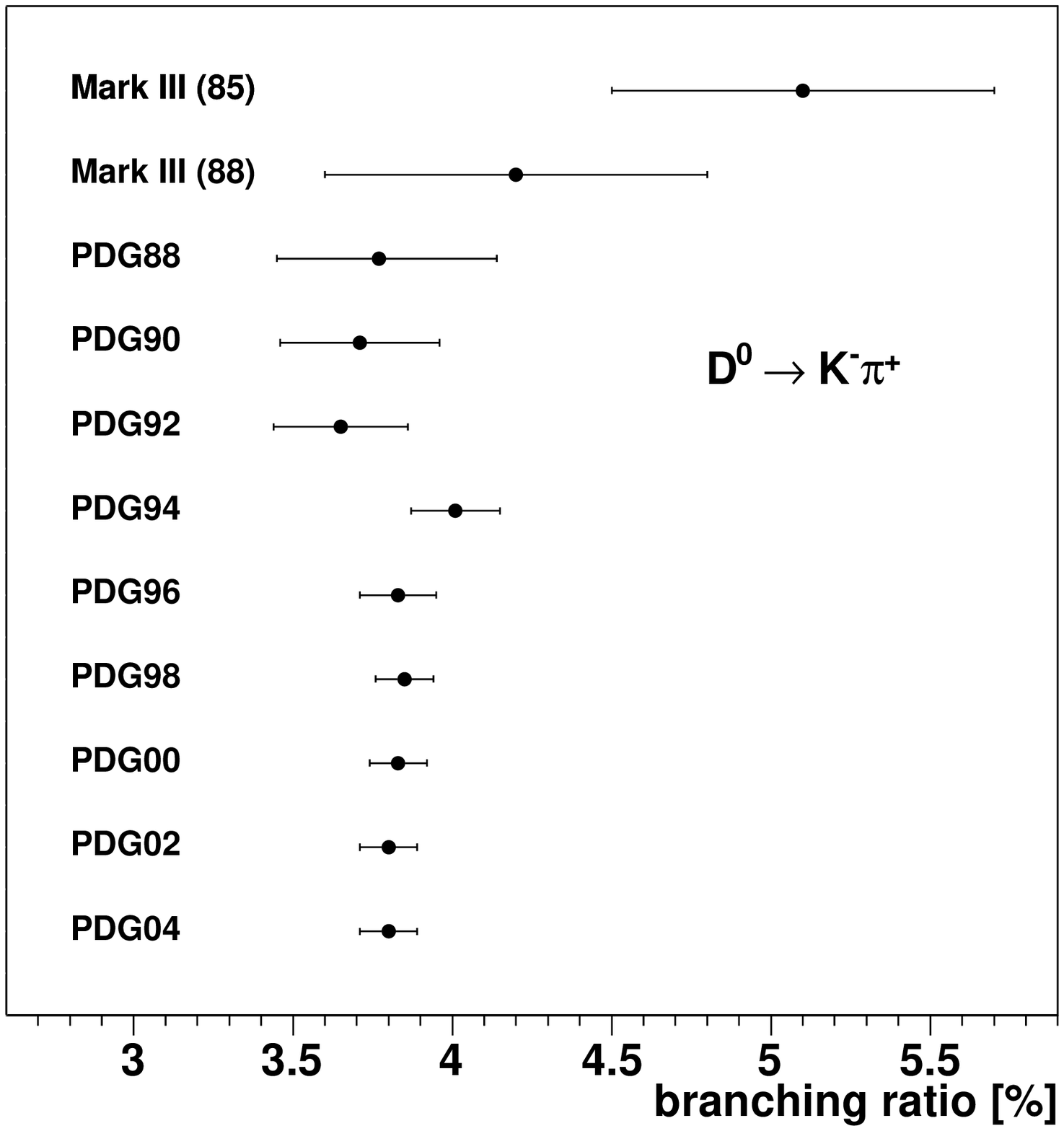}}
\resizebox{0.43\textwidth}{!}{%
\includegraphics*{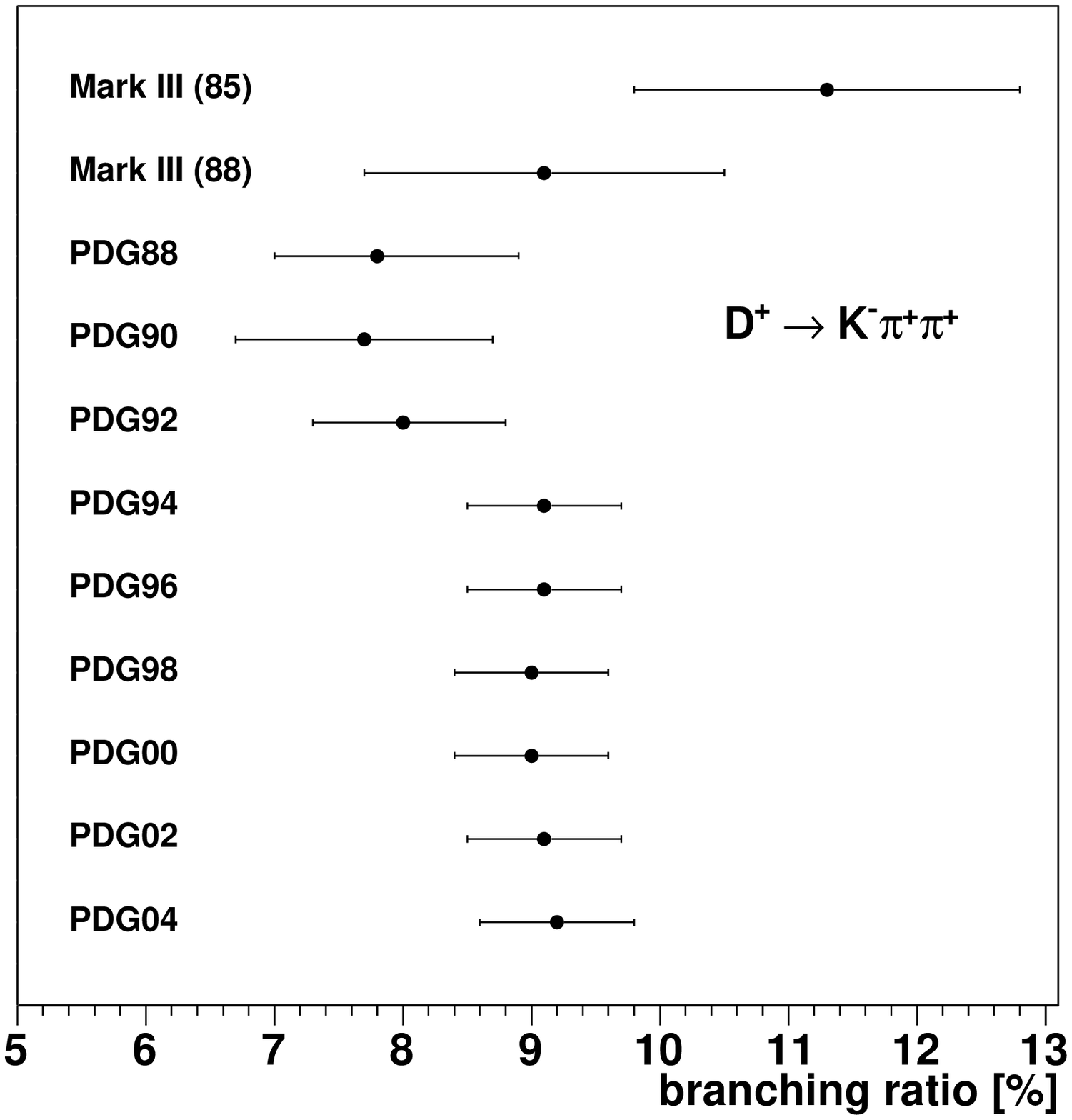}}
\caption{Time evolution of two relevant branching ratios.}
\label{fig:brTimeEv}
\end{figure}

In order to properly compare the different measurements to each other,
we applied certain corrections to some of the published values.  In
particular, whenever possible, we normalised the published
cross-sections to the latest branching ratio values, using the
Particle Data Group~2004 values~\cite{pdg04}.  Table~\ref{tab:ccBR}
summarises the decay channels of D mesons which were investigated in
each experiment.  It further lists the branching ratio used in the
original publication, together with the corresponding latest value.
Figure~\ref{fig:brTimeEv} illustrates the ``time evolution'' of two
branching ratios.

\begin{table}[h!]
\begin{center}
\begin{tabular}{|l|c|c|c|c|}\hline
\hfil Experiment\hfil & $E_\mathrm{lab}$ & $\sigma$ given & \multicolumn{2}{|c|}{$\sigma$ [$\mu$b]} \\
\cline{4-5}         & [GeV] & for & published & updated  \\ \hline
\multicolumn{5}{|c|}{\rule{0pt}{0.5cm}{p beam, $\sigma({\D}^0) +
\sigma(\overline{{\D}^0})$}}\\ \hline
\rule{0pt}{0.5cm}NA16\,(84) \hfill \cite{na16} & 360 & $x_{\rm F} > 0$ &
$10.2^{+7.9}_{-4.3}$ &$10.2^{+7.9}_{-4.3}$\\
NA27\,(88)\hfill \cite{na27p} & 400 & all
\xf & $18.3\pm 2.5$&$18.3\pm 2.5$\\
E743\,(88)\hfill \cite{e743} & 800 & all \xf &
$22^{+9}_{-7}\pm25$\% &$22^{+9}_{-7}\pm25$\% \\
E653\,(91)\hfill \cite{e653p} & 800 & all
\xf & $38\pm 3\pm 13$ &$43\pm3\pm14$ \\
E789\,(94)\hfill \cite{e789} & 800 & all \xf &
$17.7\pm0.9\pm3.4$ &$17.0\pm0.9\pm3.1$ \\
E769\,(96)\hfill \cite{e769} & 250 & $x_{\rm F}
> 0$ &$5.7\pm1.3\pm0.5$ & $6.0\pm1.4\pm0.5$ \\
HERA-B\,(05) \hfill\cite{heraB-charm} & 920 & all \xf & $56.3\pm8.5\pm9.5$ &
$56.3\pm8.5\pm9.5$\\ \hline
\multicolumn{5}{|c|}{\rule{0pt}{0.5cm}{p beam,
$\sigma({\D}^+) +\sigma({\D}^-)$}}\\ \hline
\rule{0pt}{0.5cm}NA16\,(84) \hfill
\cite{na16} & 360 & $x_{\rm F} > 0$ & $5.3^{+2.4}_{-1.6}$ &
$5.3^{+2.4}_{-1.6}$\\
NA27\,(88) \hfill\cite{na27p} & 400 & all \xf & $11.9\pm1.5$&
$11.9\pm1.5$\\
E743\,(88) \hfill \cite{e743} & 800 & all \xf & $26\pm4\pm25$\% &
$26\pm4\pm25$\%\\
E653\,(91) \hfill \cite{e653p} & 800 & all \xf & $38\pm9\pm14$ &
$37\pm9\pm12$\\
E769\,(96) \hfill \cite{e769} & 800 & $x_{\rm F} > 0$ & $3.3\pm0.4\pm0.3$
& $3.3\pm0.4\pm0.4$ \\
HERA-B\,(05) \hfill\cite{heraB-charm} & 920 & all \xf &
$30.2\pm4.5\pm5.8$& $29.9\pm4.5\pm5.7$ \\ \hline
\end{tabular}
\end{center}
\vglue-4mm
\caption{Published and updated D meson production cross-sections in
\emph{proton} induced collisions.  The NA27 errors represent the
combined statistical and systematic uncertainties.  The HERA-B values
are preliminary (see the ``Note added in proof'' at the end of this paper).}
\label{tab:ccCorr-p}
\end{table}

We have also updated the systematic errors of the published values to
reflect the smaller uncertainties of the most recent branching ratios.
Some publications have not included on their systematic errors these
uncertainties, something we must do when comparing D meson production
measured in different decay channels.  If the D mesons were searched
for in more than one decay channel, the performed corrections were
weighted according to the number of observed events in each of the
decay channels.  This procedure was not applied to the data of
experiments which searched the D mesons in topological decays, where
the search is not done in a specific decay channel, but rather by
detecting a certain number of charged or neutral final state particles
(NA16, NA27, WA75, E743).
\begin{table}[t!]
\begin{center}
\begin{tabular}{|l|c|c|c|c|}\hline
\hfil Experiment\hfil & $E_\mathrm{lab}$ & $\sigma$ given & \multicolumn{2}{|c|}{$\sigma$ [$\mu$b]} \\
\cline{4-5}           & [GeV] & for & published & updated  \\ \hline
\multicolumn{5}{|c|}{\rule{0pt}{0.5cm}{$\pi^-$ beam, $\sigma({\D}^0) +
\sigma(\overline{{\D}^0})$}}\\ \hline
\rule{0pt}{0.5cm}NA16\,(84) \hfill \cite{na16} & 360 & $x_{\rm F} > 0$ &
$7.7^{+7.2}_{-3.5}$ &$7.7^{+7.2}_{-3.5}$\\
NA27\,(86) \hfill \cite{na27pi} & 360 &
$x_{\rm F} > 0$ &$10.1\pm2.2$ & $10.1\pm2.2\pm15\%$\\
NA11\,(86) \hfill \cite{na11-1} & 200 & all \xf &
$31\pm7\pm16$ & $45\pm10\pm23$\\
\rule{0pt}{0.4cm}NA32\,(88) \hfill \cite{na32-88} & 200
& $x_{\rm F} > 0$ & $1.15\,(3.3^{+0.5}_{-0.4}\pm0.3)$ &
$4.4^{+0.7}_{-0.5}\pm0.4$ \\
NA32\,(91) \hfill \cite{na32-91} & 230 & $x_{\rm F} > 0$ &
$6.3\pm0.3\pm1.2$ & $6.6\pm0.3\pm1.0$ \\
E653\,(92) \hfill \cite{e653pi} & 600 & $x_{\rm
F} > 0$ & $22.05\pm1.37\pm4.82$ & $18.86\pm1.17\pm3.80$ \\
E769\,(96) \hfill \cite{e769} & 210 & $x_{\rm F} > 0$ & $6.4\pm0.9\pm0.3$ &
$6.8\pm1.0\pm0.3$ \\
E769\,(96)\hfill \cite{e769} & 250 & $x_{\rm F} > 0$ & $8.2\pm0.7\pm0.5$ &
$8.7\pm0.7\pm0.6$\\
WA92\,(97) \hfill \cite{wa92} & 350 & $x_{\rm F} > 0$ & $7.78\pm0.14\pm0.52$
&$7.83\pm0.14\pm0.48$ \\
\rule{0pt}{0.4cm}E791\,(99) \hfill \cite{e791} & 500 &
$x_{\rm F} > 0$ & $15.4^{+1.8}_{-2.3}$ &
$15.6^{+1.8}_{-2.3}$ \\ \hline
\multicolumn{5}{|c|}{\rule{0pt}{0.5cm}{$\pi^-$ beam, $\sigma({\D}^+) +
\sigma({\D}^-)$}}\\ \hline
\rule{0pt}{0.5cm}NA16\,(84)\hfill \cite{na16} & 360 & $x_{\rm
F} > 0$ & $4.5^{+2.2}_{-1.4}$ & $4.5^{+2.2}_{-1.4}$\\
NA27\,(86) \hfill \cite{na27pi} & 360 & $x_{\rm F} > 0$ & $5.7\pm1.5\pm0.4$ &
$5.7\pm1.5\pm1.4$\\
NA11\,(86) \hfill \cite{na11-1} & 200 & all \xf & $20\pm5\pm10$ &
$25\pm6\pm12$\\
NA11\,(87) \hfill \cite{na11-2} & 200 & all \xf & $30.0\pm3.5\pm12.6$ &
$21.3\pm2.5\pm7.3$ \\
\rule{0pt}{0.4cm}NA32\,(88) \hfill \cite{na32-88} & 200 & $x_{\rm
F} > 0$ & $1.15\,(1.7^{+0.4}_{-0.3}\pm0.1)$ & 
$1.9^{+0.5}_{-0.3}\pm0.2$ \\
NA32\,(91) \hfill \cite{na32-91} & 230 & $x_{\rm F} > 0$ &
$3.2\pm0.2\pm0.7$ & $2.7\pm0.2^{+0.5}_{-0.6}$ \\
E653\,(92) \hfill \cite{e653pi} & 600 & $x_{\rm F} > 0$ & $8.66\pm0.46\pm1.96$ &
$8.05\pm0.43\pm1.69$ \\
E769\,(96) \hfill \cite{e769} & 210 & $x_{\rm F} > 0$ &
$1.7\pm0.3\pm0.1$ & $1.7\pm0.3\pm0.1$ \\
E769\,(96) \hfill \cite{e769} & 250 & $x_{\rm F}
> 0$ & $3.6\pm0.2\pm0.2$ & $3.6\pm0.2\pm0.3$ \\
WA92\,(97) \hfill \cite{wa92} & 350 & $x_{\rm F} > 0$ & $3.28\pm0.08\pm0.29$ &
$3.24\pm0.08\pm0.28$ \\
E706\,(97) \hfill \cite{e706} & 515 & $x_{\rm F} > 0$ &
$11.4\pm2.7\pm3.3$ & $11.3\pm2.7\pm3.3$\\ \hline
\multicolumn{5}{|c|}{\rule{0pt}{0.5cm}{$\pi^+$ beam,
$\sigma({\D}^0) + \sigma(\overline{{\D}^0})$}}\\ \hline
E769\,(96) \hfill \cite{e769} & 250 & $x_{\rm F} > 0$ & $5.7\pm0.8\pm0.4$ &
$6.0\pm0.8\pm0.4$ \\ \hline
\multicolumn{5}{|c|}{\rule{0pt}{0.5cm}{$\pi^+$ beam,
$\sigma({\D}^+) + \sigma({\D}^-)$}}\\ \hline
E769\,(96) \hfill \cite{e769} & 250 & $x_{\rm F} > 0$ & $2.6\pm0.3\pm0.2$ &
$2.6\pm0.3\pm0.3$ \\ \hline
\end{tabular}
\end{center}
\vglue-4mm
\caption{Published and updated D meson production cross-sections in \emph{pion}
induced reactions.  The two charged D meson values of NA11 correspond
to different decay channels (see Table~\ref{tab:ccBR}).}
\label{tab:ccCorr-pi}
\end{table}
The neutral D mesons were searched, e.g., in ``V2'' or ``V4'' prong
decays, corresponding to neutral decays with 2 or 4~prongs, while
``C3'' is a typical topological decay of a charged D meson with
3~prongs.  Since topological decays are composed of different decay
channels, with unknown contributing fractions, we cannot correct them
for the evolution in our knowledge of the branching ratios.
Tables~\ref{tab:ccCorr-p} and~\ref{tab:ccCorr-pi} collect the
published and updated values of the production cross-sections for
fixed-target experiments using proton and pion induced collisions,
respectively.

CDF measured in Run II~\cite{cdfCharm}, the \emph{single} D meson
production cross-sections in \ppbar\ collisions at $\sqrt{s} =
1.96$~TeV, in $|y|<1$: $\sigma({\D}^0)=13.3\pm0.2\pm1.5$~$\mu$b for
$p_{\rm T}>5.5$~GeV/$c$ and $\sigma({\D}^+)=4.3\pm0.1\pm0.7$~$\mu$b
for $p_{\rm T}>6$~GeV/$c$.


With the exception of WA75, the experiments which used nuclear targets
assumed a \emph{linear} dependence on the mass number of the target
nucleus to derive the cross-section in pp (or $\pi$p) collisions,
$\sigma_{\rm pA} =A \cdot \sigma_{\rm pp}$.

We will now comment on certain specific measurements.

\begin{itemize}
\parskip=0.pt \parsep=0.pt \itemsep=0.pt

\item\textbf{NA16\,(84)} The publication does not mention if the
quoted errors include systematic uncertainties, due to the branching
ratios or to other factors.  In any case, the quoted errors are very
large (and certainly dominated by statistical uncertainties) and
increasing them would not change the overall comparison with the
other, more precise, data points. Therefore, we have not updated the
NA16 values or error bars.  This applies to all NA16 measurements,
with proton or pion beams, producing charged or neutral D mesons.

\item\textbf{E789\,(94)} The acceptance of the E789 detector was
limited to the ranges $0<x_{\rm F}<0.08$ and $p_{\rm T}<1.1$~GeV/$c$.
It is clear that the extrapolation of the measurement to full phase
space is model dependent, and a significant uncertainty should be
added to the published systematic error, before comparing it to other
measurements or to calculations.

\item\textbf{NA27\,(86)} The \emph{neutral} D mesons were searched in
the 2-prong and 4-prong decays. By weighing the contributions from
these two modes, we deduced an uncertainty of 15\,\% in the
cross-section measurement, due to the branching ratio. This value was
used to estimate the systematic error of the published cross-sections,
since no other value was explicitly quoted in the publication.

\item\textbf{NA27\,(86)} The published systematic error of the D
\emph{charged} cross-section is 0.4~$\mu$b, a value quoted as being
mostly due to uncertainties in the branching ratio.  However, the
uncertainty due to the branching ratio, on its own, would already give
an error of 1.3~$\mu$b.  We presume that there was a misprint in the
original publication and updated the systematic error to 1.4~$\mu$b.

\item\textbf{NA32\,(88)} In a later publication~\cite{na32-91} the
NA32 Collaboration explains that a normalisation error was found and
\emph{all} total cross-sections of Ref.~\cite{na32-88} should be
upscaled by 15\,\%.

\item\textbf{NA32\,(88), NA32\,(91)} The \emph{neutral} D mesons are
identified in two decay channels. Their relative contribution was
estimated according to $$N^{\mathrm{decays}} = N^{\mathrm{2body}} +
N^{\mathrm{4body}} ={\cal L} \cdot \sigma^\D\cdot (A^{\mathrm{2body}}
\cdot B^{\mathrm{2body}} + A^{\mathrm{4body}} \cdot
B^{\mathrm{4body}})~,$$ where 2body and 4body stand for the 2- and
4-body decay channels, listed in Table~\ref{tab:ccBR}.
$N^{\mathrm{decays}}$ is the total number of reconstructed D mesons,
${\cal L}$ the integrated luminosity, $\sigma^\D$ the cross-section
for D meson production, $A$ the acceptances, and $B$ the branching
ratios.  The relative populations of 2-body and 4-body decays are then
determined to be 39\,\%/61\,\% for NA32\,(88) and 23\,\%/77\,\% for
NA32\,(91).  The corrections on the branching ratios were then
performed according to these weights.\\[4mm]

\item\textbf{E653\,(92)} Several decay channels contributed to the
detection of the \emph{neutral} D mesons, but we only corrected the
branching ratios of those contributing the most, listed in
Table~\ref{tab:ccBR}.

\end{itemize}

While most of the results obtained with proton beams were published
for the full \xf\ range, those obtained with pion beams were published
for the positive \xf\ range, with the exception of NA11, which assumed
a {\em symmetric} \xf\ distribution for the extrapolation.  This
measurement should be taken with care.  NA11 was the predecessor of
NA32 and used a ``prototype version'' of the experimental apparatus.
NA32 used significantly upgraded detectors and performed a much more
accurate measurement, with the same pion beam and at the same
collision energy.

\begin{table}[h!]
\begin{center}
\begin{tabular}{|l|c|c|}\hline
\hfil Experiment\hfil & $E_\mathrm{lab}$ & $\sigma$ \\
&  [GeV]          & [$\mu$b] \\ \hline
\multicolumn{3}{|c|}{\rule{0pt}{0.5cm}{p beam, full $x_{\rm F}$}}\\ \hline
NA32\,(88) & 200 & $2\,(1.5\pm0.7\pm0.1)$\\
E769\,(96) & 250 & $11.2\pm 1.7\pm 0.8$ \\
\rule{0pt}{0.4cm}NA16\,(84) & 360 & $18.6^{+9.9}_{-5.5}$ \\
NA27\,(88) & 400 & $18.1\pm1.7$ \\
E743\,(88) & 800 & $29^{+6}_{-5}\pm5$ \\
E653\,(91) & 800 & $48\pm6\pm11$ \\
HERA-B\,(05) & 920 & $51.7\pm5.8\pm6.6$ \\ \hline
\multicolumn{3}{|c|}{\rule{0pt}{0.5cm}{$\pi^-$ beam, $x_{\rm F}>0$}}\\ \hline
NA11\,(86) & 200 & $21\pm3\pm8$ \\
\rule{0pt}{0.4cm}NA32\,(88) & 200 & $3.8^{+0.5}_{-0.3}\pm0.3$ \\
E769\,(96) & 210 & $5.1\pm0.6\pm0.2$ \\
NA32\,(91) & 230 & $5.6\pm0.2\pm0.7$ \\
E769\,(96) & 250 & $7.4\pm0.4\pm0.4$ \\
WA92\,(97) & 350 & $6.64\pm0.10\pm0.33$ \\
WA75\,(92) & 350 & $7.5\pm0.4^{+1.3}_{-1.1}$ \\
\rule{0pt}{0.4cm}NA16\,(84) & 360 & $7.3^{+4.5}_{-2.3}$ \\
NA27\,(86) & 360 & $9.5\pm1.6\pm1.2$ \\
E653\,(92) & 600 & $16.15\pm0.75\pm2.50$ \\ \hline
\multicolumn{3}{|c|}{\rule{0pt}{0.5cm}{$\pi^+$ beam, $x_{\rm F}>0$}}\\ \hline
E769\,(96) & 250 & $5.2\pm0.5\pm0.3$ \\ \hline
\end{tabular}
\caption{\ccbar\ production cross-sections in p-A and $\pi$-A
  collisions.  See the text for remarks on the NA32 \emph{proton}
  value and on the WA75 value. The HERA-B value is preliminary
  (see the ``Note added in proof'' at the end of this paper).}
\label{tab:ccbar}
\end{center}
\end{table}
In Table~\ref{tab:ccbar} we summarise the derived values for the total
\ccbar\ production cross-sections.  The p-A values are given for full
phase space while the $\pi$-A results are given for the positive \xf\
hemisphere.
To obtain the total \ccbar\ production cross-section, besides adding
the measured neutral and charged D meson values, we must take into
account the production of other charmed hadrons (D$_{\rm s}$, $\rm
\Lambda_c$ and other charmed baryons and charmonia states).  Only NA32
and E769 also measured the D$_{\rm s}$ and $\rm\Lambda_c$
hadro-production cross-sections~\cite{na32-91,e769,na32-lambdaC}, but
with very poor statistics.  Assuming that the fragmentation fractions
are universal, we can use e$^+$e$^-$ or $\gamma$p data.  World
averages of results obtained by CLEO, ARGUS, H1, ZEUS and the four LEP
Collaborations show that neutral and charged D meson production
represent $78\pm 3$\,\% of the total charm production
cross-section~\cite{fragFrac}.  This value is well reproduced by the
Pythia Monte Carlo event generator.  Note that the D meson
cross-sections compiled in Tables~\ref{tab:ccCorr-p}
and~\ref{tab:ccCorr-pi} refer to the sum of the particle and
anti-particle values.

The \emph{proton} measurement of NA32 was not listed in the
previous tables because only the total D meson cross-section was
published~\cite{na32-88}, not separating the charged and neutral
states.  The quoted number was scaled up by the factor~1.15 mentioned
above.  The factor 2 explicitly shown in the table converts the
published positive \xf\ value to full \xf, for consistency with the
other proton beam measurements.  The rather big statistical
uncertainty is due to the fact that only 9 events were observed.  Also
WA75 published~\cite{wa75} a charm production cross-section without
separating the neutral and charged contributions, from 339
pion-emulsion events observed in the window $-0.5<x_{\rm F}<0.8$.  The
published $\pi$p value, $23.1\pm1.3^{+4.0}_{-3.3}$~$\mu$b/nucleon, for
full \xf, was derived assuming an $A^\alpha$ scaling with
$\alpha=0.87$.  The value included in Table~\ref{tab:ccbar} was
recalculated assuming $\alpha=1.0$.  It was also divided by 2 so that
it roughly corresponds to the $x_{\rm F}>0$ window, for consistency
with the other measurements.  The published systematic errors of
these two measurements, by NA32 and WA75, are underestimated, since
they ignore the uncertainty on the branching ratios.  Not enough
information exists in the original publications to calculate a proper
error or to update the values using the most recent branching ratios.

\section{Data on open beauty production}
\label{sec:dataBeauty}

The available measurements on beauty production were collected over
the last 15 years.  Besides the fixed target experiments, with
energies within $200 < E_{\mathrm{lab}} < 920$~GeV, UA1 and CDF
measured the beauty cross-section at the much higher energies of
p$\bar{\rm p}$ colliders, $\sqrt{s} = 630$~GeV and 1.8--1.96~TeV,
respectively.  Since fixed-target experiments barely have the minimum
energy required for beauty production, the cross-sections are very
small, only a few nb, and very
selective triggers are needed in order to observe even a few beauty
events. In most cases this was achieved by triggering on high \pt\
single muons or dimuons. Looking at \emph{inclusive}
muonic decays, as was done by the older experiments, has the advantage
of exploring rather large branching ratios, $\sim$\,10\,\%. The new
experiments explored the beauty decay into \jpsi\ by measuring the
fraction of \jpsi's with a minimum offset with respect to the
interaction point, which has the advantage of having a well understood
reference process for normalisation purposes 
and is much more robust with respect to backgrounds: B decays are the
only source of displaced \jpsi\ mesons, while there are several
sources of single muons.
These experiments, hence, complemented a muon spectrometer with a
vertex detector to reconstruct the secondary vertices, profiting from
the B mesons' long lifetimes. 
All experiments, except one, 
reported cross-sections for a global \emph{mixture} of all
different beauty hadrons, B$^+$, B$^-$, B$^0$, $\overline{\B^0}$,
$\B_{\rm s}^0$, $\overline{\B^0_{\rm s}}$, b-baryons, etc, with a
priori unknown relative fractions.  The exception is the single B$^+$
meson cross-section obtained by CDF in Run~I, from a measurement of
\emph{charged} B mesons.

\begin{table}[ht]
\begin{center}
\begin{tabular}{|lr|lc|l|l|c|}\hline
\rule{0pt}{0.4cm} \hfil Experiment \hfil & & \multicolumn{2}{|l|}{Beam \hfill
     $E_{\mathrm{lab}}$} & \hfil Target \hfil & \hfil Phase space window \hfil 
     & \BBbar\ events\\
     &  &  &  [GeV]     &        & \hfil (\pt\ in GeV/$c$) \hfil & \\ \hline
NA10 & \cite{na10} & $\pi^-$ & 286 & W & $x_{\mathrm F}>0$; selection cuts &
14\\
WA78 & \cite{wa78} & $\pi^-$ & 320 & U & $x_{\mathrm F}>0$; selection cuts &
73 \\
E653 & \cite{e653B} & $\pi^-$ & 600 & emulsion & $x_{\mathrm
F}>-0.3$& $9\pm3$\\
E672/E706 & \cite{e672} & $\pi^-$ & 515 & Be & $x_{\mathrm
F}>0$ & $8\pm3.3$ \\
WA92 & \cite{wa92B} & $\pi^-$ & 350 & Cu & $x_{\mathrm
F}>-0.2$ & 26 \\ \hline
\rule{0pt}{0.5cm}E789 & \cite{e789B} & p & 800 & Au &
0$<$$x_{\mathrm F}^{\mathrm J/\psi}$$<$0.1, $p_{\mathrm T}^{\mathrm
J/\psi}$$<$2 & $19\pm5$ \\
\rule{0pt}{0.5cm}E771 & \cite{e771} & p & 800 & Si &
$-0.25$$<$$x_{\mathrm F}^\mu$$<$0.50, $p^\mu$$>$6 & 15 \\
\rule{0pt}{0.5cm}HERA-B & \cite{heraB-beauty2} & p & 920 & C,\,Ti,\,W &
$-0.35<x_{\rm F}^{\rm J/\psi}<0.15$& $83\pm 12$ \\ \hline
UA1 & \cite{ua1} & \multicolumn{3}{|l|}{\rule{0pt}{0.5cm}{p$\overline{\rm p}$
\hfill $\sqrt{s} = 630$~GeV}} & $|y|<1.5$, $p_{\mathrm T}^{\rm b} >6$ & 2859 \\
CDF Run~I & \cite{cdf} & \multicolumn{3}{|l|}{\rule{0pt}{0.5cm}{p$\overline{\rm
p}$ \hfill $\sqrt{s}= 1.8$~~TeV}} & $|y|<1.0$, $p_{\mathrm T}^{\B}>6$ &
$387\pm32$ \\
CDF Run~II & \cite{cdfII} &
\multicolumn{3}{|l|}{\rule{0pt}{0.5cm}{p$\overline{\rm p}$\hfill $\sqrt{s} =
1.96$~TeV}} & $|y| < 0.6$ & 38078\\ \hline
\end{tabular}
\caption{Beauty hadro-production measurements. E653 used an emulsion
  target (of average mass number 26.6).  In Run~I, CDF only measured
  B$^\pm$ mesons.}
\label{tab:bbbarOverview}
\end{center}
\end{table}

Table~\ref{tab:bbbarOverview} shows the available measurements of
beauty production, the phase space covered and the number
of identified \BBbar\ events.
Rather than updating the older measurements with the most recent
knowledge on branching ratios, an issue of minor importance when
compared to the other uncertainties involved and the large statistical
errors, we prefer to emphasise that many experiments relied on
theoretical models to determine the published cross-sections. This is
necessary because of the limited phase space coverage of the
detectors. Often very different models were used, giving different
extrapolation factors to obtain the total (full phase space)
cross-sections. Only the E653, WA92, E771 and HERA-B experiments cover
around 90\,\% of full phase space and, thus, are less sensitive to
theoretical assumptions.

The first beauty hadro-production cross-section measurement, by NA10, was published
in 1988~\cite{na10}, before QCD calculations were developed in detail, so that they
used a rather simple production model.  NA10 also provided another
cross-section value, lower by 30\,\%, obtained by using different \xf\
and \pt\ distributions for the 4$\pi$ extrapolation.  The second
beauty hadro-production measurement, made by WA78, is in good
agreement with the NA10 result, if the same production model is used~\cite{wa78}.
However, when WA78 uses a LO QCD calculation
the two results differ significantly, even though the
measurements were performed at almost the same energy.

During the past 20 years, the theoretical understanding of beauty
production has significantly evolved and the experiments profited from
this progress when analysing their data.  Since they were performed
over a long time span, each experiment used a different calculation
(Pythia Monte Carlo generator, NLO calculations with several settings,
etc), and it is highly non-trivial to ``normalise'' all the available
measurements to one common production model.  Such an update would
require knowing in detail all the kinematical cuts used and having the
measured values \emph{before} extrapolating to full phase space,
something not published by all experiments (or we would need to start
from the published full phase space cross-sections and \emph{undo} the
extrapolations made in the original analyses, if we could have access
to the codes used at that moment).  It is always better when the
experiments publish their results in a form which closely reflects the
measurements made, in terms of phase space window and particles
measured.  Extrapolations to full phase space, as well as
``deconvolutions'' of the experimental data to the bare heavy quark
level, will be biased by theoretical prejudice~\cite{CacciariPRL95}.

\begin{table}[ht]
\begin{center}
\begin{tabular}{|l|c|cr|c|}\hline
Decay channel & Experiment & BR used [\%] & & BR (PDG04) [\%] \\
\hline
\rule{0pt}{0.4cm}$b\to \mu X$ & UA1 & $10.2\pm10\%$ & \cite{argus} &
$10.95^{+0.29}_{-0.25}$ \\ \hline
$b\to {\rm J}/\psi\, X$, &
UA1 & $1.12\pm0.18$ & \cite{UA1-JPsi} & $1.16\pm0.10$ \\
& E672/706 &
$1.3\pm0.2\pm0.2$ & \cite{l3} & \\
& E789 & $1.30\pm0.17$ & \cite{pdg94} & \\ \cline{2-5}
\hspace{0.3cm}${\rm J}/\psi \to \mu^+\mu^-$ & UA1 & $6.9\pm0.9$ &
\cite{UA1-JPsi} & $5.88\pm0.10$ \\
& E672/E706 & $5.90\pm0.15\pm0.19$ & \cite{markIII-jPsi} & \\
& E789 & $5.97\pm0.25$ & \cite{pdg94} & \\ \hline
$\B^+\to {\rm J}/\psi\, {\rm K} $ & CDF I& $(10.9\pm 1.0)\times 10^{-2}$ &
\cite{pdg98} & $(10.0\pm0.4)\times 10^{-2}$\\ \hline
\rule{0pt}{0.4cm}$(b\to \D \mu X)^2$ & NA10 & $11.3^2$ & \cite{pdg86} &
$(11.95\pm0.56)^2$ \\
& WA78 & $11.6^2$ & \cite{wa78} & \\ \cline{2-5}
\hspace{0.3cm} $\D\to \mu X$ & NA10 & 11 &
\cite{na10,na10BR} & $9.2\pm0.8$\\
& WA78 & 10.4 & \cite{wa78} & \\ \hline
\end{tabular}
\caption{Decay channels of the identified \BBbar\ events and
  corresponding branching ratios, as used in the original publications and
  in the PDG~2004 tables.  In the E672/E706 case, 
although the combined uncertainty of the branching ratios
is 22\,\%, only 13\,\% was taken into account in the calculation of
the systematic error bar.}
\label{tab:BRbeauty}
\end{center}
\end{table}

\begin{table}[ht!]
\begin{center}
\begin{tabular}{|lr|c|c|c|}\hline
\multicolumn{2}{|l|}{\hfil Experiment \hfil} & Phase space window & \rule{0pt}{0.5cm}$\sigma(\Delta)$ &
$\sigma(4\pi)$ \\
\multicolumn{2}{|l|}{} & (\pt\ in GeV/$c$) & [nb] & [nb] \\ \hline
\rule{0pt}{0.4cm}NA10\,(88)&\cite{na10} & --- & --- & $14^{+7}_{-6}$ \\
WA78\,(89) & \cite{wa78} & --- & --- & $3.6\pm0.4\pm1.1$\\
E653\,(93) & \cite{e653B} & --- & --- & $33\pm11\pm6$\\
E672/706\,(95) & \cite{e672} & $x_{\rm F} > 0$ & $47\pm19\pm14$ & $75\pm31\pm26$
\\
\rule{0pt}{0.4cm}WA92\,(98) & \cite{wa92B} & --- & --- &
$5.7^{+1.3+0.6}_{-1.1-0.5}$\\ \hline
E789\,(95) & \cite{e789B}\rule{0pt}{0.5cm}& $0<x_{\rm F} <0.1$, $p_{\rm T}<2$ & 
--- & $5.7\pm1.5\pm1.3$ \\
\rule{0pt}{0.4cm}E771\,(99) & \cite{e771}
& --- & --- & $43^{+27}_{-17}\pm7$ \\
HERA-B\,(06) & \cite{heraB-beauty2} & 
$-0.35<x_{\rm F}^{{\rm J}/\psi} < 0.15$ & $13.3\pm2.9$ & $14.9\pm 2.2 \pm 2.4$ \\ \hline
\end{tabular}
\caption{\BBbar\ pair production cross-sections measured in
 fixed-target experiments, in the probed phase
 space ($\Delta$) and extrapolated to full phase space ($4\pi$). 
 NA10 quoted a second value, $10\pm5$~nb, obtained with different
 assumptions on the B meson kinematical distributions.}
\label{tab:crossBeauty}
\end{center}
\end{table}
\begin{figure}[ht!]
\centering
\resizebox{0.48\textwidth}{!}{%
\includegraphics*{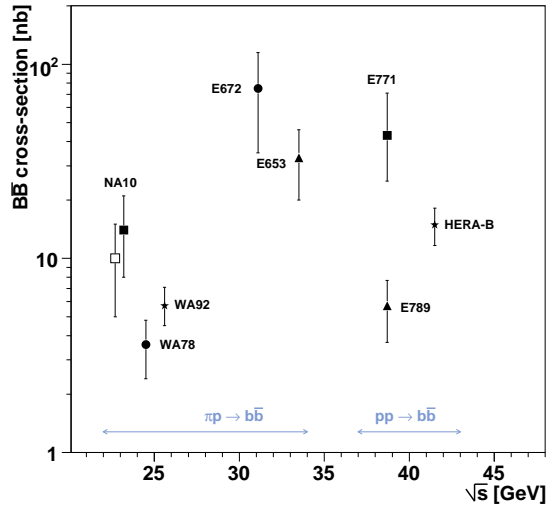}}
\caption{Fixed-target \BBbar\ cross-section measurements, including
  (open square) the cross-section NA10 derived by using alternative B
  hadron kinematical distributions.}
\label{fig:beautyData}
\end{figure}

Table~\ref{tab:BRbeauty} collects the various decay modes in which the
B mesons were looked for, with the corresponding branching ratios, as
used by the different experiments and as given in the PDG 2004 tables.
HERA-B and CDF~II used the PDG 2004 values, while E771 used the
default values of Pythia.
E653 is not listed since it reconstructed a different
decay channel for each of its 18~B mesons\ldots\ Also WA92 explored
several different event topologies, taking the branching ratios from
the PDG 1996 tables~\cite{pdg96}.

In Table~\ref{tab:crossBeauty} and Fig.~\ref{fig:beautyData} we
summarise the \BBbar\ pair production cross-sections measured in
fixed-target experiments, as taken
from the original publications.  Like in the charm sector, all
experiments assumed a linear scaling of the beauty production
cross-section with the mass number of the target nucleus to extract
the elementary hadron-nucleon production cross-section, irrespective
of the phase space window where the measurement was made.

We will now describe the models
used to evaluate detector acceptances and
efficiencies, and to extrapolate the measured cross-sections to the full
phase space values compiled in Table~\ref{tab:crossBeauty} and Fig.~\ref{fig:beautyData}.
We also mention some selection cuts applied on the
kinematics of the detected decay products, often not easy to relate
with the B mesons' kinematical distributions.

\begin{itemize}
\parskip=0.pt \parsep=0.pt \itemsep=0.pt

\item{\bf NA10\,(88)} The NA10 Collaboration measured the \BBbar\
cross-section by triggering on dimuons and looking for a third muon in
the offline analysis. This decay pattern is obtained when both B
mesons decay through the semi-muonic channel, $\B \to \D\, \mu\, X$, and
one of the D mesons also decays into a muon. An offline cut of
$p_{\mathrm T}^\mu >1.4$~GeV/$c$ was required for two out of the three
muons. At the times of NA10 (the article was submitted in 1987) the
theory of beauty production 
was not yet developed in detail and a simple model was
used with the following characteristics: the \BBbar\ pair is assumed
to be the only decay product of an intermediate state with 
the mass 
of the $\Upsilon^{\prime\prime\prime}$. This state is
produced at mid-rapidity with a gaussian distribution in mass, with a
width depending on the centre-of-mass energy~\cite{na10model1}.  The
kinematical distributions were taken to be ${\rm d}^2 \sigma/{\rm
d}{x_{\rm F}}{\rm d}p_{\rm T} \propto(1-|x_{\rm F}|)^3\, p_{\rm
T}\, \exp(-2\, p_{\rm T})$. The general features concerning the B and D
meson decays were taken from Ref.~\cite{ali}. In order to compute the
\BBbar\ cross-section, beauty branching ratios were taken from the
PDG~1986 tables~\cite{pdg86}; the value BR($\D\to \mu +X$) =
11.0\,\% was deduced from Ref.~\cite{na10BR}, assuming that D mesons
are 75\,\% neutral and 25\,\% charged. With these assumptions, a
\BBbar\ cross-section of $14^{+7}_{-6}$~nb/nucleon at $E_{\rm lab} =
286$~GeV was obtained. Alternatively, when NA10 used the \xf\ and \pt\
distributions of Ref.~\cite{na10-altModel}, a \BBbar\
cross-section of $10\pm5$~nb/nucleon is derived. In both cases only
the statistical error is given.

\item{\bf WA78\,(89)} The WA78 Collaboration explored the same decay
modes as NA10, $\pi^- + U \to \B\overline{\B}\,X$ with $\B \to 
\D\,\mu\, X$ and $\D\to \mu\, X$, triggering on two or more
muons. 68 \BBbar\ candidates (including an expected background of 5.2
events) were found in the ``like-sign'' dimuon events passing the
following requirements: $E_{\rm vis} = E_{\rm cal} + \sum{E_\mu} <
300$~GeV, $p_{\rm T,tot} = \sum{p_{\rm T}} > 2.7$~GeV/$c$, $E_{\rm
lept} = E_{\rm beam} - E_{\rm cal} > 100$~GeV, where $E_{\rm cal}$ is
the total energy deposited in the calorimeter.  Another 11 candidates
(including 1.1 estimated background events) were found in the ``three
muon sample'', with $E_{\rm vis} <270$~GeV and $p_{\rm T,tot} >
3$~GeV/$c$. Two theoretical models were used to extract the \BBbar\
production cross-section. The first one was a ``LO QCD model'', based
on the calculations outlined in Ref.~\cite{berger}, with ${\rm d}^3
\sigma /{\rm d}x_{\rm F1} {\rm d}x_{\rm F2}{\rm d}p_{\rm T}^2 = {\rm
d}^2 \sigma /{\rm d}x_{\rm F1} {\rm d}x_{\rm F2} \exp(-p_{\rm T}^2 /
c)$, where $x_{\rm F1}$ and $x_{\rm F2}$ are the Feynman $x$ of the b
and $\bar{\rm b}$ quarks and \pt\ their transverse momentum, with $c
\approx 6.9$~(GeV/$c$)$^2$. From the 2-dimensional distribution ${\rm
d}^2 \sigma /{\rm d}x_{\rm F1} {\rm d}x_{\rm F2}$ given by the model,
the single b quark distribution could be represented by ${\rm d}\sigma
/ {\rm d}x_{\rm F} \propto \exp(-(x_{\rm F} - 0.09)^2 / A_{\pi}^2)$,
with $A_{\pi}^2 \approx 0.3$. In order to evaluate the sensitivity of
the results to the mean value of \xf, its value was changed from 
$\langle x_{\rm F}\rangle=0.09$ to 0.05, increasing the cross-section
value by 25\,\%.
In order to
compare the WA78 results to those of NA10, also the ``NA10 model'',
described above, was used to extract the \BBbar\ cross-section,
yielding a value higher by a factor of $\sim$\,3.8, due to the very
different $x_{\rm F}$ distributions assumed in the
two models. The B meson decays were simulated with the Lund Monte
Carlo program.  If WA78 uses the ``NA10 model'', the two measurements
are in good agreement, but when the QCD-based production model is
used, which seems to provide a better description of the observed
$E_{\rm lept}$ distributions, the agreement suffers significantly.
The B$^0$'s in the observed event sample were an unknown mixture of
B$_{\rm d}^0$ and B$_{\rm s}^0$. Apart from using different production
models, also three different values for the mixing parameter were
explored, $\chi_{\rm B} = 0$, 0.1, 0.2. In the PDG 2004
edition~\cite{pdg04} the value for the mixing parameter for an unknown
mixture of B meson species is $\overline{\chi}_{\rm B} =
0.1257\pm0.0042$.  Therefore, out of the 8 different values given in
the publication for the \BBbar\ production cross-section, we retained
the one corresponding to $\overline{\chi}_{\rm B} = 0.1$ and the LO
QCD model, $3.6\pm0.4\pm1.1$~nb/nucleon. The systematic error, which
is \emph{higher} than the statistical one, includes uncertainties on
the acceptance and track reconstruction efficiency calculations, on
the absolute normalisation and on the branching ratios.

\item{\bf E653\,(93)} The E653 Collaboration searched for beauty
candidates by looking at topological events with muonic secondary
vertices and an associated high \pt\ muon ($p_{\rm T} >
1.5$~GeV/$c$). The \pt\ and \xf\ distributions of the reconstructed
18~beauty mesons were compared to LO~\cite{berger,e653-lo} and
NLO~\cite{nlo-nde,Beenakker,nlo-mnr1} QCD calculations, and to 
Pythia~\cite{e653-pythia}. Within statistical errors, all models gave a satisfactory
description of the experimental data and their differences resulted in
a 10\,\% systematic uncertainty in the shape of the differential
cross-section. Other sources of systematic uncertainties include the
luminosity calculation (5\,\%), the semi-muonic branching ratio
(8\,\%), the detection efficiency model (7\,\%), and others (10\,\%).

\item{\bf E672/E706\,(95)} The E672/E706 Collaboration studied the 
$\B \to {\rm J}/\psi\, X$ decay.
The evaluation of acceptances and detection efficiencies of
beauty hadron pairs was done using the MNR NLO
calculations~\cite{nlo-mnr1,nlo-mnr2}, with hadron momenta equal to that of the
parent quark, i.e.\ without smearing of the momenta due to
fragmentation.  In their phase space window, $x_{\rm F} > 0$, they
measured a \BBbar\ cross-section of $47\pm 19 \pm 14$~nb/nucleon,
based on $8\pm 3.3$ signal events and a background of $2\pm 1$.  The
systematic error includes uncertainties in the normalisation (13\,\%),
branching ratios (13\,\%), b quark production, hadronisation and decay
properties (16\,\%), and reconstruction efficiencies (18\,\%). The
extrapolation to full phase space was done using the MNR NLO
calculations~\cite{nlo-mnr2} but the systematic error also reflects
the use of alternative \xf\ distributions~\cite{e653B, berger} in the
extrapolation.

\item{\bf WA92\,(98)} The WA92 Collaboration classified the 
beauty candidates in three
samples, according to the event topologies. The first class,
composed of semi-muonic decays, had 12 candidates.  The second class
collected 12 multi-vertex events, given the large decay multiplicity of
beauty events. In the third class, the decay channel $\B \to \D X$ was
investigated, giving 5~candidates. To evaluate the acceptances, 
Pythia~\cite{pythia48} and Jetset~\cite{wa92-jetset} were used
with the EHLQ~\cite{ehlq-pdf} parton distributions for the nucleons in
the target and the Owens~\cite{owens-pdf} PDFs for the incoming
pions. The b quark fragmentation was simulated by Jetset, using the
Lund string fragmentation model~\cite{lund}. The PDG 1996~\cite{pdg96}
branching ratios were used.  The \BBbar\ cross-section was then
obtained from the three event samples individually, yielding
compatible results: $\sigma_\mu = 6.2^{+2.4+0.9}_{-1.7-0.6}$,
$\sigma_{\rm mvtx} =5.2^{+2.0+0.8}_{-1.3-0.6}$ and $\sigma_{\D} =
6.0^{+4.0+1.6}_{-2.5-0.8}$~nb/nucleon. Combining these values gave an
overall full phase space beauty cross-section of
$5.7^{+1.3+0.6}_{-1.1-0.5}$~nb/nucleon.  The systematic error includes
a 6\,\% uncertainty from the luminosity calculation and uncertainties
due to the acceptance and efficiency calculations, which were 9\,\%,
12\,\% and 17\,\% for the first, second and third event samples,
respectively. The uncertainties in the branching ratios were not taken
into account.

\item{\bf E789\,(95)} For the evaluation of acceptances and
efficiencies, E789 took
the \xf\ and \pt\ distributions of the b quarks
from the MNR NLO model~\cite{nlo-mnr3,nlo-mnr1}; the intrinsic
transverse momentum was simulated with a Gaussian of sigma 0.5~GeV/$c$.
The b quark fragmentation was
modeled with the Peterson fragmentation function~\cite{peterson},
with $\epsilon = 0.006\pm 0.002$~\cite{chrin}, and the average
lifetime of B hadrons was set to $1.537\pm0.021$~ps, as given in the
PDG 1994 tables~\cite{pdg94}, which also provided the values of the
branching ratios. The \jpsi's coming from $\B \to {\rm J}/\psi\, X$
decays were simulated with the momentum and decay angle distributions
measured by the CLEO Collaboration. The measurement was carried out in
the $0<x_{\rm F}^{{\rm J}/\psi}<0.1$, $p_{\rm T}^{{\rm J}/\psi} <
2$~GeV/$c$ kinematical region, representing 15\,\% of the full phase
space (value quoted in the original publication and reproduced by
Pythia).  Within this phase space domain, a cross-section of $\langle
{\rm d^2}\sigma / {\rm d} x_{\rm F} \, {\rm d} p_{\rm T}^2\rangle =
81\pm21\pm15$~pb/(GeV/$c$)$^2$/nucleon was measured.  The
extrapolation to full phase space yielded a total cross-section of
$5.7\pm 1.5\pm 1.3$~nb/nucleon. The systematic error is dominated by
the evaluation of the luminosity (11\,\%), efficiencies (10\,\%), b
quark production, hadronisation and decay models (8\,\%), fit of the
mass spectrum (5\,\%), and \jpsi\ branching ratio (4\,\%).

\item{\bf E771\,(99)} The E771 Collaboration measured the \BBbar\
production cross-section mainly from the semi-muonic decay of both B
mesons, $\B\overline{\B} \to \mu^+ \mu^-$. Events in which one beauty
hadron decayed through a \jpsi, 
$\B \to {\rm J}/\psi\,X \to \mu^+\mu^-\,X$,
also contributed to the evaluation of the beauty cross-section. The
muons were detected in the range $-0.25<x_{\mathrm F}^\mu <+0.5$, and
in order to pass the muon filter they needed to have $p^\mu >
6$~GeV/$c$ (or $p^\mu > 10$~GeV/$c$, in the angular region near the
beam).  We estimated, using Pythia 6.208, that this phase space window
covers 88\,\% of full phase space.  E771 evaluated acceptances and
efficiencies using Pythia 5.6 and Jetset 7.3~\cite{e771-pythia}. For
the semi-muonic decays, $\B{\overline \B} \to \mu^+\mu^-$, the default
Pythia branching ratios were taken. The main sources of systematic
uncertainties concern the production model (10\,\%), the luminosity
evaluation (5\,\%), the efficiencies (9\,\%), and the semi-muonic
branching ratio (7\,\%).

\item{\bf HERA-B\,(06)} To minimise the systematic uncertainties due
to luminosity and efficiencies, HERA-B measured the \BBbar\ production cross-section
relative to the prompt \jpsi\ production cross-section.
The \jpsi\ mesons due to decays of beauty hadrons were distinguished
from those directly produced at the target thanks to the good vertex
resolution, as done by other experiments.
The b-content of the selected sample was confirmed by a lifetime
measurement, based on the observed decay length.
The beauty events were generated with Pythia~5.7~\cite{heraB-pythia} 
and weighted with
probability distributions obtained from NLO
calculations~\cite{nlo-mnr2,nlo-nde}, using NNLL MRST PDFs~\cite{mrst}, a b
quark mass $m_{\rm b} = 4.75$~GeV/$c^2$, and a QCD renormalisation
scale $\mu = \sqrt{m_{\rm b}^2 + p_{\rm T}^2}$. The
colliding partons were assigned an intrinsic transverse momentum
distributed according to a Gaussian with sigma 0.5~GeV/$c$.
The fragmentation was described by the
Peterson function, with $\epsilon = 0.006$. The
subsequent B hadron production and decay were controlled by Pythia's
default settings.  The ``prompt'' \jpsi\ events were generated taking
into account the differential cross-sections, ${\rm d}\sigma / {\rm d}
p_{\rm T}^2$ and ${\rm d}\sigma / {\rm d} x_{\rm F}$, measured by E771
in p-Si collisions, to properly consider nuclear effects (such as the
``Cronin effect'').
The reference \jpsi\ production cross-section, at the energy of
HERA-B, was derived from a global analysis of all published \jpsi\
cross-section measurements, including the value measured by HERA-B,
with the help of non-relativistic QCD calculations.  A total beauty
production cross-section of $14.9\pm2.2\pm2.4$~nb/nucleon was obtained
by combining the analyses in the muonic and electronic channels of all
collected data, and by 
extrapolating to full phase space the measurement made in the $-0.35 <
x_{\rm F} < +0.15$ window (covering $90.6\pm 0.5$\,\% of $4\pi$).  The
overall systematic uncertainty is 14\,\%, mainly due to the
$\B\to {\rm J}/\psi\, X$ branching ratio (8.6\,\%), the trigger and
reconstruction efficiencies (5\,\%), the b production and decay model
(5\,\%), and the prompt \jpsi\ production model (3.1\,\%).

\end{itemize}

At collider energies, two experiments, UA1 and CDF, measured beauty
production cross-sections in p$\bar{\rm p}$ collisions.
Before we mention specific numerical values, we should clarify that
the experiments are not consistent in the way they quote their
measurements, referring to the production cross-sections of beauty
mesons, beauty hadrons or beauty quarks, sometimes meaning the single
flavour cross-section (only beauty, not anti-beauty), etc.  Each paper
accurately describes what was measured, but it is not trivial to get a
consistent picture of all the measurements.

The UA1 Collaboration measured the beauty production cross-section
at $\sqrt{s} = 630$~GeV, by combining four independent analyses~\cite{ua1} .
For each of these analyses the systematic error was evaluated
including uncertainties from the luminosity evaluation, acceptance and
efficiency calculations, background estimation, and additional
analysis-specific uncertainties. In order to relate the
\emph{measured} cross-sections to the production cross-sections of
\emph{beauty hadrons}, the Monte Carlo model ISAJET~\cite{isajet} was
used, as described in detail in Ref.~\cite{ua1-beauty1}. The Peterson
fragmentation function was used, with $\epsilon =0.02$ 
(softer than the standard 0.006 value). Systematic
errors due to this deconvolution include uncertainties on the
fragmentation function (6\,\%), on the branching ratios and on the
assumed shape of the b quark's transverse momentum (20\,\%). The
inclusive b \emph{quark} and B \emph{hadron} cross-sections for $|y| <1.5$ and
$p_{\rm T} > p_{\rm T, min}$, obtained in this way for each individual
analysis, are summarised in Table~2 of Ref.~\cite{ua1}. The combined
cross-section value, extracted for $p_{\rm T}^{\rm b} > 6$~GeV/$c$ and $|y|
<1.5$, was then extrapolated to full phase space using a NLO QCD
calculation~\cite{nlo-mnr3,nlo-nde}, with factorisation scale $\mu =
\sqrt{(p_{\rm T, min})^2 +m_{\rm b}^2}$ and $m_{\rm b} = 4.75$~GeV/$c^2$. 
The beauty cross-section extrapolated to $p_{\rm T, min} = 0$ was
dominated by the ``low-mass dimuon'' and ``\jpsi'' analyses, which
used $p_{\rm T}^{\mu1} > 3$, $p_{\rm T}^{\mu2} > 2$, and $p_{\rm
T}^{\mu\mu} > 5$~GeV/$c$.  It gave a b quark (beauty only, not anti-beauty)
production cross-section in the central rapidity range, $|y|<1.5$, of
$\sigma(\rm b) = 12.8\pm 4.7\rm (exp)\pm6 (th)$~$\mu$b. Using the
rapidity dependence predicted by the model, UA1 derived a total beauty
production cross-section of $\sigma({\rm b}\bar{\rm b}) = 19.3\pm7\rm
(exp)\pm9(th)$~$\mu$b.  The first error is due to the normalisation of
the theoretical QCD shape to the data; the second error is due to the
extrapolation to full rapidity, from uncertainties on the shape of the
QCD curve.

In Run~I, CDF obtained the beauty production cross-section at $\sqrt{s} = 1.8$~TeV
by reconstructing the exclusive decay $\B^\pm \to {\rm
J}/\psi \,{\rm K} \to \mu^+\mu^- \,{\rm K}$, selecting kaons of
$p_{\rm T} >1.25$~GeV/$c$. For the evaluation of the acceptances, a
Monte Carlo simulation based on a NLO QCD calculations~\cite{nlo-nde,Beenakker}
was performed, using the MRST 
PDFs~\cite{pdf-cdf} and $m_{\rm b}=4.75$~GeV/$c^2$. The
renormalisation and fragmentation scales were both set to
$\sqrt{m_{\rm b}^2+ p_{\rm T}^2}$. The hadronisation into B mesons was
modeled with the Peterson fragmentation function, with $\epsilon =
0.006$.  The $\B \to {\rm J}/\psi \,{\rm K}$ decays were
simulated using a modified version of the CLEO Monte Carlo
program~\cite{cleo-MC}. The B$^+$ meson production cross-section was
measured in the phase space domain $p_{\rm T}^\B > 6.0$~GeV/$c$ and
$|y| <1.0$, and was \emph{not} extrapolated to full phase space. The
value measured was $\sigma_{\B^+}(p_{\rm T}^\B > 6.0, |y| < 1.0) =
3.6\pm 0.4 \rm (stat \oplus syst_{{\it p}_{\rm T}}) \pm 0.4 \rm
(syst)$~$\mu$b, the first error being the quadratic sum of the
statistical and \pt-dependent systematic errors, while the second
is the same for all \pt\ bins.  
The main contribution to the systematic error is the uncertainty
on the (combined) branching ratio (10.2\,\%). Other sources of
systematic uncertainties include the luminosity evaluation, the
influence of in-flight kaon decays, and \pt-dependent trigger
efficiencies.  From the theoretical side, \pt\ dependent uncertainties
were estimated due to the QCD renormalisation scale and to the
Peterson parameter, $\epsilon$.  Meanwhile, the
uncertainty on the combined branching ratio decreased from 10.2\,\% to
4.3\,\%, so that in this case an update would decrease the overall
systematic error (see Table~\ref{tab:BRbeauty}).
Note that the published value is the B$^+$ production cross-section,
obtained dividing by 2 the sum of the B$^+$ and B$^-$ measurements.
The production cross-section of beauty \emph{hadrons}, of a
\emph{single} flavour (only beauty or only anti-beauty) would be a
factor 2.5 higher, according to the PDG 2004 tables~\cite{pdg04},
where the B$^+$ appears as being 39.7\,\% of all beauty hadrons
produced, the rest being B$^0$ (39.7\,\%), B$_{\rm s}^0$ (10.7\,\%),
and b-baryons (9.9\,\%).

In Run~II, at $\sqrt{s} = 1.96$~TeV, CDF searched for beauty decay
topologies in 299\,800 events with a reconstructed \jpsi, leading to
38\,078 beauty hadrons. For the acceptance evaluations, the simulated
distributions ($\eta$, \pt, $z$-vertex, offset, \ldots) were compared
to the reconstructed data and the input distributions tuned until
there was perfect agreement between data and MC. For cases where the
quality of the data is good enough, such an iterative procedure is the
best possible method to evaluate acceptances, since it does not rely
on theoretical inputs. Furthermore, since CDF did not extrapolate the
measured cross-section to full phase space, no theoretical model was
used at all. The systematic errors are, hence, entirely of
experimental origin.  Very detailed descriptions of the \jpsi\ and
beauty analyses are given in Ref.~\cite{cdfII}.  The inclusive single
B hadron cross-section is $\sigma({\rm p}\bar{\rm p} \to \B X,
|y|<0.6) = 17.6\pm 0.4^{+2.5}_{-2.3}~\mu$b.  To be precise, this is
the cross-section for \emph{one} B hadron to be within the $|y|<0.6$
window, irrespective of where the second B is.  We emphasise that this
is the cross-section for the production of any beauty \emph{hadron}
(mesons and baryons; charged or neutral), but only beauty or only
anti-beauty, not both flavours.  Hence, this is the same numerical
value as the \BBbar\ pair cross-section, $\sigma(\B\overline{\B})$.  It is
also the same as the b quark pair cross-section, $\sigma(\rm b\bar{\rm
  b})$, since bottomonium production is negligible.

CDF observed~\cite{cdf} that the Run~I measurement of the B$^+$
production cross-section, for $p_{\rm T}^\B > 6.0$~GeV/$c$, was higher
than the NLO calculations~\cite{nlo-nde,Beenakker}, by a factor $2.9\pm0.2\rm
(stat \oplus syst_{{\it p}_{\rm T}})\pm 0.4(syst)$.  However, Cacciari
and Nason~\cite{cacciari1} pointed out that the discrepancy is not
really due to the perturbative calculations of the cross-section but
rather to the use of the Peterson fragmentation function with
$\epsilon = 0.006$.  Using smaller values of $\epsilon$ in the
framework of a NLL (next-to-leading-log) calculation reduced the
discrepancy to $1.7\pm0.5\rm (experiment)\pm0.5(theory)$ (including a
better treatment of the theoretical error bars). For more information
on this topic, see Refs.~\cite{cacciari1,cacciari2}.  It is worth
mentioning that the Run~II and Run~I measurements agree with each
other, when compared for $p_{\rm T}^\B > 6.0$~GeV/$c$.

To clarify the source of the disagreement between the CDF data and the
NLO QCD calculation, the Tevatron ran for nine days at
$\sqrt{s}=630$~GeV~\cite{cdf630}, so that CDF could measure the ratio
between the beauty quark production cross-sections at 630 and
1800~GeV.  The UA1 measurement, at 630~GeV, had not shown a
significant departure from expectations, but was affected by a larger
uncertainty.  It is obvious that, both experimentally and
theoretically, it is much more accurate to obtain a \emph{ratio} of
cross-sections at two different energies than each of the absolute
values, since many uncertainties cancel out.  Given the very short
beam time available, a highly efficient beauty finding algorithm had
to be developed, using single moderate-\pt\ muon triggers, combining
the muon with a high-\pt\ track and selecting the high-mass muon-track
combinations with positive lifetime.  The two data sets (630 and
1800~GeV) were as similar as possible, with all data collected between
December 1995 and February 1996, and with identical online and offline
event selections.  The cross-section ratio was reported for \pt\ of
the beauty quark above 10.75~GeV/$c$, 
because Monte Carlo simulations showed that 90\,\% of the
reconstructed and identified beauty events have a \pt\ above this
threshold.
The measured ratio, between the beauty quark
production cross-sections for \pt\ above 10.75~GeV/$c$ at 630 and
1800~GeV, is $0.171\pm0.024\rm (stat)\pm0.012\rm (syst)$, in very good
agreement with the theoretical NLO QCD
prediction~\cite{nlo-nde,nlo-mnr3}, as shown in Figs.~3 and 4 of
Ref.~\cite{cdf630}.  The conclusion is that the disagreement between
the CDF data and the NLO QCD calculation is essentially the same at
630~GeV and at 1800~GeV.  It can also be seen that the CDF beauty
cross-section at 630~GeV, derived by combining the ratio with the
absolute value measured at 1800~GeV,
and extrapolated from $|y|<0.6$ to $|y|<1.5$,
is about a factor 2 higher than
the value of UA1, for roughly the same \pt\ range, as shown in Fig.~5
of Ref.~\cite{cdf630}.

\section{Pythia calculations versus data}
\label{sec:calculations}

After having collected and revised the open charm and open beauty
production cross-sections measured by many experiments, we will now
compare them with theoretical calculations made using the Pythia Monte
Carlo event generator~\cite{Pythia6}.

Our first motivation is to describe the energy dependence of the
measured production cross-sections, so as to calculate the values
expected for energies relevant to recent or on-going experiments at
the CERN SPS (NA60) and at RHIC (PHENIX and STAR).  However, the yield
of D or B mesons to be expected in a given detector depends not only
on the production cross-section but also on the kinematical
distributions of the single mesons, on the pair correlations, etc.
Therefore, we will also see in this section how the results of Pythia
compare to the available differential data, and how sensitive they are
to changes in certain settings.

Pythia is an easily accessible tool which has been and continues to be
extensively used by many experiments interested in heavy flavour
production (to evaluate acceptances, efficiencies and other important
elements of the data analyses procedures).  It may very well be,
however, that better calculations of heavy flavour production, in one
or another aspect, are provided by other computer codes, such as
Herwig~\cite{Herwig}, ISAJET~\cite{isajet}, MC@NLO~\cite{MCatNLO}, or
MNR NLO~\cite{nlo-mnr2,nlo-nde}.  Surely, a work similar to the one we present
in this section could and should be made using those other theoretical
models.

In some of our calculations, we varied the set of PDFs, the mass of
the charm quark and the definition of the $Q^2$ scale, as will be
explained below.  The intrinsic transverse momentum of the colliding
partons, $k_\mathrm{T}$, was generated according to a Gaussian
distribution with a width determined by the parameter PARP(91):
$\exp(-k_{\rm T}^2/{\rm PARP(91)}^2)$.  This expression is often
written as $\exp(-k_{\rm T}^2/\langle k_{\rm T}^2\rangle)$, defining
PARP(91) as $\sqrt{\langle k_{\rm T}^2\rangle}$.
We used the Lund string fragmentation scheme~\cite{lund}, modified for
heavy flavoured quarks by Bowler~\cite{bowler}, which provides a good
description of the high-precision data obtained by the SLD, ALEPH,
DELPHI and OPAL Collaborations~\cite{kerzel}.

We generated charm and beauty events setting ${\rm MSEL} = 4$ and~5,
respectively, to ensure that every generated event has a \ccbar\ (or
\bbbar) pair.  This is particularly important when simulating
collisions at relatively low energies, where heavy flavour production
is a rare process.  Besides the LO diagrams (quark-antiquark
annihilation and gluon fusion, see Section~\ref{sec:hvinpqcd}), the
calculations also include initial and final state radiation.  However,
they do not include ``flavour excitation'' and ``gluon splitting''
diagrams.
These can also be generated by Pythia, but only in the context of a
full minimum bias generation (${\rm MSEL} = 1$), a very inefficient
operation mode at low collision energies (see
Refs.~\cite{norrbinI,norrbinII} for details).
The absence of higher-order diagrams is (partially) compensated with
scaling up the calculations by empirical \emph{K-factors}, under the
hypothesis that the kinematical distributions do not change
significantly between leading and higher-order diagrams.

\subsection{Absolute \ccbar\ production cross-sections}
\label{sec:fitSigma}

As a first step, we used Pythia's default values for the mass of the c
quark ($m_{\rm c} = 1.5$~GeV$/c^2$) and for the definition of the
$Q^2$ scale, while varying the set of parton distribution
functions. For the proton PDFs we used the CTEQ6L (2002)~\cite{cteq6},
MRST~LO (2001)~\cite{mrst} and GRV~LO (1998)~\cite{grv} sets. Since
NLO PDFs have also been used with Pythia in the past, despite the
inconsistency resulting from the fact that Pythia does not include
higher-order diagrams, we have also used the CTEQ6M (2002) and
MRST~c-g (2001) sets for the extraction of the K-factors.  To describe
the parton densities in pions we used the following sets:
GRV-P~LO~(1992), SMRS-P2~(1992), GRV-P~HO~(1992) and ABFKW-P~Set~1
(1989). As indicated by the year, the available pion PDFs are
considerably older than the proton PDFs.  Furthermore, there is only
one set of LO pion PDFs.  For the simulations with $\pi$ beams we have
used MRST~LO (2001) to describe the target nucleons, since they were
fit with a $\Lambda_{\rm QCD}$ value, 220~MeV, similar to the values
used in the $\pi$ PDFs, $\Lambda_{\rm QCD}=190$--231~MeV.

\begin{figure}[!ht]
\centering
\resizebox{0.48\textwidth}{!}{%
\includegraphics*{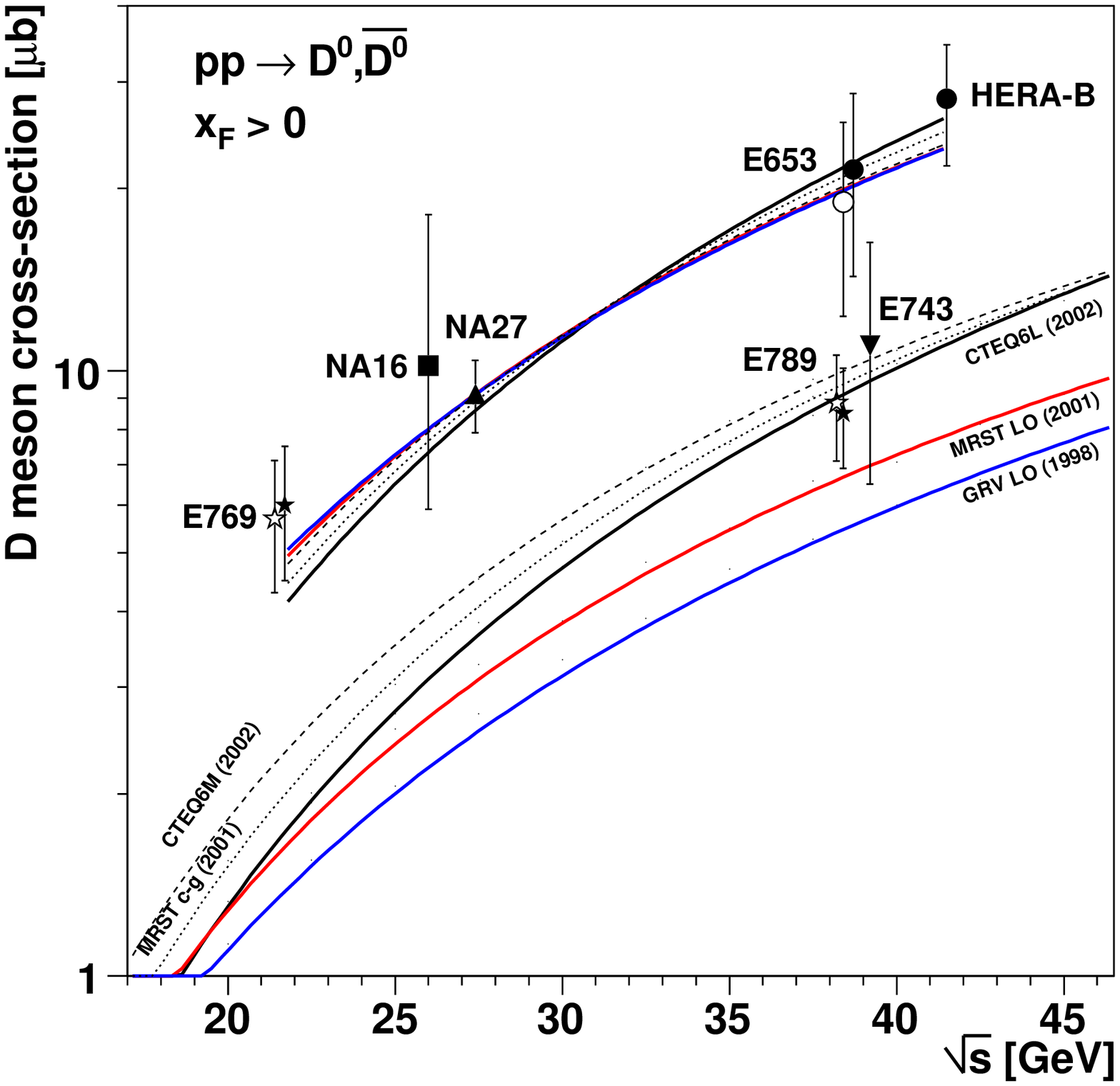}}
\resizebox{0.48\textwidth}{!}{%
\includegraphics*{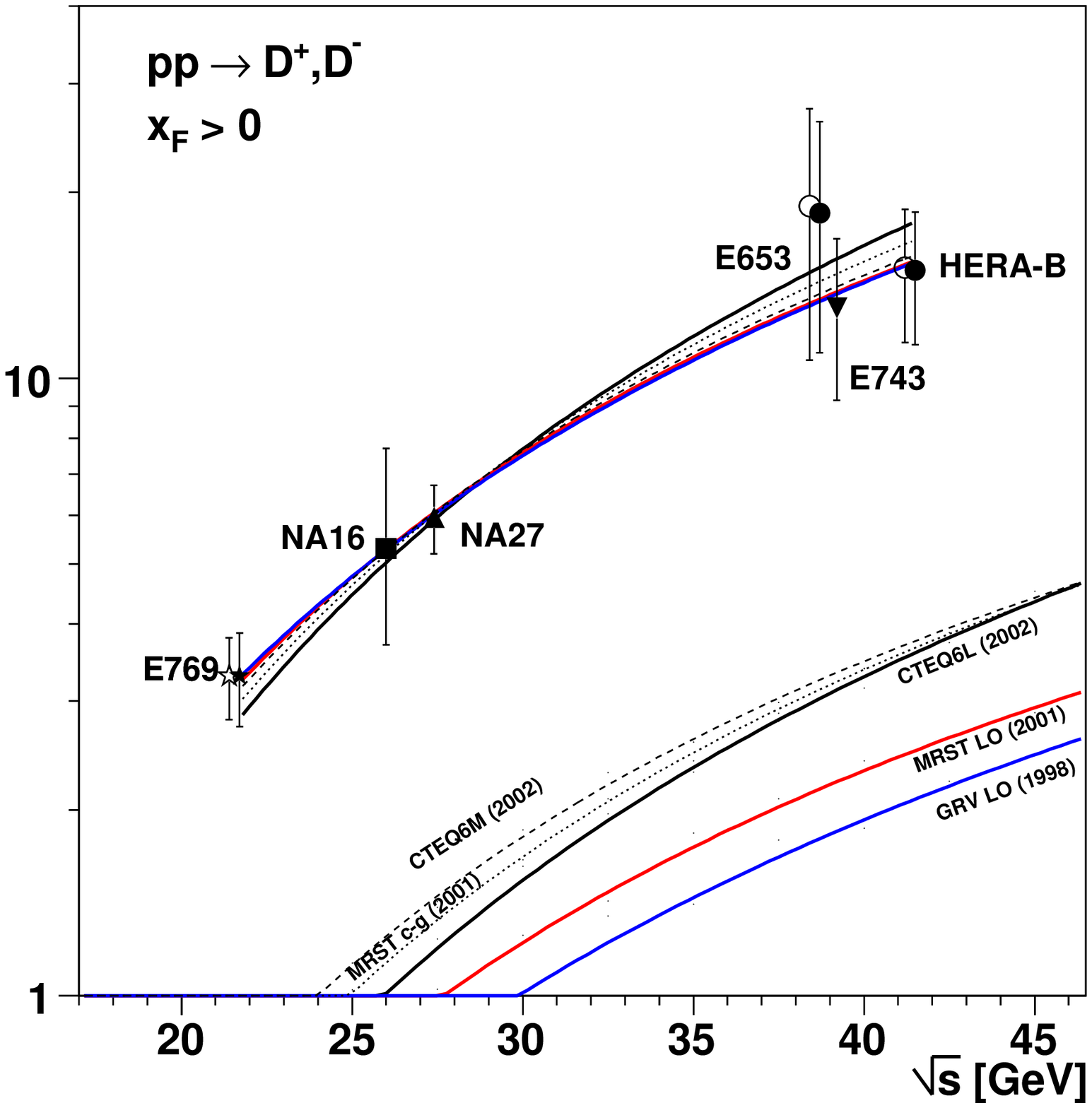}}\\
\resizebox{0.48\textwidth}{!}{%
\includegraphics*{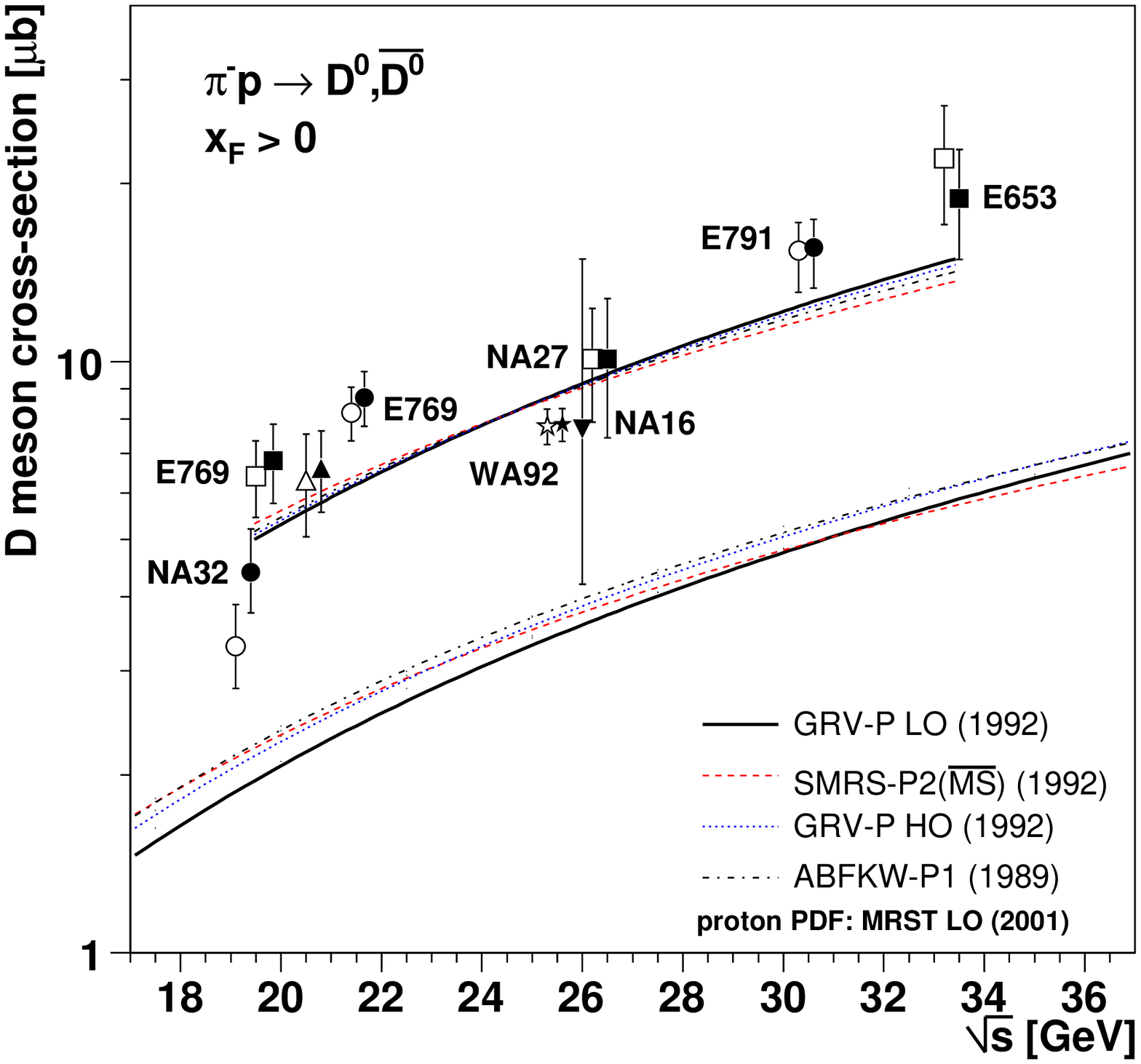}}
\resizebox{0.48\textwidth}{!}{%
\includegraphics*{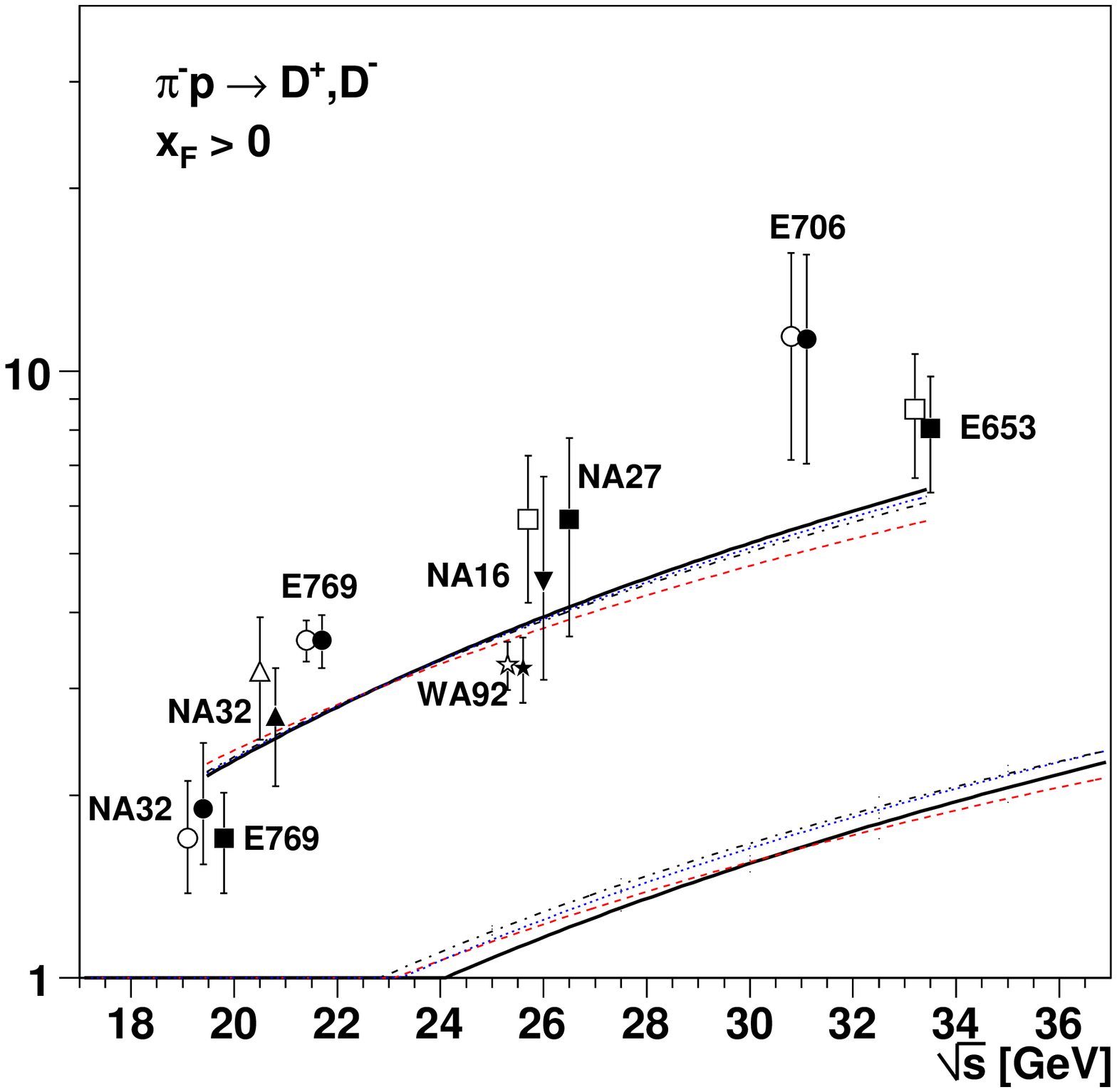}}
\caption{Energy dependence of the production cross-section of neutral
(left) and charged (right) D mesons, in proton (top) and pion (bottom)
induced collisions. The upper curves were scaled by K-factors fitted
to the data (excluding the E789 point).}
\label{fig:Pythia}
\end{figure}

To easily compare the results obtained, we parameterised the $\sqrt{s}$
dependence of the \ccbar\ production cross-section with the expression
\begin{equation} \label{eq:charmProd}
\sigma_{{\rm c\bar{c}}}(\sqrt{s}) = p_0\cdot
\bigg(1-\frac{p_1}{\sqrt{s}\,^{p_3}}\bigg)^{p_2}\quad, 
\end{equation}
inspired by a formula commonly used to describe the energy dependence
of J/$\psi$ production~\cite{carlos}.
The parameter $p_3$ was fixed to 0.35 in pp collisions and to 0.3 in
$\pi^-$p collisions, while the other parameters were adjusted to each
specific set of PDFs.  The calculated curves for neutral and charged D
meson production in pp and $\pi^-$p collisions are presented in
Fig.~\ref{fig:Pythia}, where they are compared to the corresponding
experimental measurements, Tables~\ref{tab:ccCorr-p}
and~\ref{tab:ccCorr-pi}, normalised to $x_{\rm F}>0$.

In several cases, the same measurement is represented twice, before
(open marker) and after (solid marker) the corrections explained in
Section~\ref{sec:dataCharm}. Note that the measurements of E743, E653
and E789 were all performed at $E_{\rm lab}=800$~GeV; to improve their
visibility, they were slightly displaced in Fig.~\ref{fig:Pythia}
(without affecting any calculations).  The same is valid for the NA16
and NA27 $\pi^-$p measurements, both performed at 360~GeV.  Each curve
appears twice in the figures, as directly calculated by Pythia and
after being scaled up by a K-factor fitted to the data. The solid
lines correspond to the LO PDF sets. The E789 p-Au measurement of the
neutral D meson cross-section was not considered in the fitting of the
K-factors, since it would have significantly deteriorated the quality
of the fits, maybe because this measurement was made in an
exceptionally small phase space window ($0<x_{\rm F} < 0.08$, $p_{\rm
T} < 1.1$~GeV/$c$).  We remind that the HERA-B measurement is still
preliminary.

\begin{table}[ht]
\begin{center}
\begin{tabular}{|l|c|c|} \hline
\rule{0pt}{0.45cm} p beam 
 & $\sigma(\D^0) + \sigma(\overline{\D^0})$
 & $\sigma(\D^+) + \sigma({\D}^-)$ \\ \hline
CTEQ6L (2002)  & 2.4 & 5.0 \\
MRST LO (2001) & 3.0 & 6.2 \\
GRV LO (1998)  & 3.6 & 7.4 \\
CTEQ6M (2002)  & 2.0 & 4.2 \\
MRST c-g (2001)& 2.2 & 4.6 \\ \hline
\rule{0pt}{0.45cm} $\pi^-$ beam 
 & $\sigma(\D^0) + \sigma(\overline{\D^0})$
 & $\sigma(\D^+) + \sigma({\D}^-)$ \\ \hline
GRV-P LO (1992) & 2.6 & 3.4 \\
SMRS-P2 ($\overline{\rm MS}$) (1992) & 2.4 & 3.1 \\
GRV-P HO (1992) & 2.4 & 3.1 \\
ABFKW-P1 (1989) & 2.3 & 3.0 \\ \hline
\end{tabular}
\caption{K-factors obtained from fitting the experimental data points
with Pythia's default settings, for several sets of PDFs. The
K-factors of the proton and pion data have a relative uncertainty of 9
and 5\,\%, respectively. The MRST~LO (2001) PDFs were used for the
target nucleons in the $\pi^-$ induced collisions.}
\label{tab:cckFac}
\end{center}
\vglue-5mm
\end{table}

Using the appropriate
K-factor for each curve, we can see from Fig.~\ref{fig:Pythia} that
all of them describe the \sqrts\ dependence of the data points within
the experimental error bars.  The reduced $\chi^2$ values of the fits
are around 1 or lower for the proton data and around 2 for the pion
data.

Table~\ref{tab:cckFac} shows the K-factors which best describe the
four sets of data.  To reproduce the measured data points, the
K-factor required by the charged D mesons is significantly higher than
the one needed by the neutral mesons. For instance, in pp collisions
the CTEQ6L and MRST~LO sets require K-factors around 5 and 2.5 for the
charged and neutral D mesons, respectively.
The significant difference between the charged and neutral D meson
K-factors, required by the experimental data, indicates a problem in
the probabilities used by Pythia when fragmenting the charm quarks
into D hadrons: in $\sim$\,63\,\% of the cases, Pythia fragments the
charm quarks into D$^0$ or $\overline{\D^0}$ mesons, while the
probability of producing D$^+$ or D$^-$ mesons is only 20\,\% (the
remaining charm quarks hadronise essentially into D$_{\rm s}$ or
$\Lambda_{\rm c}$).  This is due to the larger feed-down contribution
from D$^*$ decays to neutral D mesons.  It is easy to derive the ratio
between the charged and neutral D meson yields as given by Pythia; it
follows directly from the assumption that the c quark has equal
probabilities of fragmenting into the neutral and charged D (singlet)
or D$^*$ (triplet) states, and from the numerical values of the
branching ratios of the D$^*$ to D decays~\cite{pdg04}:
\begin{equation}
\begin{split}
\frac{\sigma(\D^+)}{\sigma(\D^0)} 
 &= \frac{\sigma(\D^+_{\rm direct}) 
+ \sigma(\D^{*+})\cdot B(\D^{*+}\to\D^+) 
+ \sigma(\D^{*0})\cdot B(\D^{*0}\to\D^+)}
{\sigma(\D^0_{\rm direct})
+ \sigma(\D^{*+})\cdot B(\D^{*+}\to\D^0) 
+ \sigma(\D^{*0})\cdot B(\D^{*0}\to\D^0)}=\\
 &= \frac{0.25 + 0.75\cdot 0.323 + 0.75\cdot 0.0}
{0.25 + 0.75\cdot 0.677 + 0.75\cdot 1.0} = 0.33
\quad .
\end{split}
\end{equation}

\begin{figure}[h!]
\centering
\resizebox{0.48\textwidth}{!}{%
\includegraphics*{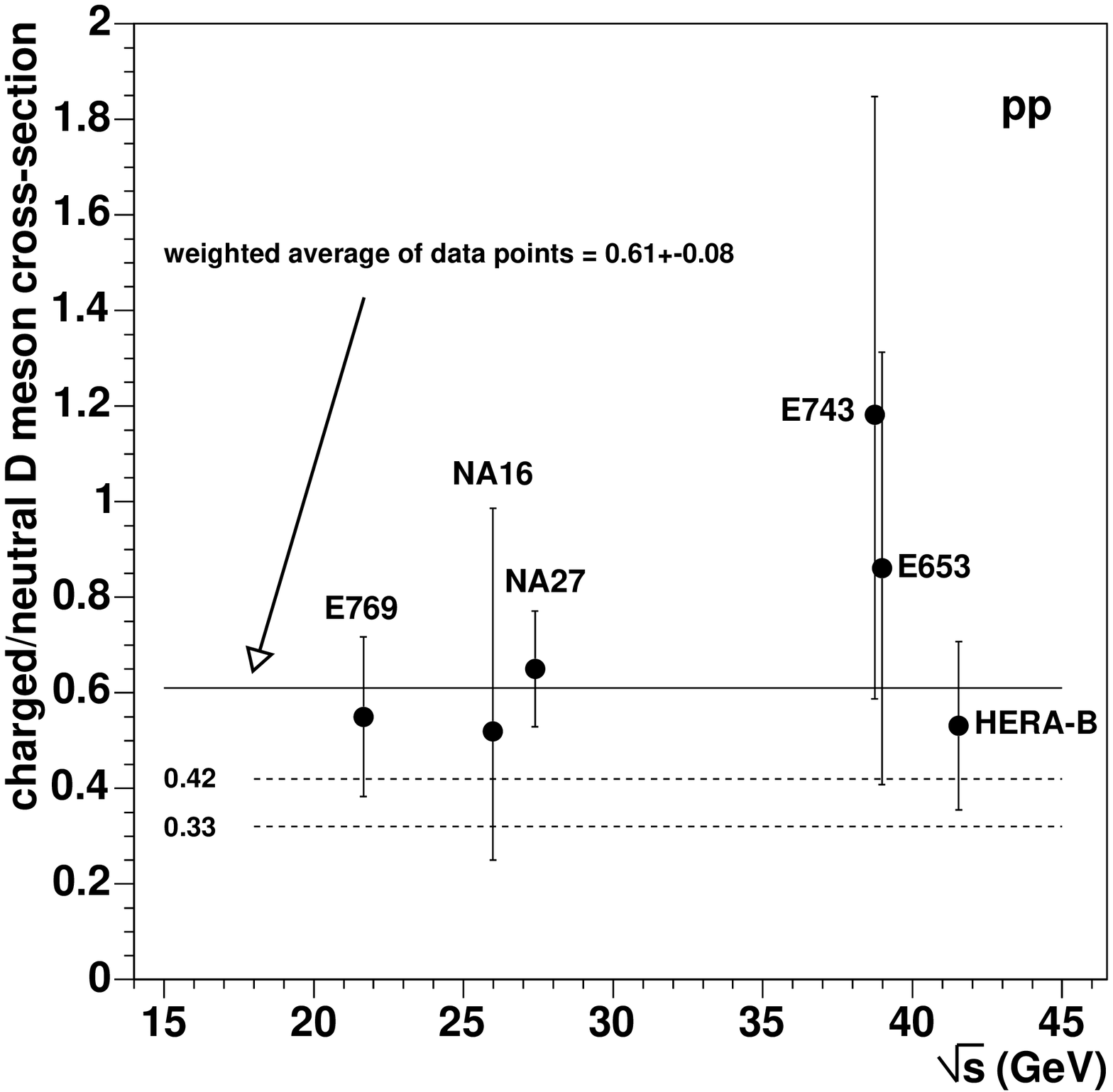}}
\resizebox{0.48\textwidth}{!}{%
\includegraphics*{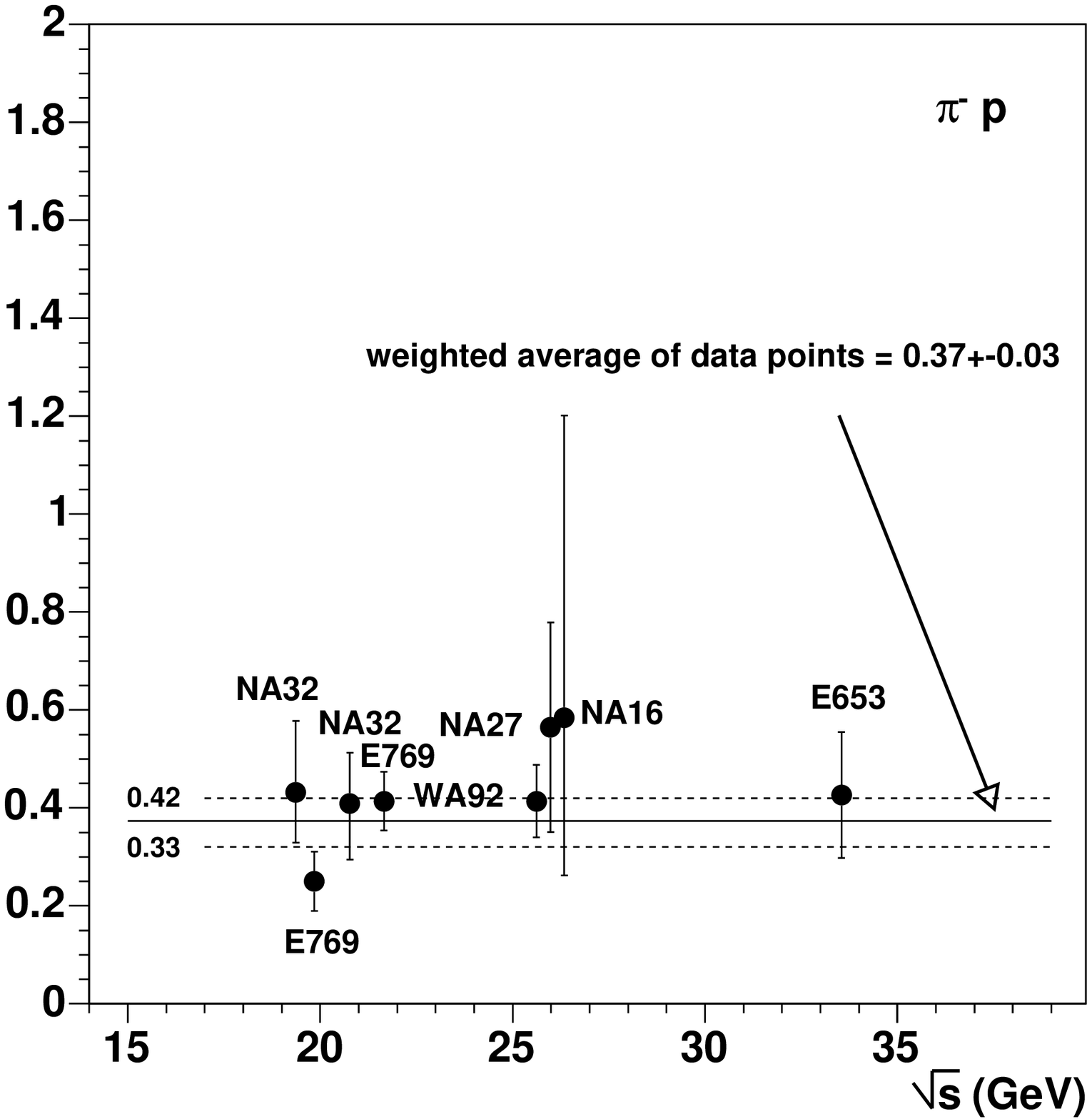}}
\caption{Ratio between charged and neutral D meson cross-sections, for
pp (left) and $\pi^-$p (right) collisions, compared to the values 0.33
and 0.42, expected when using $P_{\rm V}=0.75$ and 0.6, respectively.}
\label{fig:charOverNeu}
\end{figure}

\begin{figure}[t!]
\centering
\resizebox{0.8\textwidth}{!}{%
\includegraphics*{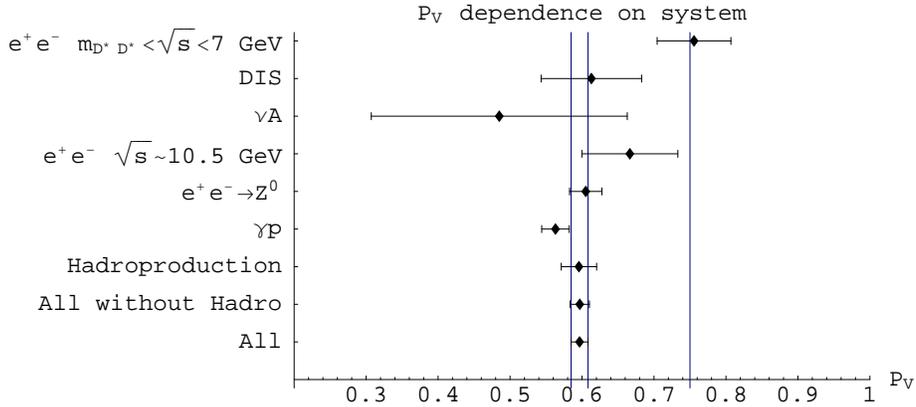}}
\caption{Fraction of directly produced vector D$^*$ states to the
total, vector and pseudo-scalar states, averaged over the measurements
made for each of the indicated production systems. The central band
represents the world average $0.596\pm0.012$, quite different from the
spin-counting expectation, 0.75.  Figure taken from Ref.~\cite{andre}.}
\label{fig:pv}
\end{figure}

Figure~\ref{fig:charOverNeu} shows that this ``expected value'' is
considerably lower than the ratios given by the hadro-production
experiments which measured both charged and neutral D mesons.
The average values of the measured ratios are $0.61\pm0.08$ in p-A,
$0.37\pm0.03$ in $\pi^-$p collisions, and $0.41\pm0.03$ if we merge
both data sets.  This discrepancy is due, at least in part, to the
fact that (na\"{\i}ve) spin counting is not applicable to charm
production: it would imply that the charm quark fragments into a
vector state with a probability $P_{\rm V}=3/(3+1)=0.75$, while all
available measurements (in hadro-production, photo-production, at LEP,
etc) give a combined value of $P_{\rm V}=0.59\pm0.01$~\cite{andre},
as illustrated in Fig.~\ref{fig:pv}.  This is presumably because the
mass difference between the D and D$^*$ mesons cannot be neglected,
making the lighter D mesons more probable to be produced.
In Pythia, the value of $P_{\rm V}$ is given by the parameter
PARJ(13), which is 0.75 by default.  Setting it to 0.6 provides a much
better agreement with all the presently available measurements, as
just mentioned, and increases the ratio between the charged and
neutral D meson production cross-sections from 0.33 to 0.42.

We will now assume that the ratio between charged and neutral D meson
production cross-sections is the same in proton and pion induced
collisions, as suggested by the $P_{\rm V}$ universality seen in
Fig.~\ref{fig:pv} (and consistent with the left panel of
Fig.~\ref{fig:charOverNeu}, given the large error bars of the proton
data).  This means that we can use a common K-factor for charged and
neutral D mesons, as long as we set $P_{\rm V}=0.6$.
Table~\ref{tab:commonKfactors} gives the K-factors extracted in this
way, for several sets of PDFs.

\begin{table}[htb]
\begin{center}
\begin{tabular}{|l|c|} \hline
\hfil PDF set \hfil & K-factor \\ \hline
CTEQ6L (2002) & 3.0 \\
MRST LO (2001) & 3.8 \\
GRV LO (1998) & 4.6 \\ \hline
CTEQ6M (2002) & 2.6 \\
MRST c-g (2001) & 2.8 \\ \hline
\end{tabular}
\caption{K-factors obtained from simultaneously fitting the charged
  and neutral D meson p-A data with Pythia calculations, setting $\rm
  PARJ(13)=0.6$.  These values have an uncertainty, from the fit, of
  $\sim$\,8\,\%.}
\label{tab:commonKfactors}
\end{center}
\vglue-4mm
\end{table}

It is unfortunate that, until now, Pythia uses the same parameter,
PARJ(13), to also define the $P_{\rm V}$ value for beauty production,
where the value 0.75 seems to be appropriate.  Indeed, measurements
made by ALEPH, DELPHI, OPAL and L3 give an average value of $P_{\rm
V}=0.75\pm0.04$~\cite{andre}.  It would be more useful to have
different parameters for charm and for beauty, so that
they could be given different values (as done for the u/d and s
quarks, through the PARJ(11) and PARJ(12) parameters).
This is not so crucial if the charm and beauty events are
independently generated (with $\rm MSEL=4$ or 5, respectively) but
becomes mandatory if we generate both heavy flavours simultaneously,
as frequently done.

\begin{figure}[h!]
\centering
\resizebox{0.48\textwidth}{!}{%
\includegraphics*{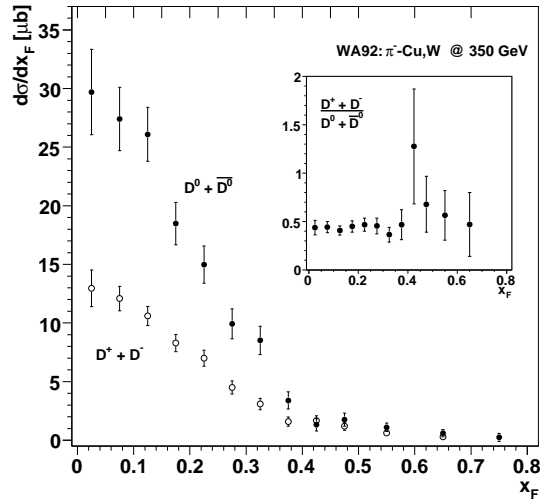}}
\caption{$x_{\rm F}$ distributions for charged and neutral D mesons,
as measured by WA92 in pion induced collisions. The inset shows their
ratio.}
\label{fig:wa92xF}
\end{figure}

Figure~\ref{fig:wa92xF} shows that the ratio between charged and
neutral D meson cross-sections does not seem to depend on $x_{\rm F}$,
according to the WA92 measurements~\cite{wa92}.  This ratio also seems
to be independent of the collision energy, when comparing measurements
made by many experiments (see Fig.~\ref{fig:charOverNeu}).

We will now concentrate on the total \ccbar\ production cross-sections
for pp collisions, in our evaluation of the influence of certain
parameters on Pythia's results.

\begin{figure}[ht!]
\centering
\resizebox{0.48\textwidth}{!}{%
\includegraphics*{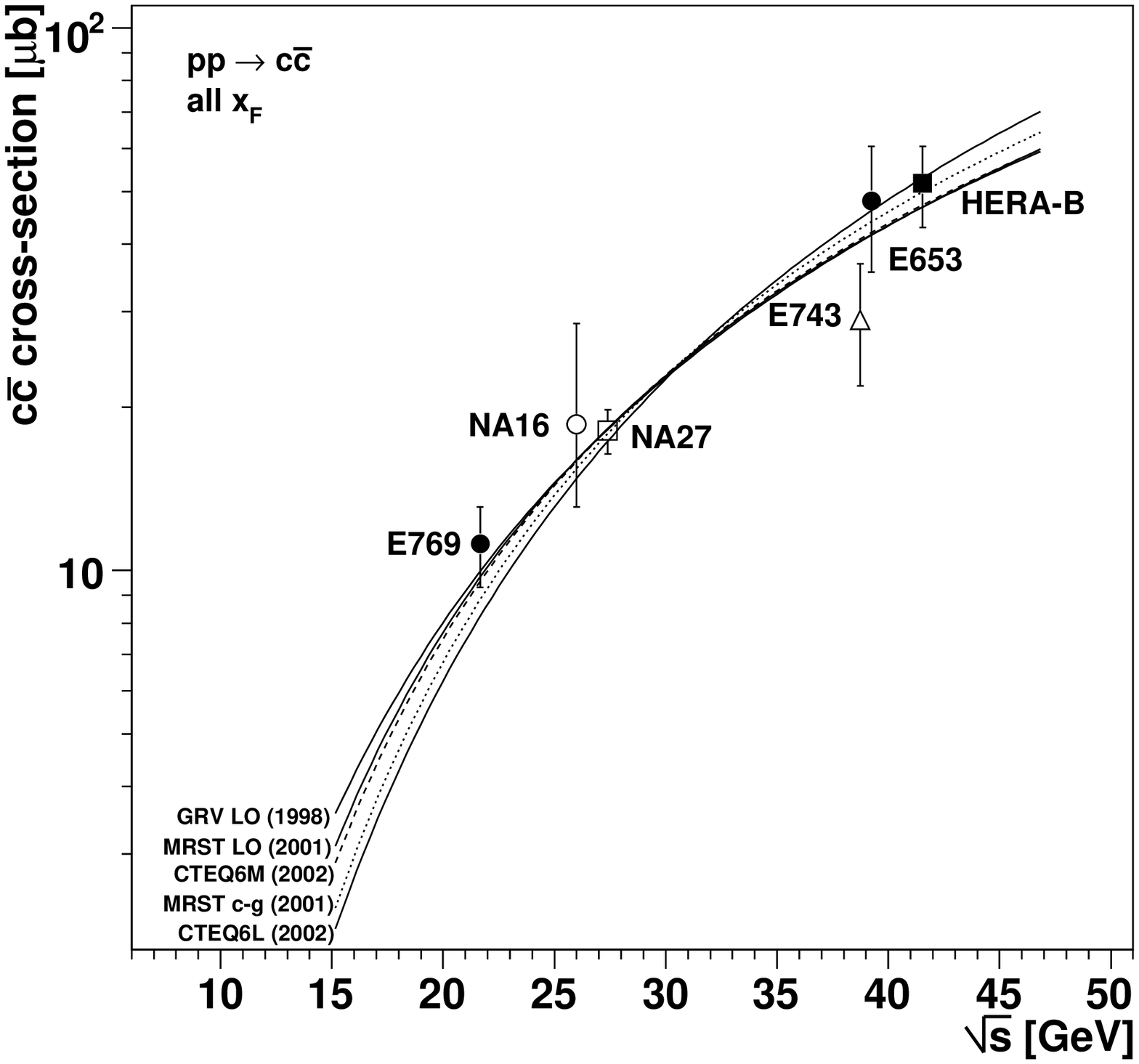}}
\resizebox{0.48\textwidth}{!}{%
\includegraphics*{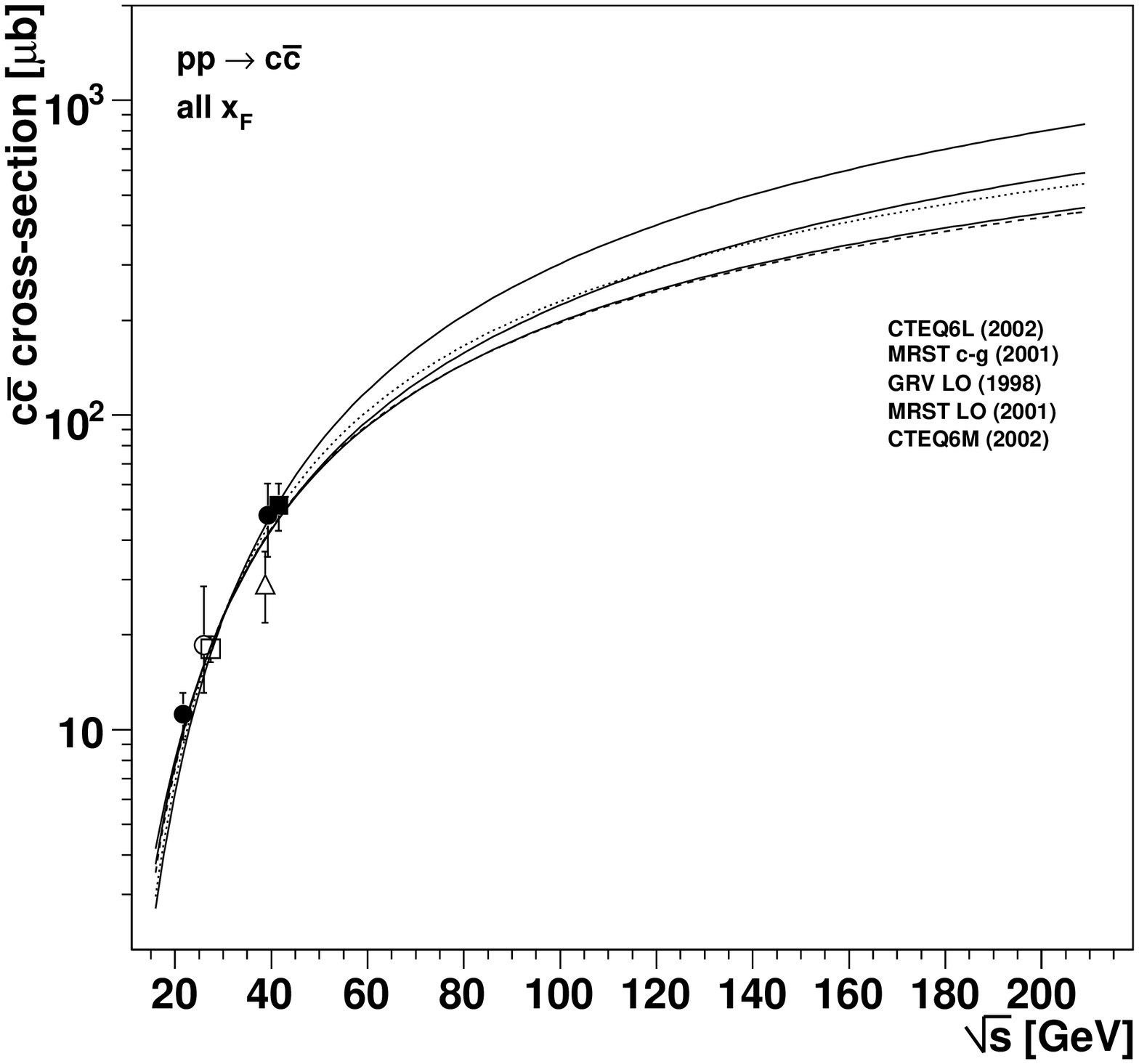}}
\caption{Total \ccbar\ production cross-sections for fixed-target
  energies (left) and up to $\sqrt{s} = 200$~GeV (right).  Open
  symbols indicate the pp measurements.}
\label{fig:pdfVar}
\end{figure}
\begin{table}[h!]
\begin{center}
\begin{tabular}{|c|c|c||c|c|c|} \cline{4-6}
\multicolumn{3}{}{} & \multicolumn{3}{|c|}{$\sigma_{\rm c\bar{c}}~[\mu$b]}
\\  \hline 
PDF set & K-factor & $\chi^2$/ndf & $E_{\rm lab} = 158$ & $E_{\rm
  lab} = 400$ &  $\sqrt{s} = 200$ \\ \hline
CTEQ6L (2002)  & 3.0 & 1.4 & 3.6 & 17.5 & 803 \\
MRST LO (2001) & 3.8 & 0.7 & 4.8 & 18.4 & 439 \\
GRV LO (1998)  & 4.5 & 0.7 & 5.2 & 18.3 & 563 \\ \hline
CTEQ6M (2002)  & 2.5 & 0.7 & 4.6 & 18.4 & 427 \\
MRST c-g (2001)& 2.7 & 1.0 & 4.0 & 18.0 & 524 \\ \hline
\end{tabular}
\caption{K-factors which provide the best description of the \ccbar\
  data in pp and p-A collisions, for each PDF set. The values have a
  relative uncertainty of around 7\,\%.  The last three columns give
  the elementary pp cross-sections calculated by Pythia with these
  K-factors, for three different energies, given in GeV.}
\label{tab:ccbarPDF}
\end{center}
\end{table}

Figure~\ref{fig:pdfVar} compares the measured \ccbar\ production
cross-sections with Pythia's curves, obtained with five different PDF
sets.  The curves are labelled in the order, from top to bottom, in
which they occur at $\sqrt{s}=15$ and 200~GeV, in the left and right
panels, respectively.  The extracted K-factors for each PDF set are
summarised in Table~\ref{tab:ccbarPDF}, together with the
corresponding pp cross-sections for typical SPS and RHIC energies. At
$\sqrt{s}=200$~GeV, the calculated pp cross-sections vary between 440
and 800~$\mu$b, if we only consider LO PDFs.

\begin{figure}[ht!]
\centering
\resizebox{0.5\textwidth}{!}{%
\includegraphics*{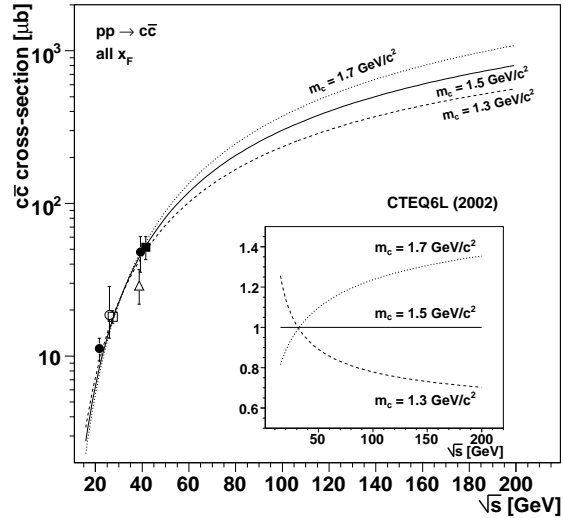}}
\caption{LO \ccbar\ production cross-sections for $m_{\rm c}=1.3$,
1.5 and 1.7~GeV$/c^2$, scaled up by the appropriate K-factor. The
inset shows the corresponding ratios.  See the text for details.}
\label{fig:mcVar}
\end{figure}

In a second step we studied the influence of varying the mass of the c
quark by $\sim$\,15\,\% with respect to Pythia's default.
We used the CTEQ6L PDFs for these calculations, but other proton PDFs
give similar results.  Figure~\ref{fig:mcVar} shows the calculated
\ccbar\ production cross-section in pp collisions, using $m_{\rm c}
=1.3$, 1.5 and 1.7~GeV$/c^2$, scaled up by the appropriate K-factor.
While smaller $m_{\rm c}$ values lead to higher calculated
cross-sections, in particular at energies close to threshold, after
the curves are normalised using the available fixed target data, and
given the somewhat different shapes of the calculated curves, it turns
out that at $\sqrt{s} = 200$~GeV the estimated \ccbar\ cross-section
is 35\,\% \emph{higher} with $m_{\rm c} =1.7$~GeV$/c^2$ and 30\,\%
\emph{lower} with $m_{\rm c} =1.3$~GeV$/c^2$, with respect to the
default value.  The results are
summarised in Table~\ref{tab:ccbarMass}.

\begin{table}[ht!]
\begin{center}
\begin{tabular}{|c|c|c||c|c|c|} \cline{4-6}
\multicolumn{3}{}{} & \multicolumn{3}{|c|}{$\sigma_{{\rm c\bar{c}}}~[\mu$b]}
\\  \hline 
$m_{\rm c}$ [GeV$/c^2$] & K-factor & $\chi^2$/ndf & $E_{\rm lab} = 158$ & $E_{\rm
  lab} = 400$ &  $\sqrt{s} = 200$ \\ \hline
1.3 & 1.2 & 0.9 & 4.5 & 18.1 & 569 \\ 
1.5 & 3.0 & 1.4 & 3.7 & 17.5 & 811 \\
1.7 & 6.6 & 2.0 & 3.2 & 16.9 & 1100 \\ \hline
\end{tabular}
\caption{Same as previous table, when varying the mass of the c quark,
  $m_{\rm c}$.  The CTEQ6L PDFs are used.  See the text for details.}
\label{tab:ccbarMass}
\end{center}
\end{table}

\begin{figure}[htb]
\centering
\resizebox{0.5\textwidth}{!}{%
\includegraphics*{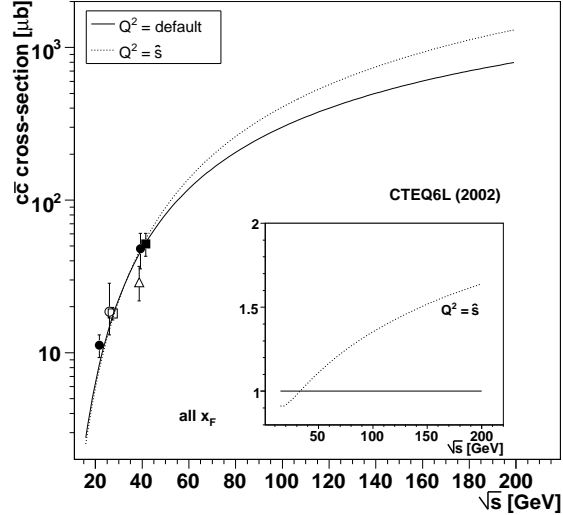}}
\caption{\ccbar\ production cross-sections, for two different $Q^2$
  definitions, scaled up by the appropriate K-factors. The inset shows
  the ratio of the curves, taking Pythia's default as the reference.}
\label{fig:q2Var}
\end{figure}

Different definitions of the squared energy-momentum transfer, $Q^2$,
can be used. 
To evaluate the influence of this setting on our results, we replaced
Pythia's default, equivalent to $Q^2 = \hat{m}_\mathrm{T}^2$ in the
processes we are studying, by $Q^2=\hat{s}$,
the choice of Refs.~\cite{pbm,phenix02}.
Figure~\ref{fig:q2Var} shows the effect of using these two different $Q^2$
definitions on the \ccbar\ cross-section, keeping $m_{\rm c} =
1.5$~GeV$/c^2$ and using the CTEQ6L PDFs.

We see that using $\hat{s}$ as the $Q^2$ definition leads to
significantly lower cross-sections with respect to the values obtained
when using Pythia's default setting.  The difference is energy
dependent: at low energies the cross-sections obtained with the
$\hat{s}$ definition are around 3 times lower, while at $\sqrt{s} =
200$~GeV the difference reduces to a factor of~2. Once the curves are
scaled up to describe the data, the steeper rise with \sqrts\ of the
$Q^2 = \hat{s}$ curve leads to 60\,\% higher cross-sections at
$\sqrt{s}=200$~GeV, with respect to the values obtained with the
default setting. The results are summarised in
Table~\ref{tab:ccbarQ2}.  Calculations with other PDF sets give
comparable results.

\begin{table}[h!]
\begin{center}
\begin{tabular}{|c|c|c||c|c|c|} \cline{4-6}
\multicolumn{3}{}{} & \multicolumn{3}{|c|}{$\sigma_{\rm c\bar{c}}~[\mu$b]}
\\  \hline 
$Q^2$ definition & K-factor & $\chi^2$/ndf & $E_{\rm lab} = 158$ & $E_{\rm
  lab} = 400$ &  $\sqrt{s} = 200$ \\ \hline
default & 3.0 & 1.4 & 3.8 & 17.5 & 808 \\ 
$\hat{s}$ & 8.8 & 1.9 & 3.4 & 16.9 & 1327 \\ \hline
\end{tabular}
\caption{Same as previous tables, when varying the $Q^2$ definition,
  keeping $m_{\rm c} = 1.5$~GeV$/c^2$ and using the CTEQ6L PDFs.
  Note the high K-factor required by the $Q^2=\hat{s}$ choice.}
\label{tab:ccbarQ2}
\end{center}
\end{table}

These calculations show that the \ccbar\ production cross-section at
$\sqrt{s} = 200$~GeV, as derived from Pythia's calculations normalised
by the existing SPS, FNAL and HERA-B measurements, can vary by $\pm
30$\,\% due to the use of different sets of PDFs and by around $\pm
30$\,\% if the c quark mass is changed by $\pm 15$\,\%.  Furthermore,
using $Q^2 = \hat{s}$, as done by some experiments, leads to a 60\,\%
higher \ccbar\ cross-section at $\sqrt{s} = 200$~GeV. From
Table~\ref{tab:ccbarPDF}, where we used the default c quark mass and
$Q^2$ definition, and only considering the values corresponding to the
LO PDF sets, we obtain $600\pm 30\,\%~\mu$b as our best estimate for
the \ccbar\ cross-section at $\sqrt{s} = 200$~GeV.  The uncertainties
of the calculations for SPS energies are smaller.  At $E_{\rm lab} =
158$ and 400~GeV, we expect total \ccbar\ cross-sections of $4.5\pm
20\,\%~\mu$b and $18\pm 5\,\%~\mu$b, respectively.

\subsection{Single D meson kinematical distributions}
\label{sec:kinem}

In this section we compare measured D meson kinematical distributions
to calculations done with Pythia. These distributions are sensitive to
non-perturbative effects, such as the intrinsic transverse momentum
which the partons carry before they collide and the fragmentation of
the heavy quarks into hadrons.

Table~\ref{tab:Dkin} gives an overview of the experiments
which measured \pt\ and \xf\ distributions of D mesons produced in p-A
and $\pi$-A collisions.
\begin{table}[p]
\begin{center}
\begin{tabular}{|l|lc|l|l|c|}\hline
Experiment & \multicolumn{2}{l|}{Beam \hfill $E_{\mathrm{lab}}$} & Target
& particle & events \\ 
    & & [GeV] & & &\\ \hline
NA27 \hfill{\cite{na27p}} & p & 400 & p & \rule{0pt}{0.45cm}D$^0$,
$\overline{{\rm D}^0}$ & 29, 22 \\ 
 & & & & D$^+$, D$^-$ & 24, 27 \\
 & & & & D$^+ + \D^0$ (non-leading) & 53 \\
 & & & & D$^- + \overline{\D^0}$ (leading) & 49 \\
NA32 \hfill{\cite{na32-88}} & p & 200 & Si & D$^0 + \overline{{\rm
    D}^0} + \D^\pm$ & 9 \\
E743 \hfill{\cite{e743}} & p & 800 & p & D$^0 + \overline{{\rm
    D}^0}+\D^\pm$ & 31 \\ 
E653 \hfill{\cite{e653p}} & p & 800 & emulsion & D$^0+\overline{{\rm
    D}^0}+\D^\pm$ & 146 \\
WA82 \hfill{\cite{wa82kin}} & p & 370 & Si,\,W & D$^0+\overline{{\rm
    D}^0}+\D^\pm$ & $266\pm28$ \\ 
E789 \hfill{\cite{e789}} & p & 800 & Au & D$^0+\overline{{\rm D}^0}$ &
$\sim 2200$ \\ 
E769 \hfill{\cite{e769kin}} & p & 250 & Be,\,Al,\,Cu,\,W & $\D^0+
\overline{{\rm D}^0} + \D^\pm + \D_{\rm s}^\pm$ & $320\pm 26$ \\
CDF \hfill{\cite{cdfCharm}} & $\overline{\rm p}$ & 1960 & p & D$^0+
\overline{{\rm D}^0}$ & $36\,804\pm 409$ \\
 & & & & D$^\pm$ & $28\,361\pm 294$ \\
 & & & & D$^{*\pm}$ & $5\,515\pm 85$\\
 & & & & D$_{\rm s}^\pm$ & $851\pm 43$\\ \hline
NA27 \hfill{\cite{na27kin}} & $\pi^-$ & 360 & p & D$^0 + \overline{{\rm
    D}^0} + \D^\pm$ & 57\rule{0pt}{0.45cm}\\
NA11 \hfill{\cite{na11-1}} & $\pi^-$ & 200 & Be & D$^0 + \overline{{\rm
    D}^0} + \D^\pm$ & 29 \\
NA11 \hfill{\cite{na11-2}} & $\pi^-$ & 200 & Be & D$^-$ (leading) & 44\\
 & & & & D$^+$ (non-leading) & 30\\
NA32 \hfill{\cite{na32-88}} & $\pi^-$ & 200 & Si & D$^- + \D^0$
(leading) & 54\\ 
 & & & & D$^+ + \overline{{\rm D}^0}$ (non-leading) & 60\\
 & & & & D$^{*\pm}$ & 46\\
NA32 \hfill{\cite{na32-91}} & $\pi^-$ & 230 & Cu &
D$^0$+$\overline{{\rm D}^0}$, D$^\pm$ & 543, 249\\
 & & & & D$_{\rm s}^\pm$, D$^{*\pm}$ & 60, 147\\
 & & & & D$^- + \D^0+ \D^{*-}$ (leading) & 427\\
 & & & & D$^+ +\overline{{\rm D}^0}+ \D^{*+}$ (non-l.) & 425\\
E653 \hfill{\cite{e653pi}} & $\pi^-$ & 600 & emulsion & D$^0+
\overline{{\rm D}^0} + \D^\pm$ & 676 \\
WA75 \hfill{\cite{wa75}} & $\pi^-$ & 350 & emulsion &  D$^0+
\overline{{\rm D}^0}+\D^\pm$ & 459 \\
WA82 \hfill{\cite{wa82kin}} & $\pi^-$ & 340 & Si,\,Cu,\,W &  D$^0 +
\overline{{\rm D}^0}+\D^\pm$ & $2\,214\pm 70$ \\
WA82 \hfill{\cite{wa82kinII}} & $\pi^-$ & 340 & Si,\,Cu,\,W &  
D$^+$ & $322\pm 20$ \\
 & & & & D$^-$ & $449\pm 23$ \\
E769\hfill{\cite{e769kin}} & $\pi^\pm$ & 250 & Be,\,Al,\,Cu,\,W &D$^0 + 
\overline{{\rm D}^0}+ \D^\pm + \D_{\rm s}^\pm$ & $1\,665\pm 54$ \\WA92 
\hfill{\cite{wa92}} & $\pi^-$ & 350 & Cu,\,W & D$^0 + \overline{{\rm
    D}^0} + \D^\pm$ & $7\,172\pm 108$ \\
E706 \hfill{\cite{e706}} & $\pi^-$ & 515 & Be,\,Cu & D$^\pm$ & 110 \\
E791 \hfill{\cite{e791}} & $\pi^-$ & 500 & C & D$^0 + \overline{\D^0}$
& $88\,990\pm 460$ \\ \hline 
\end{tabular}
\caption{Experiments measuring \xf\ and \pt\ distributions of D
  mesons. $\D^\pm$ stands for $\D^+ + \D^-$ (and $\D_{\rm s}^\pm$
  likewise). E789 and CDF only provide \pt\ distributions, measured in
  $0<x_{\rm F}<0.08$ (E789) or in $|y|<1.0$ and $p_{\rm T} > 5.5$ (or
  6)~GeV/$c$ (CDF).}
\label{tab:Dkin}
\end{center}
\end{table}
Some experiments separately give the distributions of ``leading'' and
``non-leading'' particles. A D meson is called ``leading'' if its c or
$\bar{\rm c}$ quark combines with one of the non-interacting beam or
target valence quarks. These quarks have in general a much higher
momentum than a light quark from the sea, so that the formed D meson
will have a visible (forward or backward) boost. In pp and p-A
reactions the D$^-$ and $\overline{{\rm D}^0}$ are the leading
particles.  In $\pi^-$ induced reactions, for $x_{\rm F} > 0$, the
D$^-$ and D$^0$ are leading.  This purely non-perturbative
(hadronisation) effect is properly described by the Lund string
fragmentation model, as shown in Refs.~\cite{norrbinI, norrbinII}.
See end of next section for further information.

\bigskip

\begin{figure}[ht!]
\centering
\resizebox{0.48\textwidth}{!}{%
\includegraphics*{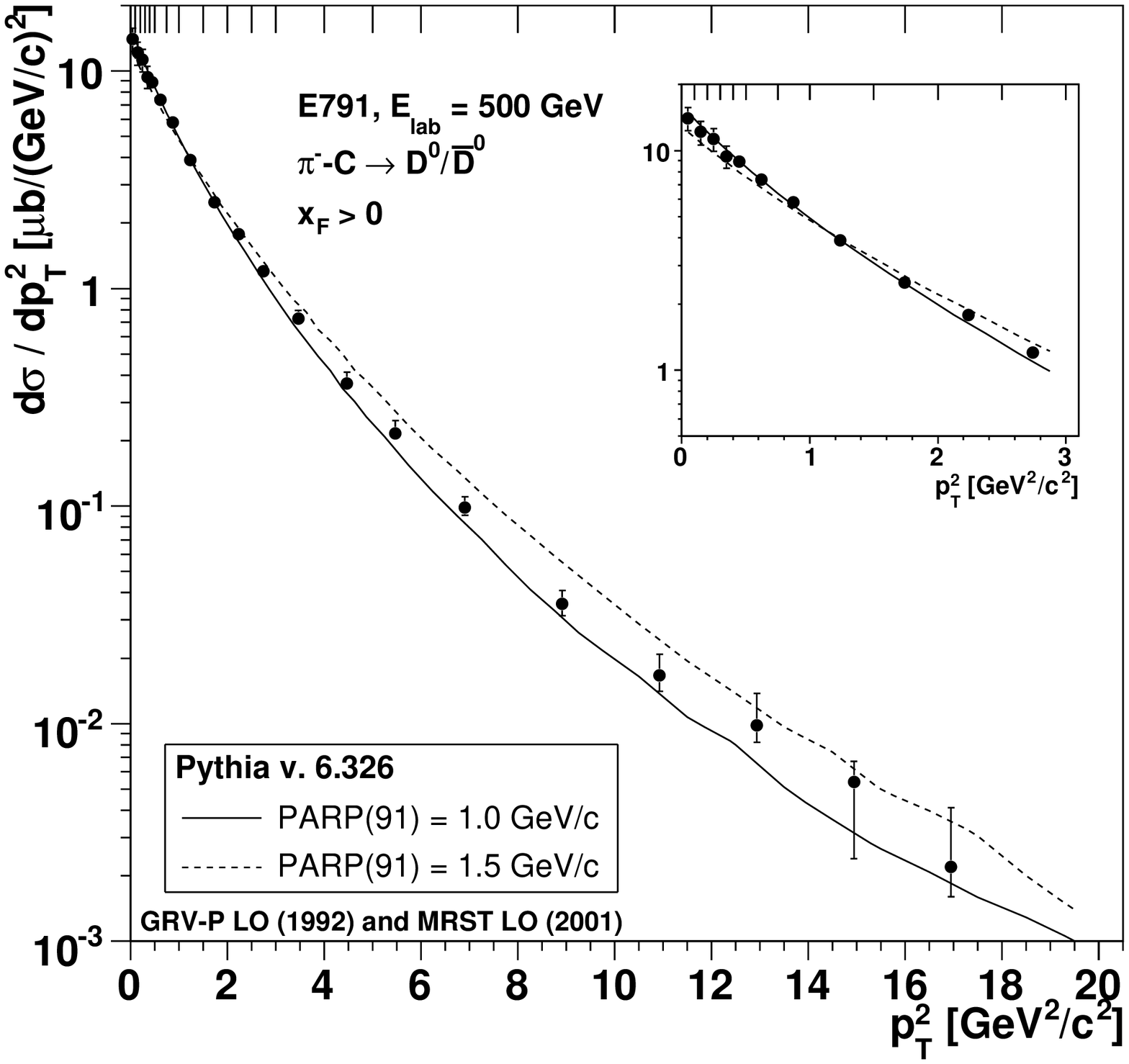}}
\resizebox{0.48\textwidth}{!}{%
\includegraphics*{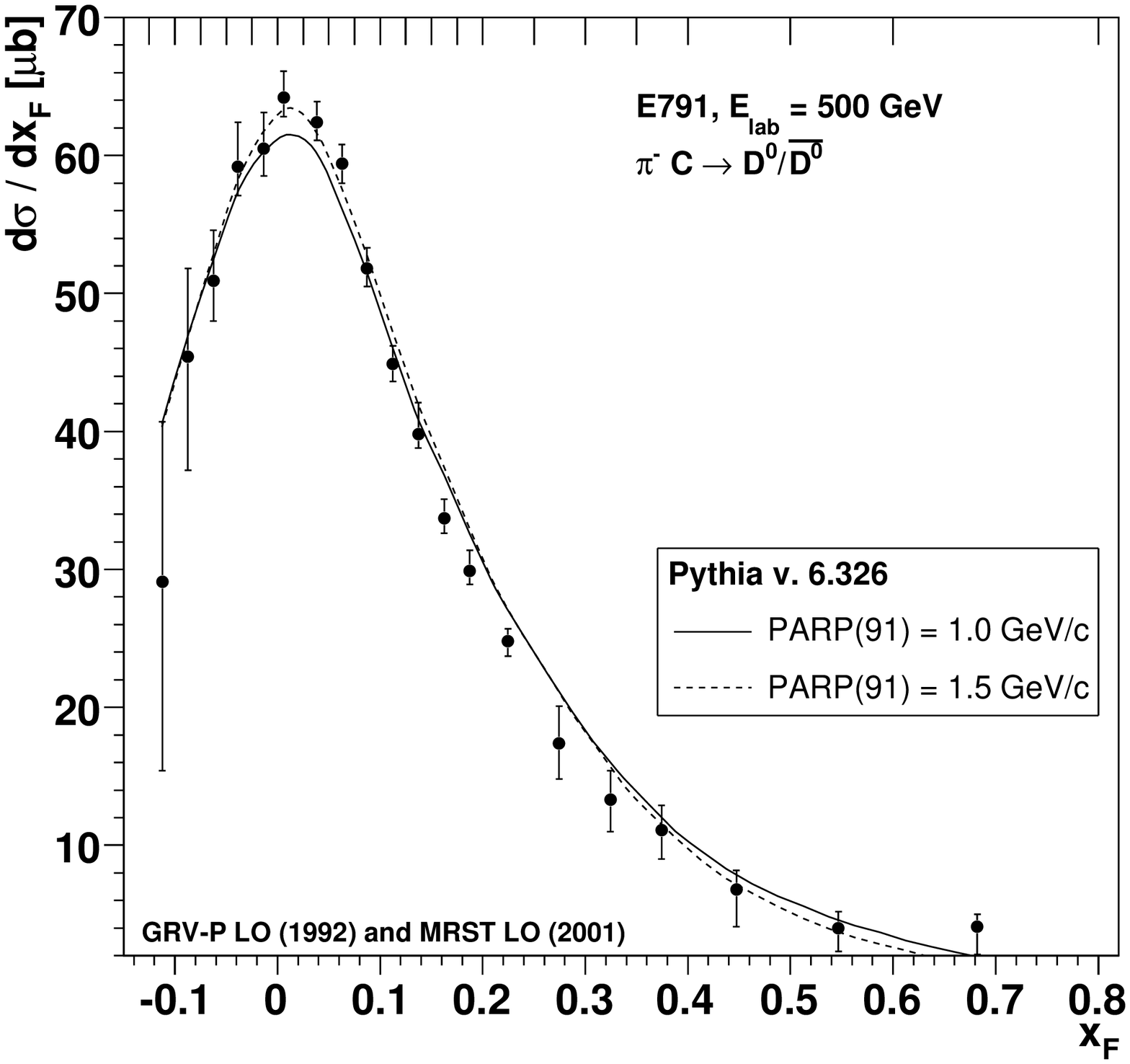}}
\caption{$p_\mathrm{T}^2$ (left) and $x_\mathrm{F}$ (right)
 distributions, as measured by the E791 Collaboration in 500~GeV
 $\pi^-$-C collisions and as calculated by Pythia, version 6.326.}
\label{fig:kinE791}
\end{figure}

We will now focus on the \pt\ and \xf\ distributions of single D
mesons.  The E791 Collaboration measured the $p_{\rm T}^2$ and \xf\
distributions of neutral D mesons, produced in 500~GeV $\pi^-$-C
collisions~\cite{e791}. This measurement represents by far the largest
D meson data sample at fixed-target energies, with $\sim 90\,000$
fully reconstructed neutral D mesons. Figures~5 and 6 of
Ref.~\cite{e791} show comparisons between the measured distributions
and curves calculated with Pythia 5.7 and Jetset
7.4~\cite{heraB-pythia}. With these code versions, of 1994, the
calculations were unable to reproduce the measured distributions. On
the contrary, the most recent version of Pythia, 6.326, describes
quite well\,\footnote{When we made calculations with version
6.325 we noticed that the resulting \pt\ distributions were abnormally
hard.  T.~Sj\"ostrand, main author of Pythia, immediately identified
the problem, introduced in version 6.319, and sent us the solution,
later on implemented in Pythia 6.326.} the $p_{\rm T}^2$ and \xf\
distributions of E791, with PARP(91) between 1 and 1.5~GeV/$c$, as can
be seen in Fig.~\ref{fig:kinE791}.

Note that from version 6.314 (Oct.~2004) onwards the default value of
PARP(91) changed from 1.0 to 2.0~GeV/$c$.  The E791 calculations, made
with version 5.7, used an even lower PARP(91) value, 0.44~GeV/$c$.
Also the default value of the c quark mass has changed, from
1.35~GeV/$c^2$ in version 5.7 to 1.5~GeV/$c^2$ in the latest versions.
The data points in Fig.~\ref{fig:kinE791} were placed at $x$ values
weighted by the function given in the original
publication~\cite{e791}, rather than at the bin centre, following the
procedure explained in Ref.~\cite{lafferty}.  The vertical lines at
the top of the figures indicate the bin edges. The Pythia curves were
scaled up by 2.6, the K-factor extracted from the fit to the neutral D
meson measurements (see Table~\ref{tab:cckFac}).

\begin{figure}[ht!]
\centering
\resizebox{0.48\textwidth}{!}{%
\includegraphics*{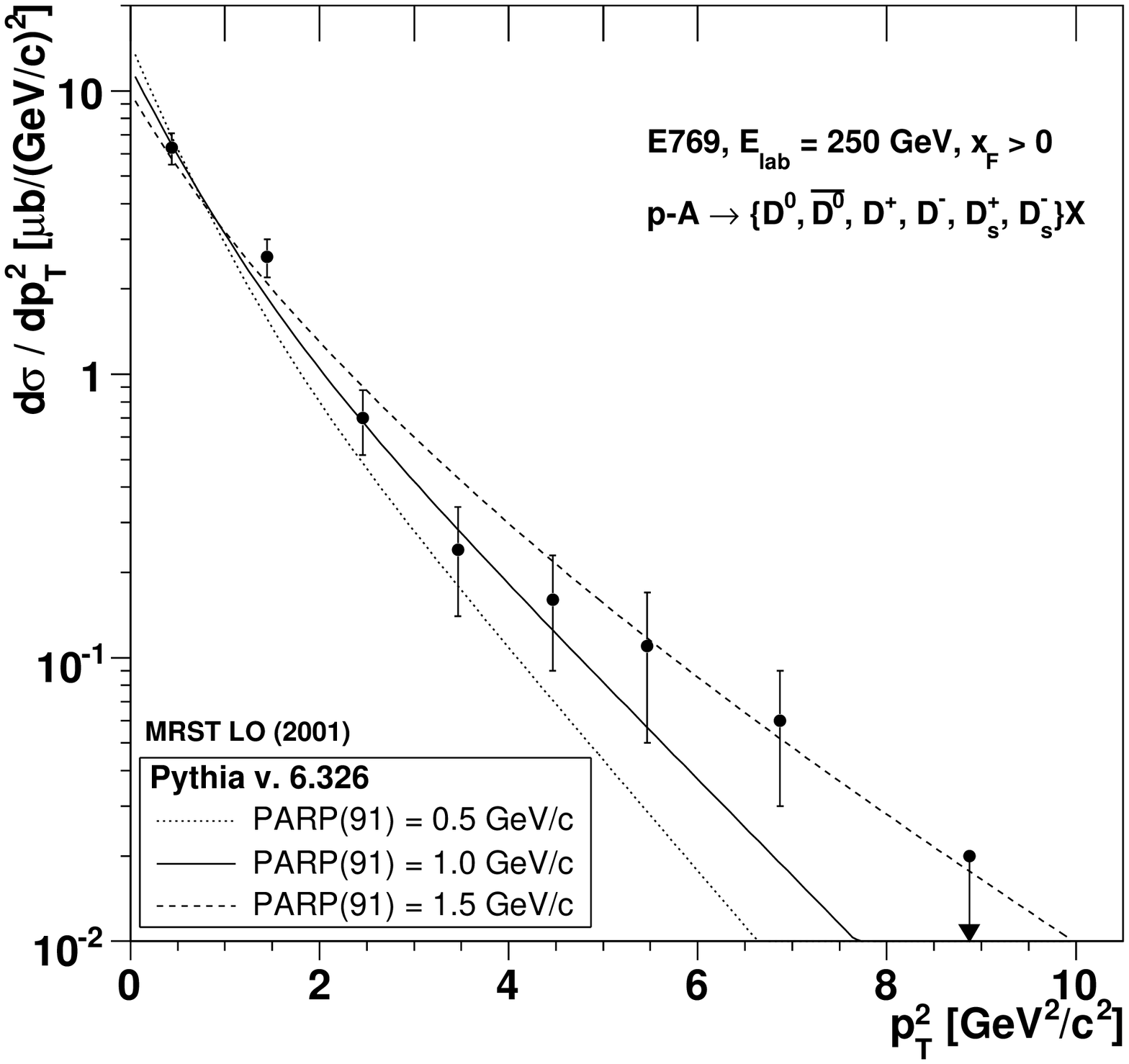}}
\resizebox{0.48\textwidth}{!}{%
\includegraphics*{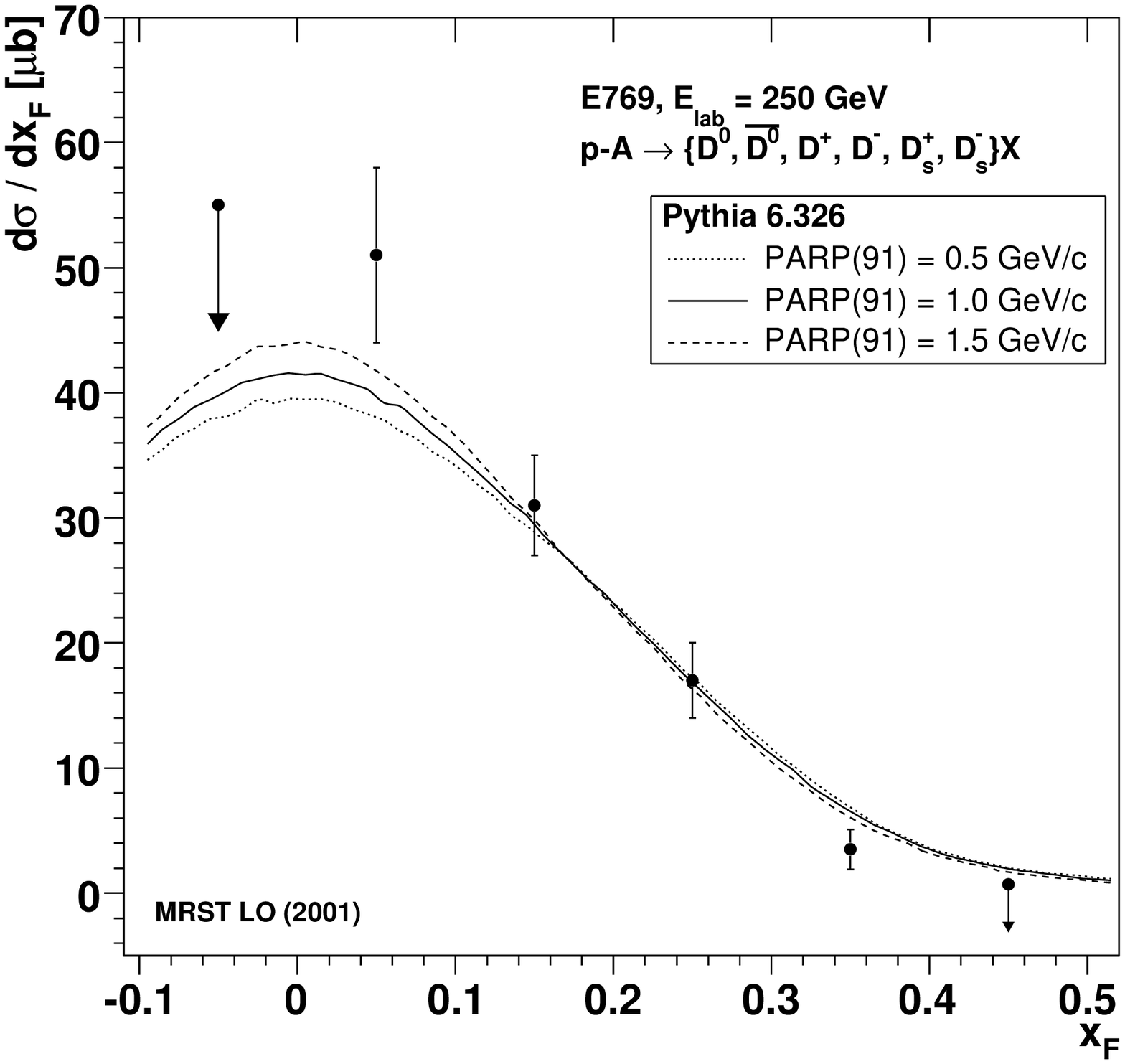}}
\caption{$p_{\rm T}^2$ (left) and \xf\ (right) distributions, as
measured by the E769 Collaboration in p-A collisions and as calculated
with Pythia, version 6.326, using $\rm PARP(91) = 0.5$, 1.0
and 1.5~GeV/$c$.}
\label{fig:kinE769}
\end{figure}

The highest statistics $p_{\rm T}^2$ and \xf\ distributions available
for \emph{proton} induced collisions were collected by the E769
Collaboration~\cite{e769kin}, at $E_{\rm lab}=250$~GeV. The event
samples are rather small, nevertheless, and the distributions shown in
Fig.~\ref{fig:kinE769} add together all measured D mesons (D$^0$,
D$^+$, D$_{\rm s}^+$ and corresponding anti-particles), in data sets
taken with Be, Al, Cu and W targets. The measured distributions are
compared with calculations done with Pythia, version 6.326, using
the default settings, except for PARP(91), which was set to 0.5, 1.0
and 1.5~GeV/$c$ (and $\rm PARJ(13)=0.6$).  The K-factor was set to 3.8
(see Table~\ref{tab:commonKfactors}).

\begin{figure}[ht!]
\centering
\resizebox{1.0\textwidth}{!}{%
\includegraphics*{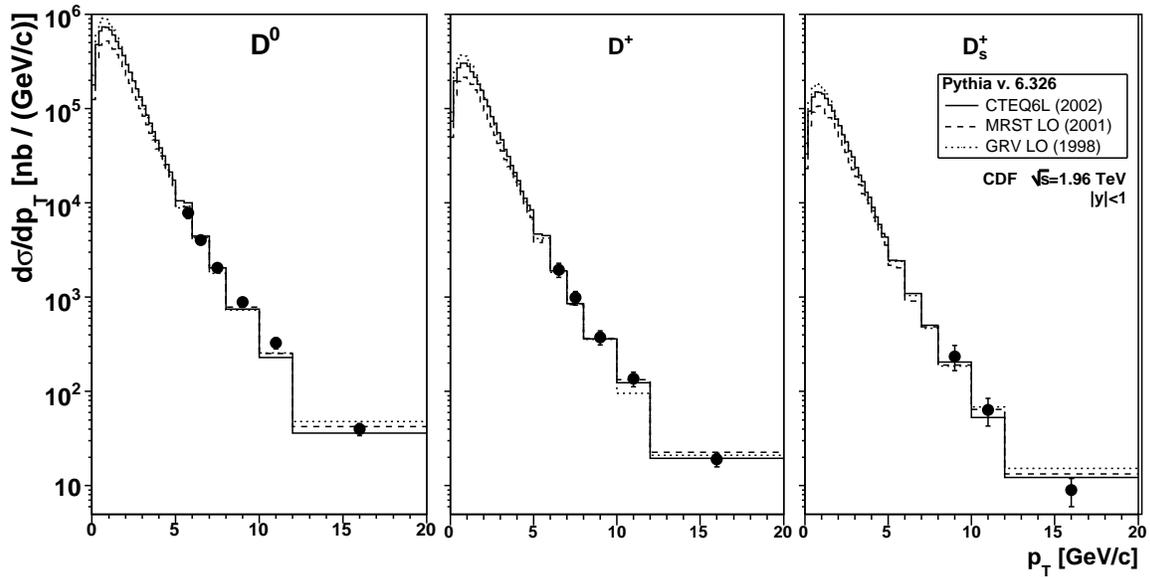}}
\caption{Single D meson \pt\ distributions measured by CDF in \ppbar\
  collisions at $\sqrt{s} = 1.96$~TeV, within $|y|<1$, compared to
  Pythia calculations.}
\label{fig:kinCDF-charm}
\end{figure}

Figure~\ref{fig:kinCDF-charm} shows the D$^0$, D$^+$ and D$^+_{\rm s}$
\pt\ differential cross-sections measured by CDF in \ppbar\ collisions
at $\sqrt{s}=1.96$~TeV.  Both D and $\overline{\rm D}$ mesons
contribute to these \emph{single} D meson differential cross-sections.
The measurements are compared with Pythia curves, calculated with
three different sets of PDFs, $\rm PARP(91)=1$~GeV/$c$ and $\rm
PARJ(13)=0.6$.  The K-factors globally fitted to the three data sets
are 3.3 (CTEQ6L), 5.8 (MRST~LO) and 5.5 (GRV~LO).  In the case of the
MRST~LO PDFs, this represents a 50\,\% increase of the K-factor with
respect to the value fitted from the fixed-target measurements, 3.8.
This observation might indicate that at Tevatron energies the
higher-order diagrams missing in our calculation are relatively more
important than at the lower energies of the fixed-target data.
However, the increase is only 10--20\,\% when we use the CTEQ6L and
GRV~LO PDF sets, which give production cross-sections with a steeper
energy dependence.  This exercise shows that the K-factors of
Table~\ref{tab:commonKfactors} can be used in the estimates of charm
production at the intermediate energies of the RHIC experiments,
$\sqrt{s}=200$~GeV, where such higher order effects, if any, must be
even smaller than at the Tevatron energies.

In summary, we can say that the existing data on kinematical
distributions of D mesons can be reasonably well reproduced by
calculations made with the latest version of Pythia (6.326).

\subsection{D meson pair correlations}

LO diagrams lead to back-to-back \ccbar\ pairs, with 180\degree\ as
the difference of azimuthal angles of the quarks, $\Delta \phi({\rm
c\bar{c}})$, and zero pair transverse momentum, $p_{\rm T}({\rm
c\bar{c}})$~\cite{frixione}. However, higher order diagrams,
intrinsic $k_{\rm T}$ and fragmentation lead to smeared
$\Delta\phi(\D{\overline{\D}})$ and $p_{\rm T}(\D{\overline{\D}})$
distributions.

In hadro-production, D meson correlations have been measured in the
experiments NA27~\cite{na27p, na27Corr}, WA75~\cite{wa75, wa75Corr},
NA32~\cite{na32Corr}, E653~\cite{e653Corr}, WA92~\cite{wa92,wa92Corr} and
E791~\cite{e791Corr}. The correlation variables $\Delta x_{\rm F}$,
$\Delta y$, $\Delta \phi$ and $\Delta p_{\rm T}^2$ are defined as the
difference between the D and $\overline{\rm D}$ values: $\Delta x_{\rm
F} = x_{\rm F}({\rm D}) - x_{\rm F}(\overline{\rm D})$, etc. Also the
\emph{pair} variables $p_{\rm T}^2({\rm D} \overline{\rm D})$, $x_{\rm
F}({\rm D} \overline{\rm D})$ and $M({\rm D} \overline{\rm D})$ have
been studied. Out of the quoted experiments, E791 is the only one
which fully reconstructed \emph{both} D mesons in one of the channels
$\D \to {\rm K}n\pi$ ($n$=1,2,3). In order to increase the statistical
significance of their measurements, the other experiments looked for
events with secondary vertices, characteristic of long-lived
particles, irrespective of the reconstruction of the decay products.
In our comparisons with calculations, we have only used data sets with
reasonable statistics, as provided by NA32, WA92 and E791.

The WA92 Collaboration provided $\Delta \phi$, $\Delta y$, $\Delta
x_{\rm F}$, $M({\rm D} \overline{\rm D})$, $p_{\rm T}^2({\rm D}
\overline{\rm D})$ and $x_{\rm F}({\rm D} \overline{\rm D})$
distributions~\cite{wa92,wa92Corr}, for 475~events collected in $\pi^-$-Cu
interactions at 350~GeV. One of the D mesons, with $x_{\rm F} > 0$,
was fully reconstructed in one of the channels $\D \to {\rm K}n\pi$
($n$=1,2,3).  The second one, which could have any \xf, was often only
partially reconstructed either due to undetected neutral decay
products or to the limited detector acceptance.  To calculate the
correlation variables which require the momenta of the D mesons, the
influence of the neutral decay products was estimated by imposing the
D meson mass in the reconstruction step and by connecting the
secondary and primary vertices.  A Monte Carlo simulation showed that
this estimation of the missing information gave a correct calculation
of the correlation variables.

In NA32 \emph{both} D mesons were only partially reconstructed.  A
purely topological analysis method selected $\sim$\,500 events with
two reconstructed secondary vertices well displaced with respect to
the primary vertex, resulting in $\Delta \phi$, $\Delta \eta$, $\Delta
y$, $M({\rm D} \overline{\rm D})$, $p_{\rm T}^2({\rm D} \overline{\rm
D})$ and $x_{\rm F}({\rm D} \overline{\rm D})$ distributions, for
$x_{\rm F} > 0$, from data collected in 230~GeV $\pi^-$-Cu
interactions~\cite{na32Corr}.  When there were neutral decays
products, an algorithm similar to the one used by WA92 estimated the
momenta of the D mesons, needed to calculate the correlation
variables, except for $\Delta \phi$ and $\Delta \eta$.  The error on
the estimated momentum was evaluated by a Monte Carlo simulation to be
$\sim$\,15\,\%, having little impact on the reconstructed correlation
variables of the D meson pairs.

The E791 experiment measured D meson correlations from data collected
in 500~GeV $\pi^-$-C and (a smaller fraction) in $\pi^-$-Pt
collisions~\cite{e791Corr}, reconstructing \emph{both} D mesons in one
of the channels $\D \to {\rm K}n\pi$ ($n$=1,2,3). The published
$\Delta \phi$, $\Delta y$, $\Delta x_{\rm F}$, $\Delta p_{\rm T}^2$,
$M({\rm D} \overline{\rm D})$ and $p_{\rm T}^2({\rm D} \overline{\rm
D})$ distributions are based on $791\pm 44$ events.

We will now compare the three experimental data sets with each other,
for each of the measured correlation variables. We prepared Pythia
curves for each data set, simulating $\pi^-$-p collisions at the
appropriate energy and properly applying the phase space cuts specific
of each experiment. The WA92 results were given for full phase space,
NA32 required that both D mesons were produced at $x_{\rm F} > 0$, and
E791 required that both D mesons were produced within $-0.5 < y^* <
2.5$, where $y^*$ is the rapidity in the centre-of-mass reference
frame.

\begin{figure}[ht]
\centering
\resizebox{1.0\textwidth}{!}{%
\includegraphics*{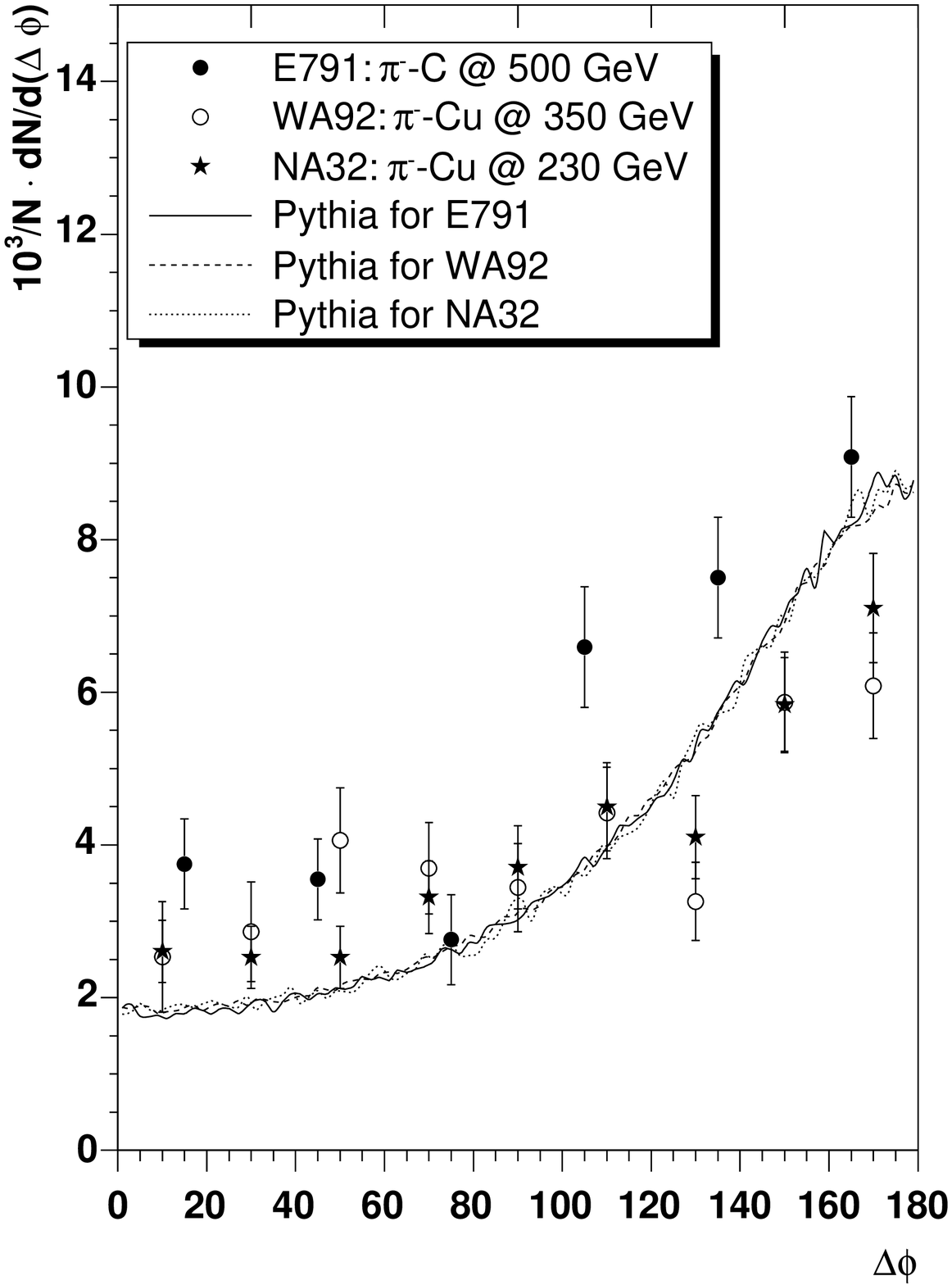}
\includegraphics*{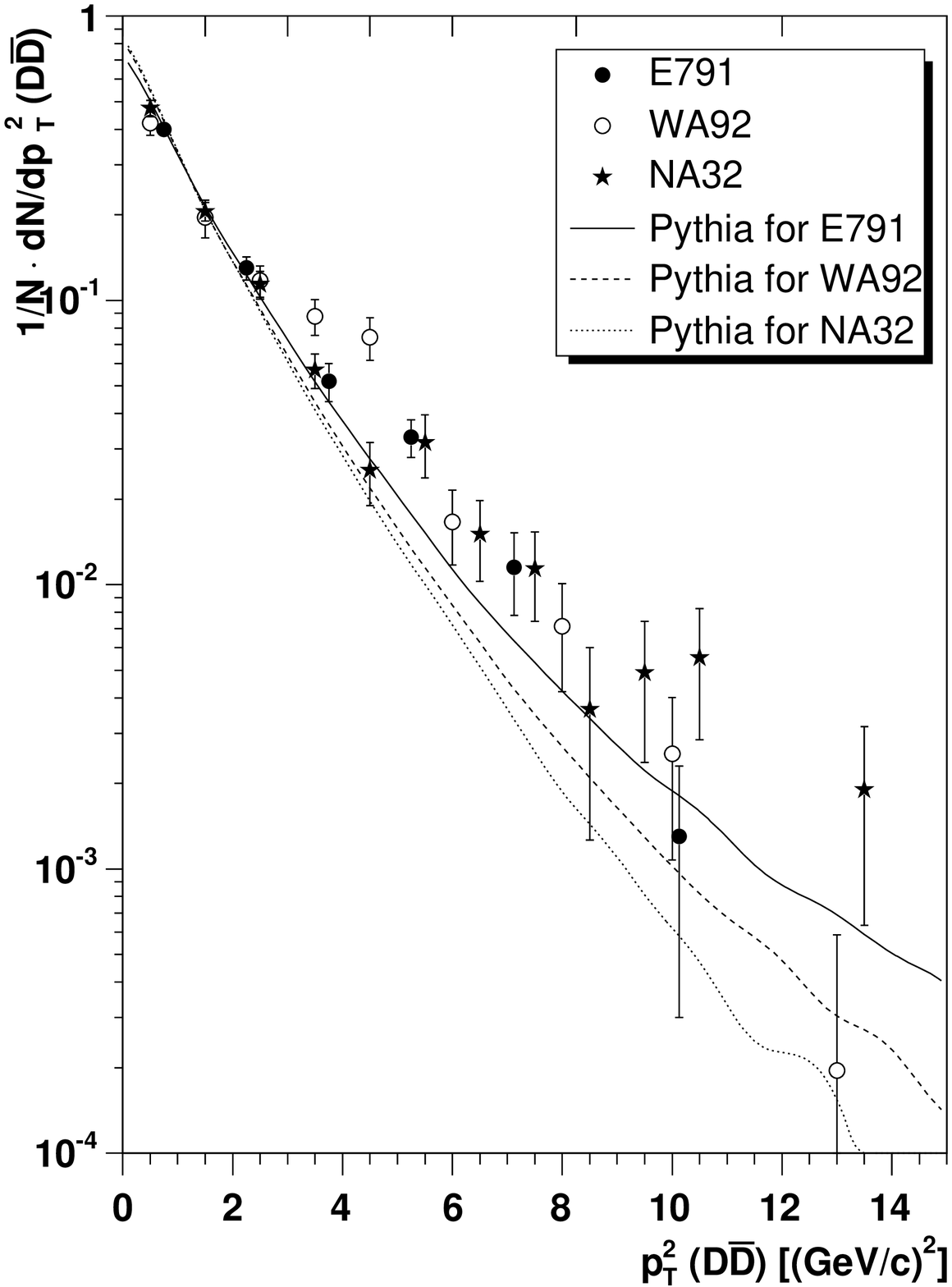}
\includegraphics*{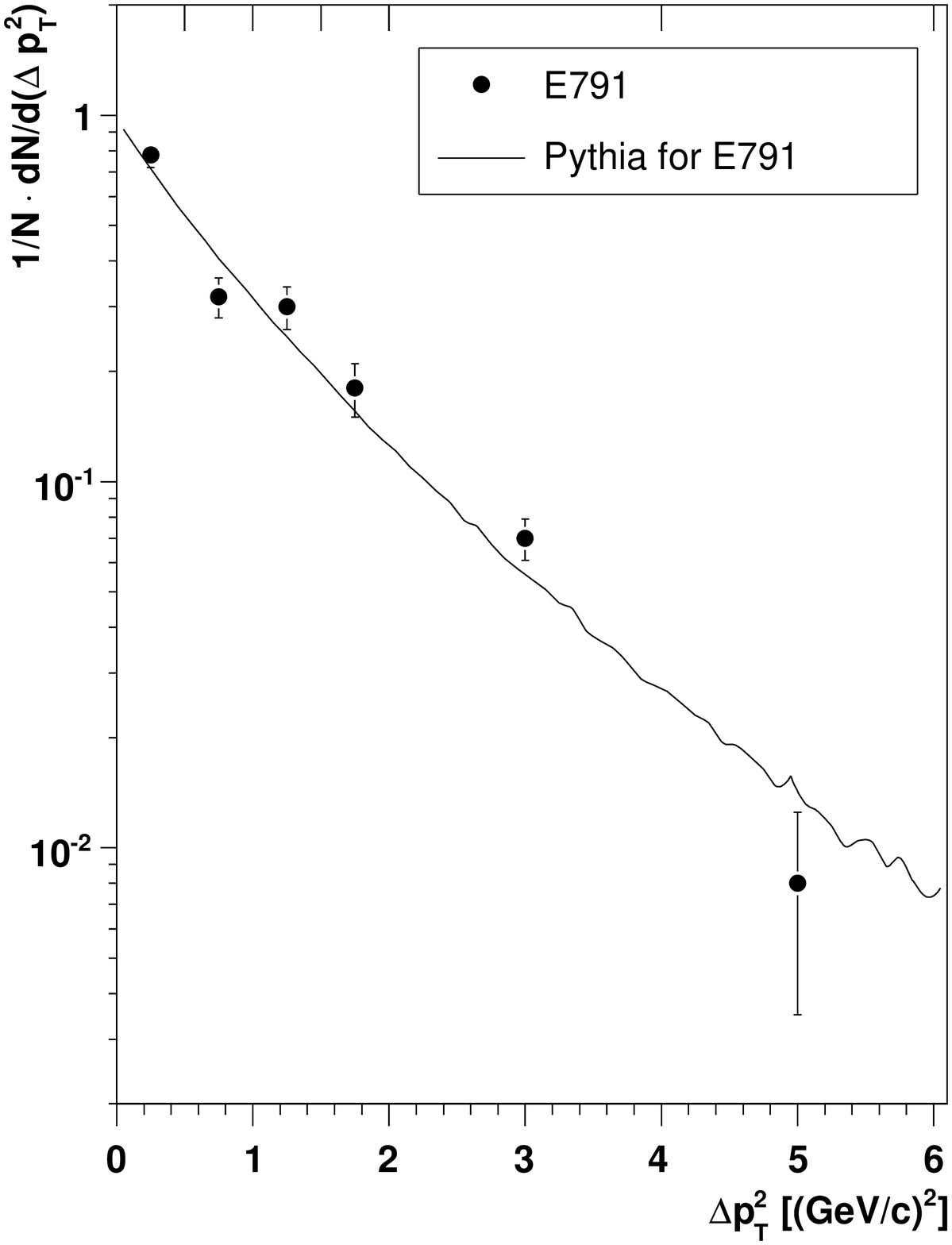}}
\caption{$\Delta \phi$, $p_{\rm T}^2({\rm D}
\overline{\rm D})$ and $\Delta p_{\rm T}^2$ distributions measured
by NA32, WA92 and E791 in $\pi^-$ induced collisions, compared to
calculations made with Pythia 6.326.}
\label{fig:deltaPhi_pT2}
\end{figure}

\begin{figure}[ht!]
\centering
\resizebox{0.35\textwidth}{!}{%
\includegraphics*{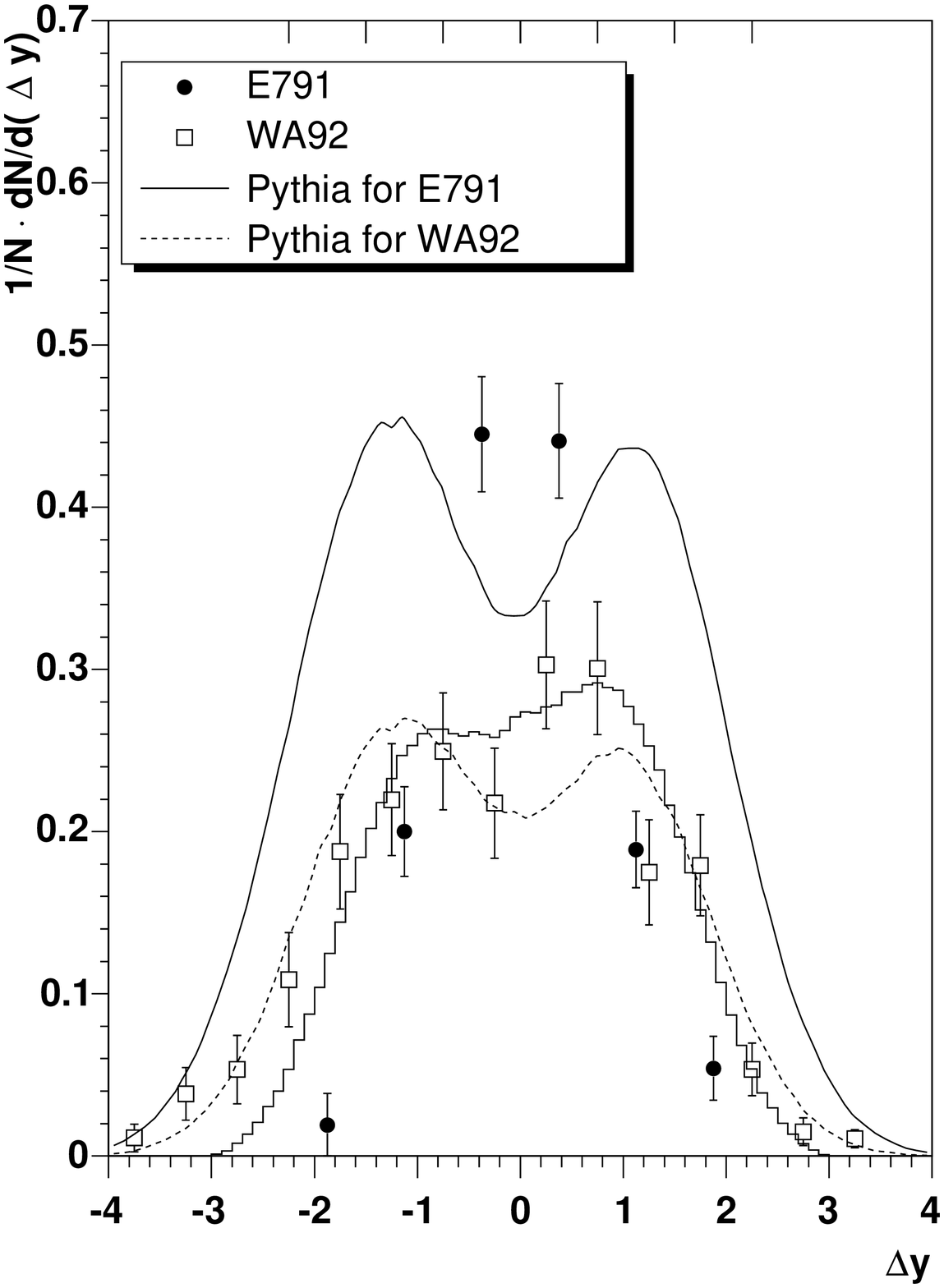}}
\resizebox{0.35\textwidth}{!}{%
\includegraphics*{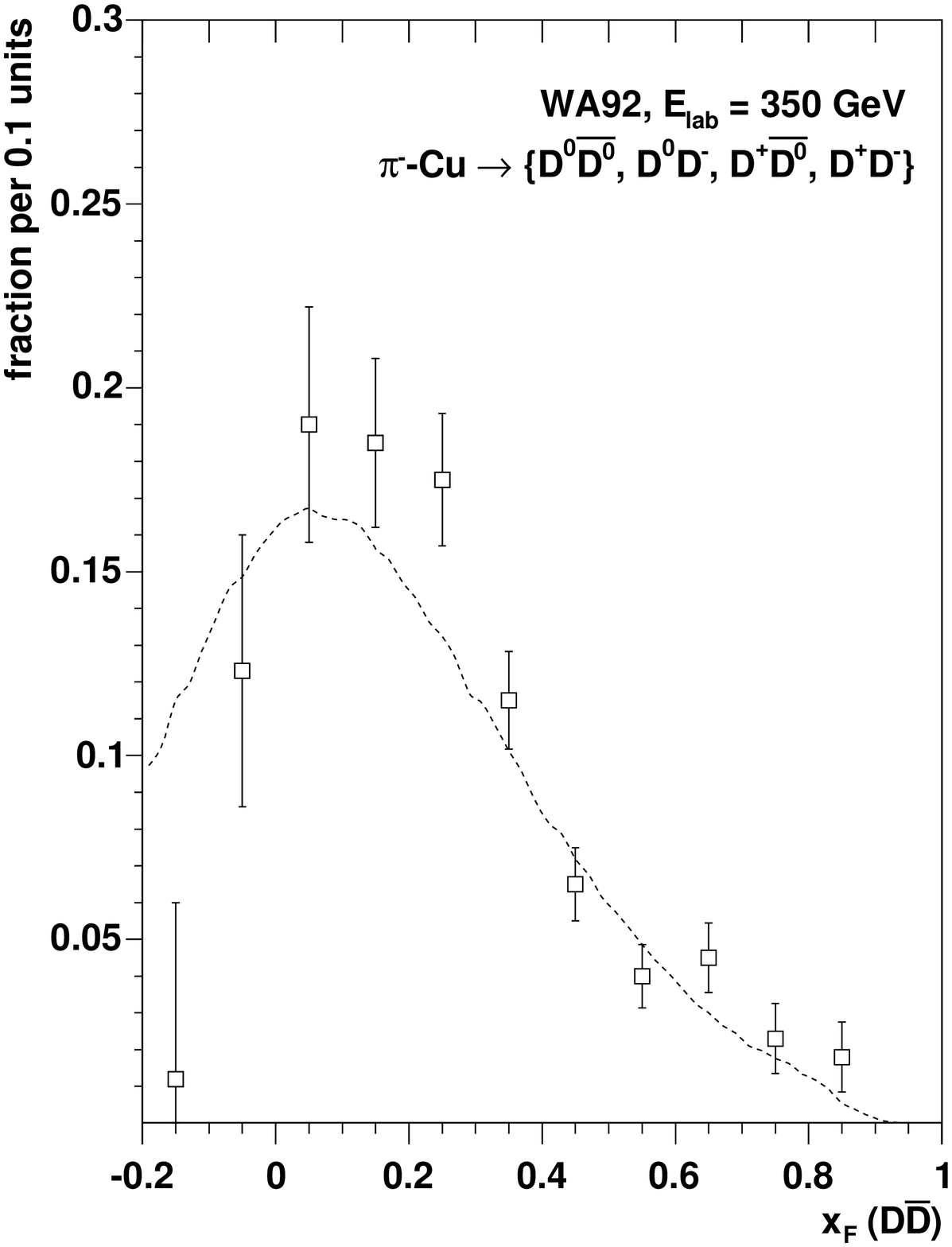}}
\caption{$\Delta y$ and $x_{\rm F}({\rm D} \overline{\rm D})$
  distributions measured by E791 and WA92 in $\pi^-$ induced
  collisions, compared to calculations made with Pythia for full phase
  space (solid and dashed lines) and after applying the phase space
  cuts (only for E791, histogram).}
\label{fig:deltaYxF}
\end{figure}
\begin{figure}[ht!]
\centering
\resizebox{1.0\textwidth}{!}{%
\includegraphics*{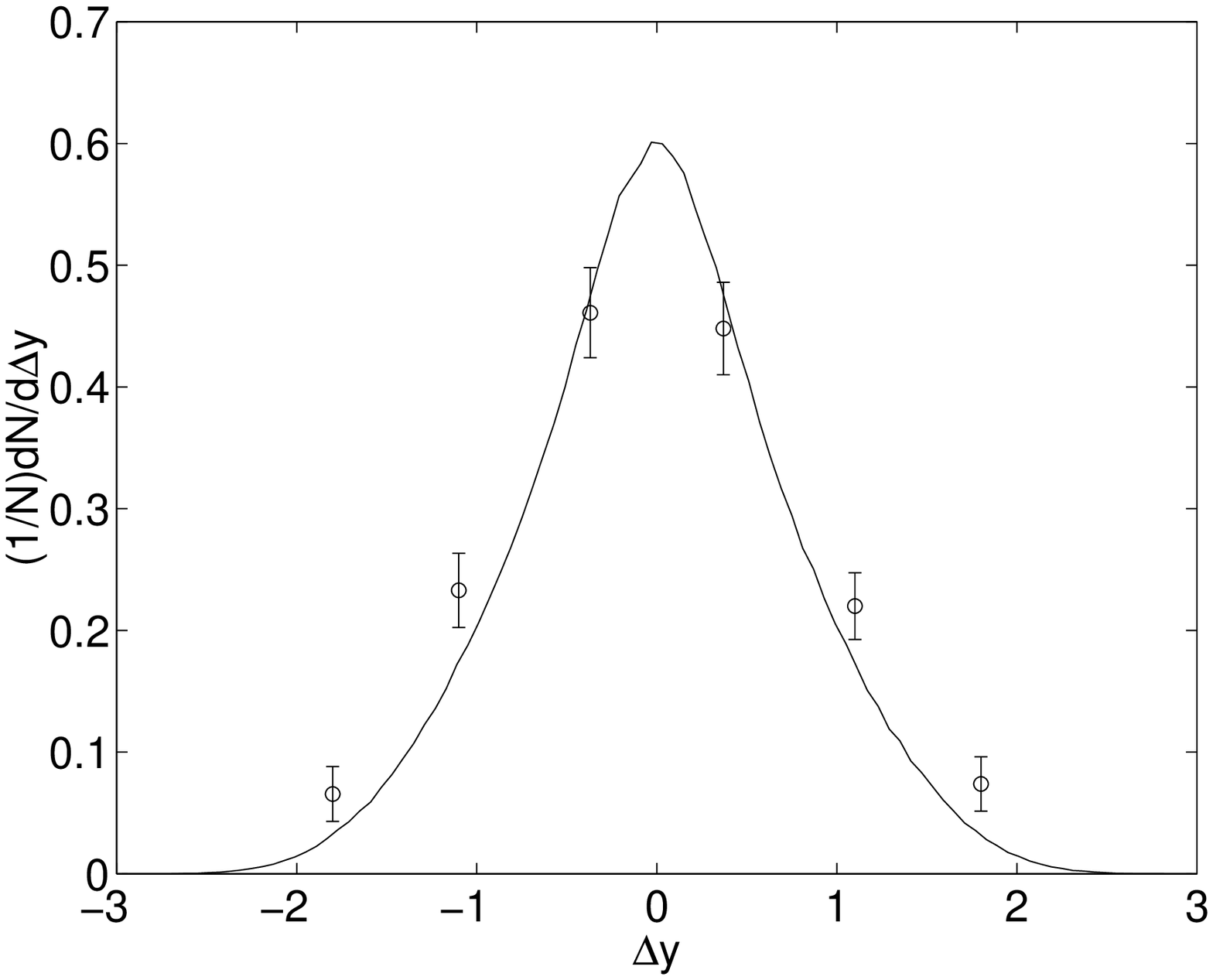}
\includegraphics*{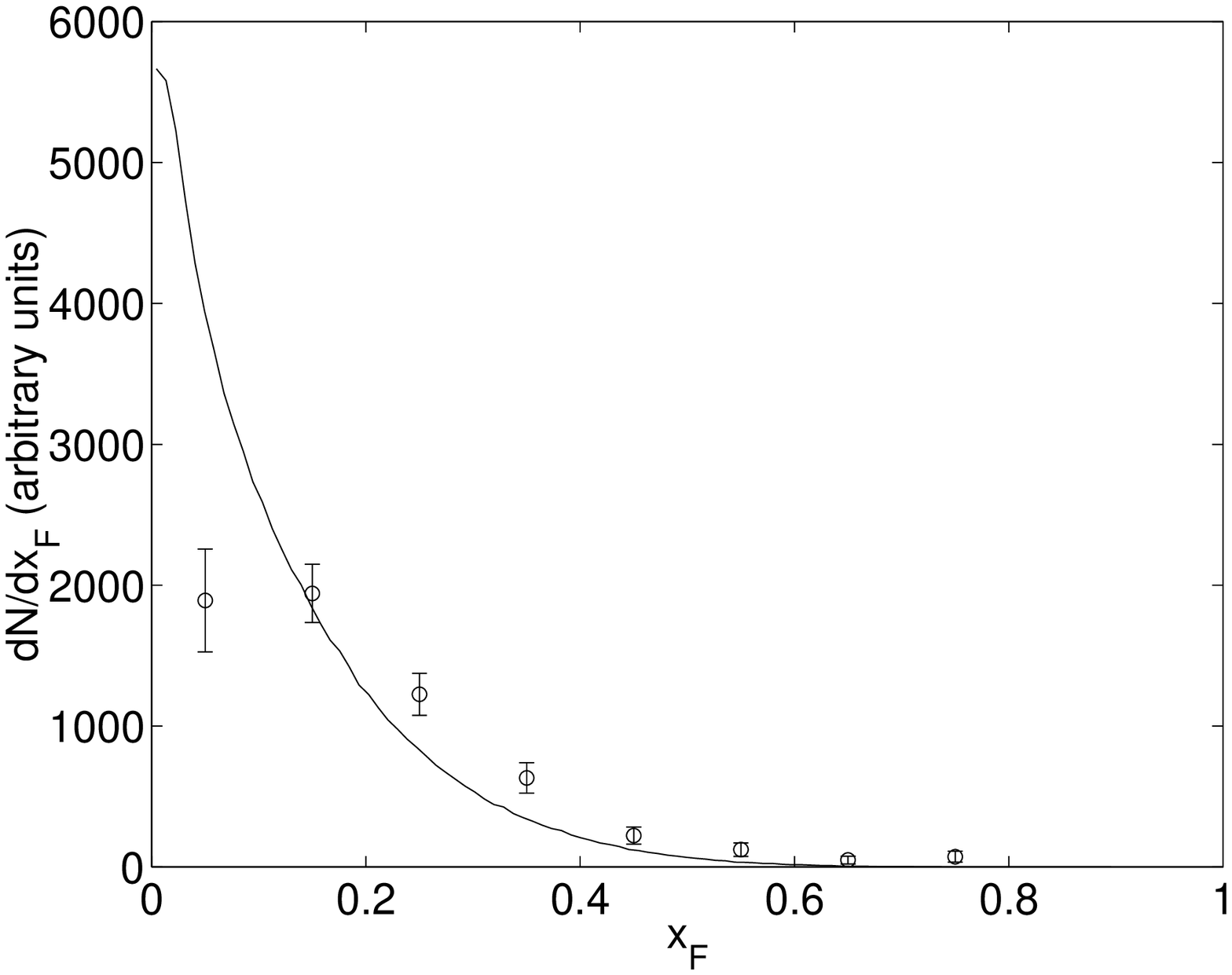}
\includegraphics*{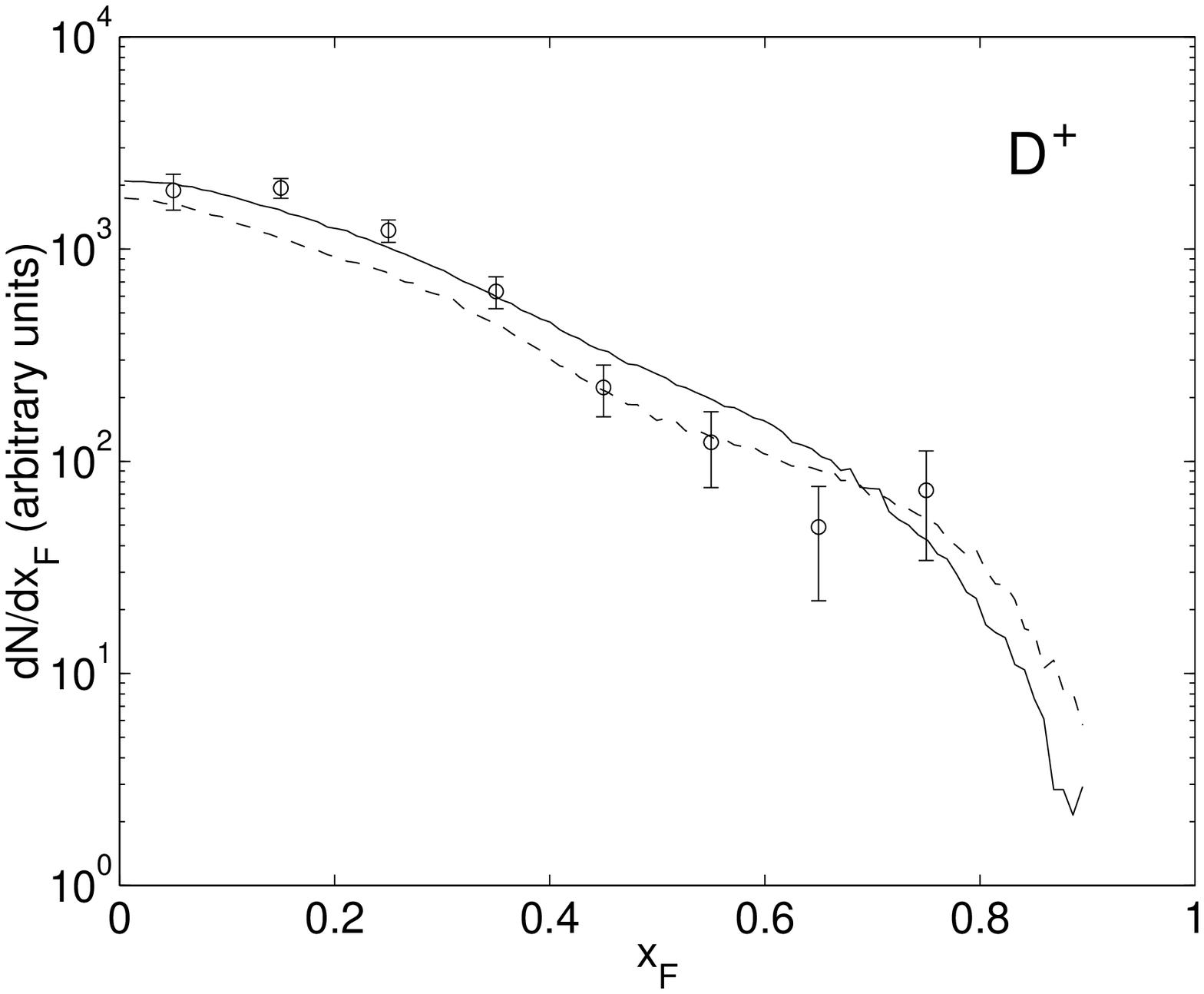}}
\caption{E791 $\Delta y$ distribution of \DDbar\ pairs (left) and WA82
\xf\ distribution of D$^+$ mesons (centre and right) versus Pythia
curves calculated with the Peterson fragmentation function (left and
centre) and with the Lund scheme (right).  Figures taken from
Ref.~\cite{norrbinIII}.}
\label{fig:Norrbin}
\end{figure}

Figure~\ref{fig:deltaPhi_pT2} shows the transverse correlation
variables, $\Delta \phi$, $p_{\rm T}^2({\rm D} \overline{\rm D})$ and
$\Delta p_{\rm T}^2$, measured in the three experiments.  Since we are
only interested in a shape comparison, all data sets and respective
curves were normalised to each other, taking the E791 measurement as
reference. All measurements are compatible with each other, in terms
of shape, and can be described reasonably well by the Pythia curves.

In Fig.~\ref{fig:deltaYxF} we compare the measured longitudinal
correlation variables, $\Delta y$ and $x_{\rm F}({\rm D}\overline{\rm
D})$, with the corresponding calculated curves.  These variables are
significantly affected by the hadronisation step and, therefore, test
the modelling of this non-perturbative effect.  The curves for WA92
(full phase space), depicted as dashed lines, are in fair agreement
with the measurements.  For the E791 case, we show the Pythia
calculation before (solid line) and after (histogram) applying the
phase space cuts; the histogram fails to describe the data points.
Clearly, the phase space window of E791 limits the $\Delta y$ range to
$\pm 3$, but the measured distribution seems to be too narrow, with
respect to Pythia's curve.

It has been observed by one of the authors of Pythia~\cite{norrbinIII}
that the $\Delta y$ and $\Delta x_{\rm F}$ distributions measured by
E791 could be described quite well (see Fig.~\ref{fig:Norrbin}-left)
if Pythia's default hadronisation scheme would be replaced by the
Peterson fragmentation function (with $\epsilon = 0.05$), so that the
non-interacting valence quarks would not influence the kinematics of
the produced charm quark.  However, this modified model would fail to
reproduce the single D meson kinematical distributions measured by
WA82~\cite{wa82kinII}, as shown in Fig.~\ref{fig:Norrbin}-centre,
which are well described by the standard scheme
(Fig.~\ref{fig:Norrbin}-right).

\begin{figure}[h!]
\centering
\resizebox{1.0\textwidth}{!}{%
\includegraphics*{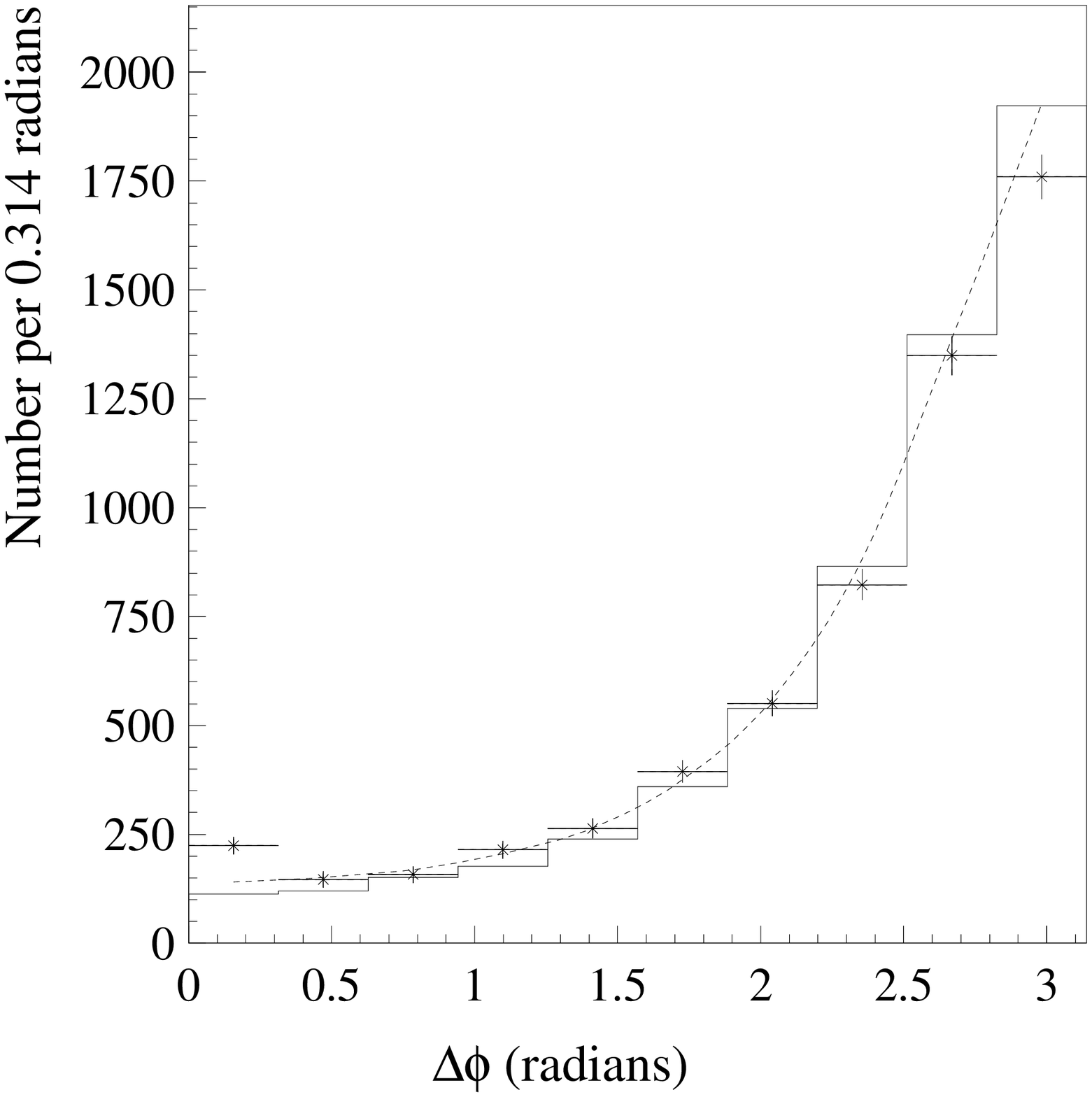}
\includegraphics*{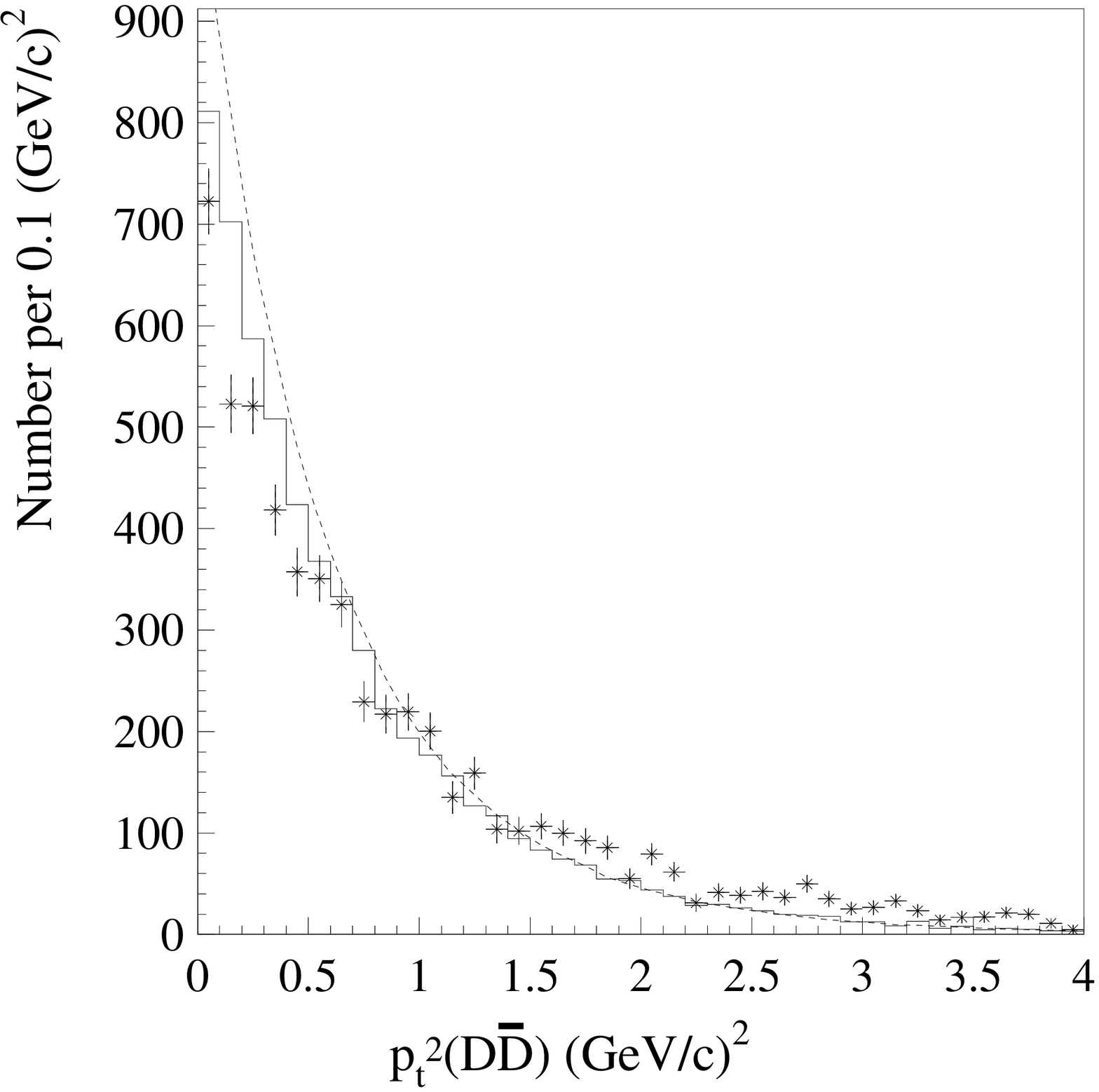}
\includegraphics*{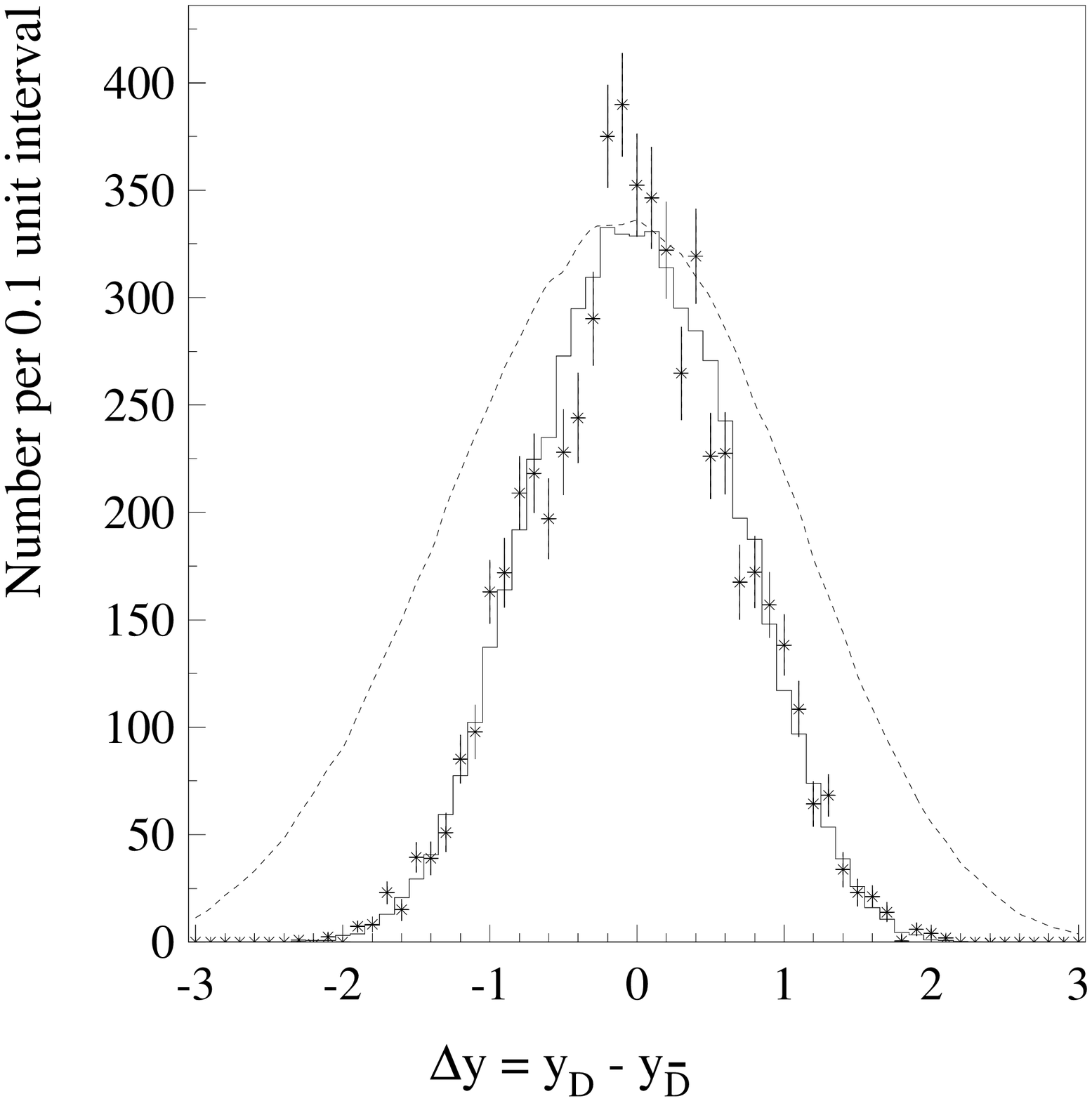}
\includegraphics*{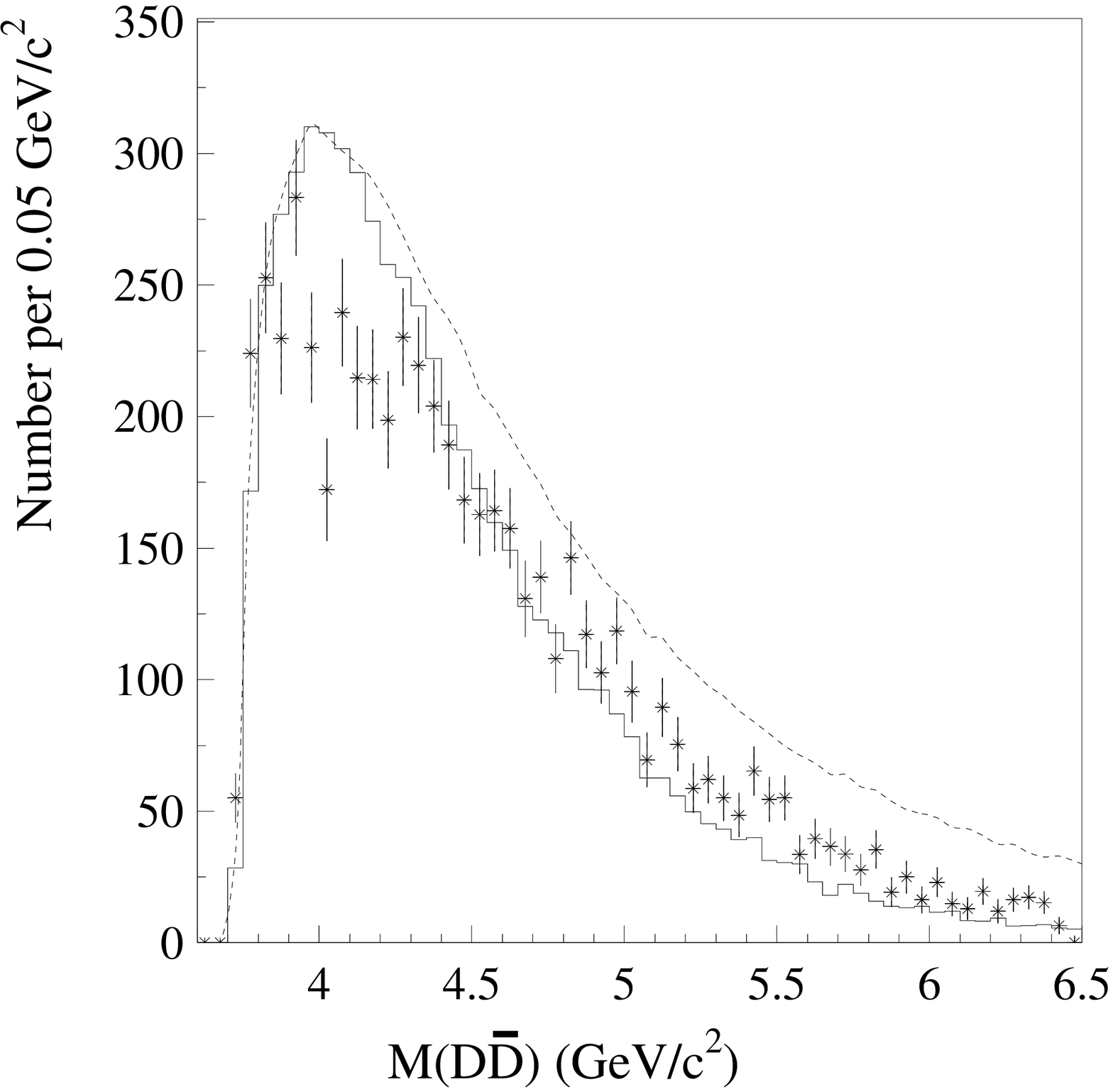}}
\caption{Correlations for fully reconstructed ${\rm D}\overline{\rm
    D}$ pairs from the photo-production experiment FOCUS.  For
    comparison, events generated with Pythia 6.203 (dotted lines) were
    tracked and reconstructed by the simulation code (histograms),
    being affected by acceptances, efficiencies and smearing effects,
    as the data. Figures taken from Ref.~\cite{focus}.}
\label{fig:focus}
\end{figure}

In the context of D meson correlations it is worth mentioning the
photo-production Fermilab experiment ``FOCUS'' (E831). On the basis of
$\sim$\,7000 fully reconstructed D meson pairs, they
showed~\cite{focus} that Pythia~6.203 describes very well the
correlation variables, $\Delta \phi$, $p_{\rm T}^2({\rm
D}\overline{\rm D})$, $\Delta y$ and $M({\rm D}\overline{\rm D})$.
Contrary to all other experiments, FOCUS compared the measured data
with Pythia events propagated through the simulation, reconstruction
and analysis algorithms, so that the calculated curves become affected
by acceptance and efficiencies.  This method also takes into account
smearing effects, and should lead to a more robust comparison between
data and theory.  In Fig.~\ref{fig:focus} (taken from the FOCUS
publication) we can see a remarkable narrowing of the $\Delta y$
distribution, due to the detector acceptance.

\begin{figure}[h!]
\centering
\resizebox{1.0\textwidth}{!}{%
\includegraphics*{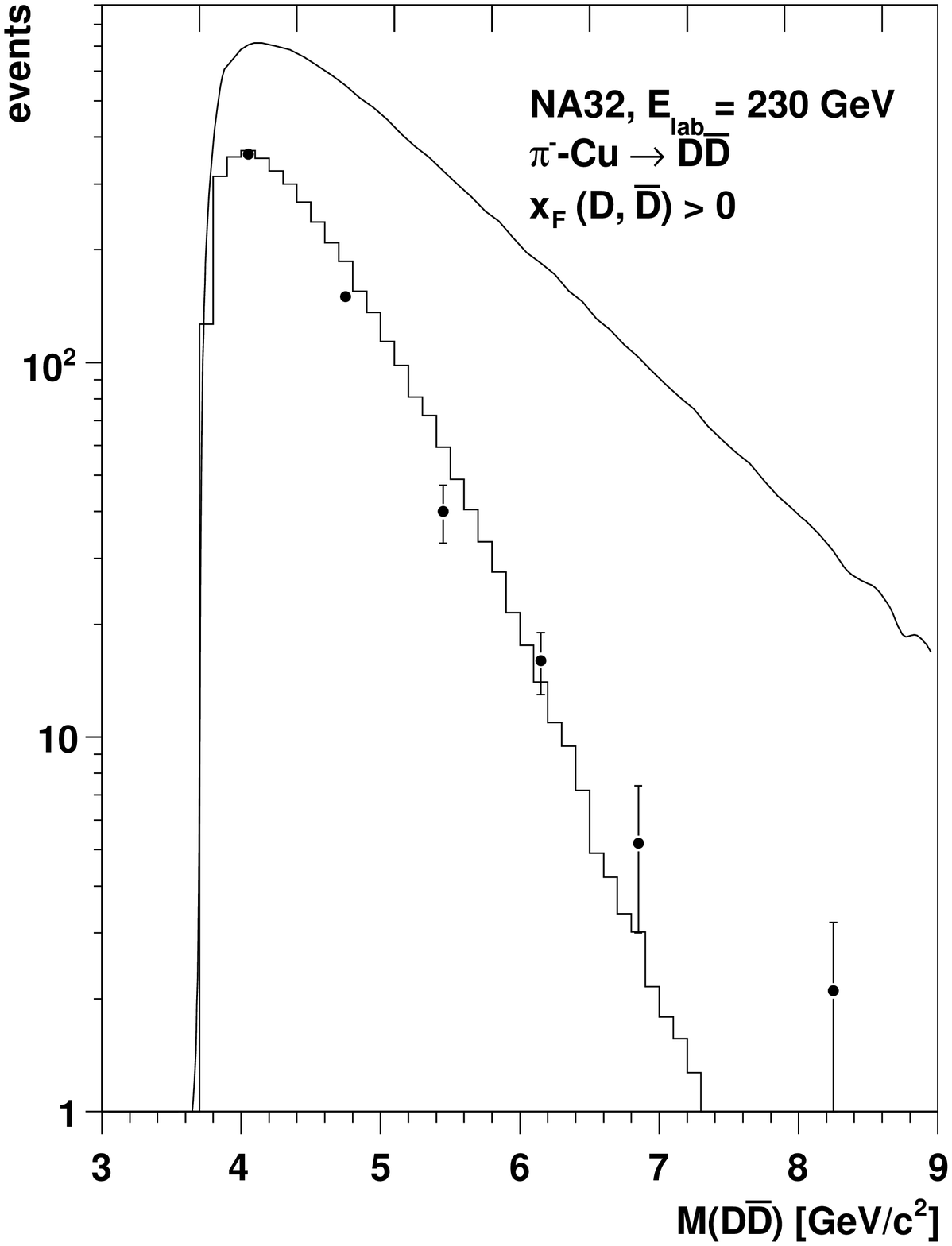}
\includegraphics*{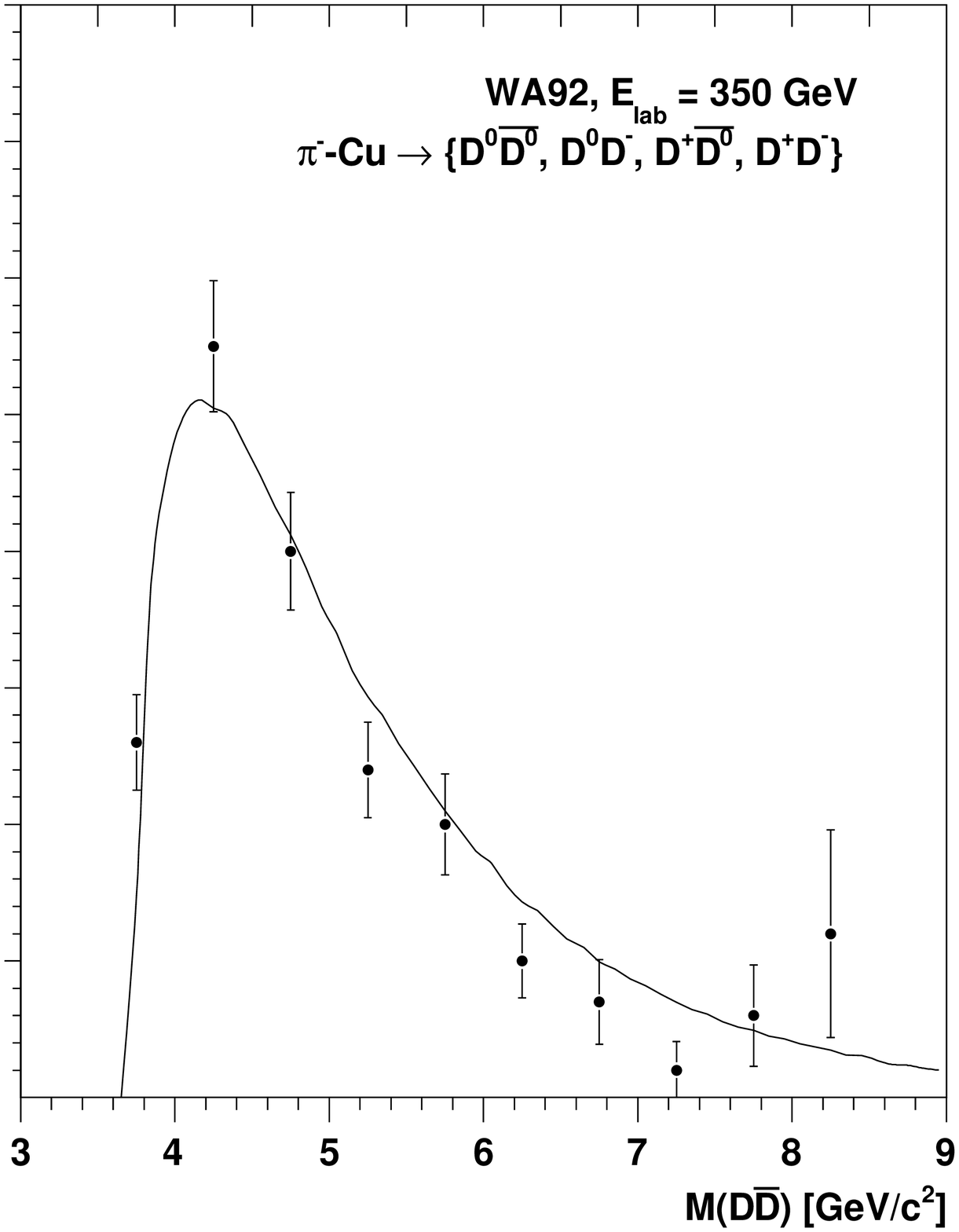}
\includegraphics*{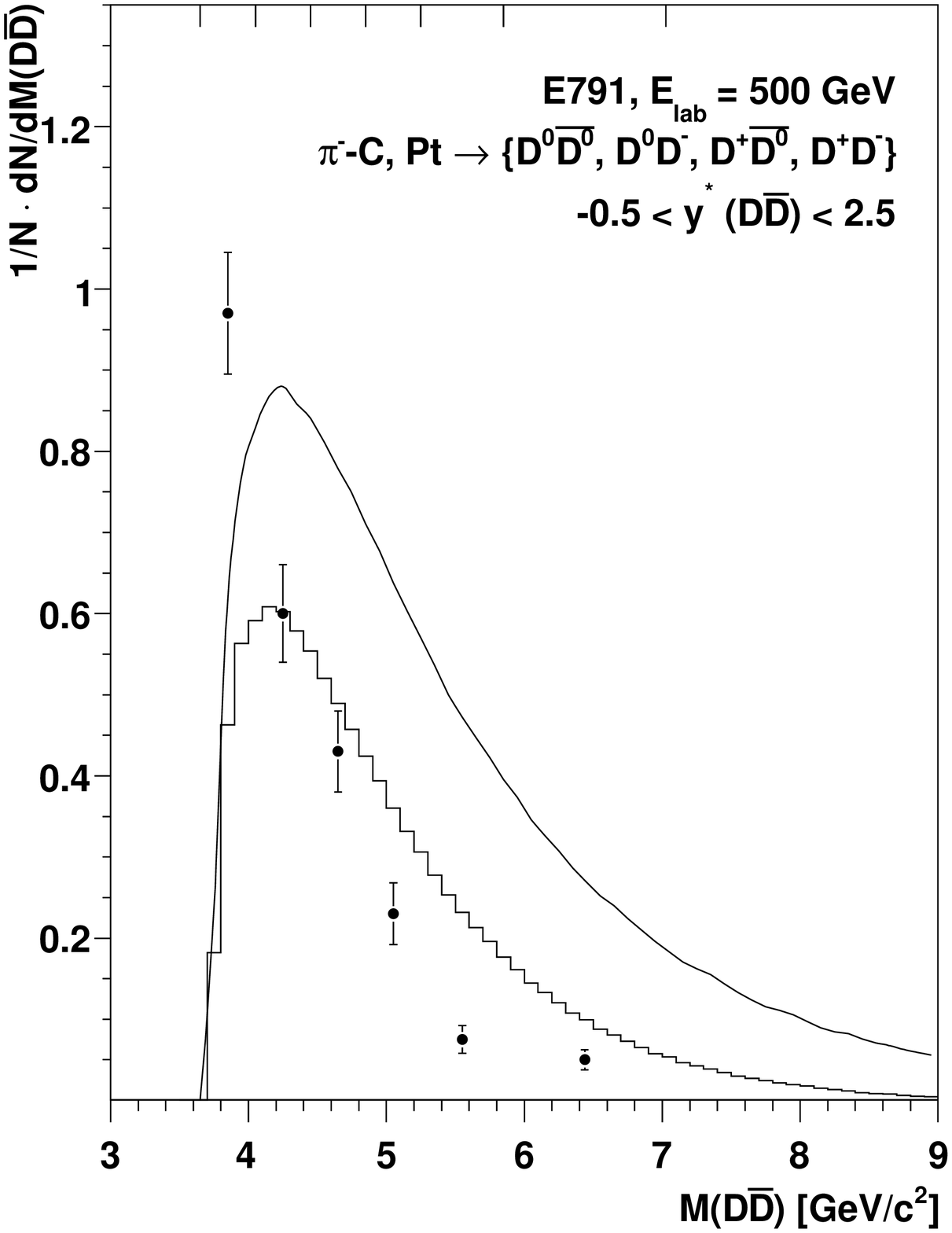}}
\caption{${\rm D} \overline{\rm D}$ invariant mass distributions
    measured by NA32, WA92 and E791, in $\pi^-$ induced collisions,
    compared to calculations made with Pythia, version 6.326.
    The lines show Pythia calculations prepared for full
    phase space while the histograms show the Pythia curves with the
    correct phase space cuts applied.}
\label{fig:mass}
\end{figure}

NA32, WA92 and E791 also measured the $\D \overline{\D}$ invariant
mass distributions, as reported in Fig.~\ref{fig:mass}. While the NA32
and WA92 distributions are nicely reproduced by Pythia's curves, the
E791 points show a spectral shape quite different from the
calculation, with a very unexpected rise when approaching the $\D
\overline{\D}$ production threshold, in apparent conflict with phase
space considerations.

To summarise, the latest version of Pythia (6.326), with default settings
(except for PARP(91), which should be kept between 1 and 1.5~GeV/$c$),
is able to reproduce reasonably well the available D meson pair
correlation measurements, at least as provided by the NA32, WA92 and
FOCUS Collaborations. However, there is a significant disagreement in
what concerns the $\Delta y$, $\Delta x_{\rm F}$ and $M({\rm
D}\overline{\rm D})$ distributions measured by E791, which might be
due to a problem in the 8-dimensional acceptance correction procedure.
The highest statistics data set available for D meson correlations,
collected by FOCUS, is in good agreement with Pythia events propagated
through the simulation of the apparatus.

\bigskip

\begin{figure}[h!]
\centering
\resizebox{0.5\textwidth}{!}{%
\includegraphics*{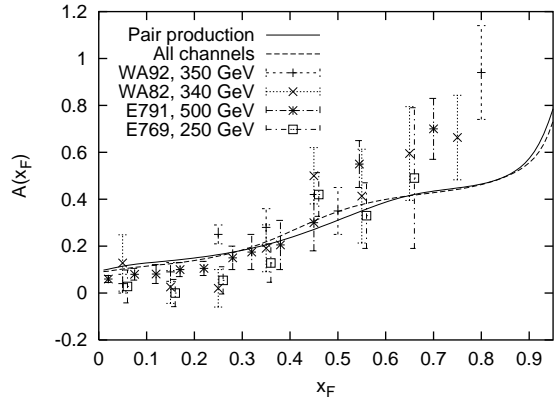}}
\caption{Asymmetry of charm particle production. Figure taken from
    Ref.~\cite{norrbinII}.}
\label{fig:asymm}
\end{figure}

We finish this section by mentioning a different kind of pair
correlation, the relative abundance of $\D^-$ with respect to $\D^+$,
evaluated through the \emph{asymmetry}
\begin{equation}\label{eq:asymm}
A(x_{\rm F}) = \frac{\sigma({\rm D}^-) - \sigma({\rm
    D}^+)}{\sigma({\rm D}^-) + \sigma({\rm D}^+)} \quad .
\end{equation}
$A(x_{\rm F})$ has been measured in $\pi^-$ induced collisions by the
WA82~\cite{wa82kinII}, WA92~\cite{wa92,wa92Corr}, E769~\cite{e769} and
E791~\cite{e791:asymm} Collaborations. In Fig.~\ref{fig:asymm} we show
a comparison (made by the authors of Pythia~\cite{norrbinII}) between
the available data and a Pythia calculation specifically made for
340~GeV $\pi^-$p collisions.
The solid line, labelled ``pair production'', corresponds to a
calculation only including the standard LO processes of charm
production, plus initial and final state radiation, while the dashed
line labelled ``all channels'' also includes ``flavour excitation''
and ``gluon splitting'' diagrams.

These measurements, and the calculations, clearly indicate that the
$\rm D^-$ mesons are produced at more forward rapidities than the $\rm
D^+$ mesons, confirming that the $\rm D^-$ is the leading particle in
$\pi^-$ induced collisions.

\subsection{Nuclear dependence of charm production}
\label{sec:nuclData}

Most experiments used nuclear targets and published the D meson
cross-sections for p-N and $\pi$-N collisions, assuming a linear
dependence with the mass number of the target nucleus, $A$. However,
as explained in Section~\ref{sec:nucl}, we expect that anti-shadowing
in the parton distribution functions will affect charm production at
energies below $\sqrt{s}\sim 150$~GeV.  Therefore, the p-N and $\pi$-N
values obtained by extrapolating the p-A and $\pi$-A measurements
using the linear $A$ scaling should be somewhat higher than the values
directly measured in pp and $\pi$-p collisions.  From
Fig.~\ref{fig:nucl-ccbb} we can see that the nuclear anti-shadowing
given by the EKS98 model increases with energy between $\sqrt{s}=20$
and 40~GeV, the energy range where the data have been collected.  But
even at $\sqrt{s}=40$~GeV, and for the Pb nucleus, the expected
increase in the charm production cross-section due to the nuclear
anti-shadowing effect, with respect to the linear extrapolation of pp
collisions, is only around 10\,\%, a value too small to be visible in
the presently available measurements, given their rather large
uncertainties.


\begin{table}[ht!]
\centering
\begin{tabular}{|c|c|c|c|}\hline
Exp.\rule{0pt}{0.5cm}& \elab [GeV] \hfill Target  & Observed &
$\alpha$\\ 
& Phase space & D mesons & \\ \hline
\multicolumn{4}{|c|}{\rule{0pt}{0.5cm}{p-A collisions}}\\ \hline
E789 & 800 \hfill Be,\,Au & Be: $1360~\D^0$ & $\D^0: 1.02\pm0.03\pm0.02$ \\
\cite{e789} & 0$<$$x_{\rm F}$$<$0.08, $p_{\rm T}$$<$1.1 & Au: $1040~\D^0$ & \\ \hline
\multicolumn{4}{|c|}{\rule{0pt}{0.5cm}{$\pi^-$-A collisions}}\\ \hline
WA82 & 340 \hfill Si,\,Cu,\,W& ~Si:\,$102~(\D^0,\D^+)$ & $\D^0$+$\D^+ : 0.92\pm0.06$ \\
\cite{wa82}& $x_{\rm F}>0.0$ & Cu:\,$528~(\D^0,\D^+)$ & $\D^0\to {\rm K}\pi : 1.03\pm0.11$ \\
& & ~W:\,$1017~(\D^0,\D^+)$ & $ \D^0\to {\rm K}\pi\pi\pi : 0.93\pm0.11$ \\
& & & $ \D^+\to {\rm K}\pi\pi : 0.84\pm0.08$\\ \hline
E769 & 250 \hfill Be,\,Al,\,Cu,\,W& all targets: & $ \D^0$+$\D^+ : 1.00\pm0.05\pm 0.02$ \\
\cite{e769Adep} &  $x_{\rm F}>0.0$& $650~\D^0$ & $ \D^0 : 1.05\pm0.15\pm 0.02$ \\
& & $776~\D^+$ & $ \D^+ : 0.95\pm0.06\pm 0.02$ \\ \hline
WA92 & 350 \hfill Cu,\,W &
Cu:\quad\quad\quad W: & $\D^0$+$\D^+: 0.93\pm0.05\pm0.03$ \\ 
\cite{wa92} &  $x_{\rm F}>0.0$ & $3245~\D^0$, $628~\D^0$ & $ \D^0 : 0.92\pm0.07\pm0.02$\\
& & $2753~\D^+$, $546~\D^+$ & $\D^+: 0.95\pm0.07\pm0.03$ \\ \hline
E706 & 515 \hfill Be,\,Cu & Be+Cu: $110~\D^+$ & $ \D^+ : 1.28\pm0.33$ \\ 
\cite{e706} &  $x_{\rm F}$$>$$-0.2$, 1$<$$p_{\rm T}$$<$8 & & \\ \hline
\end{tabular}
\caption{Nuclear target dependence in proton and pion induced
collisions.  Note that $\D^0$ and $\D^+$ mean $\D^0 + \overline{\D^0}$
and $\D^+ + \D^-$, respectively. \pt\ in GeV/$c$.}
\label{tab:alpha}
\end{table}

\begin{figure}[ht!]
\centering
\resizebox{0.5\textwidth}{!}{%
\includegraphics*{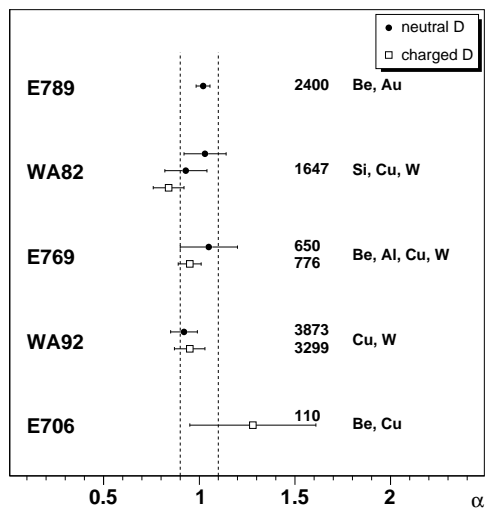}}
\caption{Values of the $\alpha$ parameter extracted from data
collected in pion and proton induced collisions. The two vertical
lines indicate a $\pm 10\,\%$ range around $\alpha = 1$.}
\label{fig:alpha}
\end{figure}

Some experiments made measurements with two or more targets and fitted
their data to the $A^\alpha$ form, extracting the $\alpha$ values
compiled in Table~\ref{tab:alpha} and Fig.~\ref{fig:alpha}.  In the
third column we give the number of observed D mesons for each target,
except if only the total number was provided.  The errors of the WA82
measurements include systematic uncertainties, which are small with
respect to the statistical error.  All the published values are
consistent with $\alpha = 1$, within their large errors.

In a dedicated study~\cite{e769Adep}, using the \emph{relative}
cross-sections of four nuclear targets (Be, Al, Cu and W), E769
extracted $\alpha=1.00\pm0.05\pm0.02$ as the best description of the
measurements, made with a pion beam.  Within the statistical accuracy
of the data, they saw no change between the values obtained with the
$\pi^+$ and $\pi^-$ beams,
or any dependence of $\alpha$ on the D meson's $p_\mathrm{T}$ and
$x_\mathrm{F}$.  Also WA82 and WA92 have not seen any dependence of
$\alpha$ on \xf\ or \pt, but with even larger uncertainties.  The
existence of such a dependence (seen for many other particles,
including the \jpsi) cannot be excluded from the present measurements
and, if observed by a future high statistics experiment, could imply a
revised extrapolation of the E706 and E789 measurements to elementary
$\pi$-N/p-N cross-sections, given their limited \pt\ coverages.

The WA78 Collaboration also studied the nuclear dependence of the
charm production cross-section, by measuring the yield of prompt
single muons in a beam-dump experiment, both with 320~GeV $\pi^-$
and with 300~GeV proton beams, interacting on Al, Fe and U
targets.  The resulting $\alpha$ values (for $x_\mathrm{F}$ above
$\sim$\,0.1) were
$\alpha(\mu^+)=0.79\pm0.12$ and 
$\alpha(\mu^-)=0.76\pm0.13$ for the p-A data~\cite{wa78alphap}, and 
$\alpha(\mu^+)=0.76\pm0.08$ and 
$\alpha(\mu^-)=0.83\pm0.06$ for the $\pi$-A data~\cite{wa78alphapi}.
These results are significantly lower than 1, raising doubts on
whether the prompt single muons can be cleanly ascribed to
semi-leptonic decays of charmed particles.

We close this discussion with a final remark concerning the
$\sigma_{\rm pA} = \sigma_0 \cdot A^\alpha$ parameterisation.
Figure~\ref{fig:Pythia} collects the D meson production cross-sections
measured in proton and pion induced collisions.  Three measurements
were made with hydrogen targets (NA16, NA27 and E743) while the other
values were measured with nuclear targets and divided by $A$.
Considering the error bars of the data points, we do not see any
significant difference between the $\sigma_{\rm pA}/A$ values and the
$\sigma_{\rm pp}$ values.  In Ref.~\cite{macdermott} (from 1987) it
was argued that the elementary charm production cross-sections derived
from p-A measurements (performed with nuclear targets of $A\geq9$)
should be parameterised as $\sigma_{\rm pA} = K_0 \cdot \sigma_{\rm
pp} \cdot A^\alpha$, following observations made with light flavour
data~\cite{barton} (where $K_0$ is around 1.5--2.0).  Is was noted, in
particular, that $K_0=1.5$ would result in $\alpha=1.0$ at $x_{\rm
F}=0$ when comparing NA11 ($\pi$-Be) and NA27 ($\pi$-p)
data~\cite{macdermott}.  Hence, the $\sigma_{\rm pA}/A$ values should
be \emph{higher} than the pp values, or $\sigma_0^{c\bar{c}} \neq
\sigma_{\rm pp}^{c\bar{c}}$, as recently recalled in
Ref.~\cite{Antinori}.  However, looking at the error bars in Fig.~1\,b
of Ref.~\cite{macdermott} we see that $K_0=1.5$ implies
$\alpha=1.0^{+0.2}_{-0.4}$, a very poor statistical significance.  In
fact, setting $K_0=1.0$ would lead to $\alpha=1.2^{+0.2}_{-0.4}$,
perfectly compatible with 1.0, within errors\dots\ The claim of
Ref.~\cite{macdermott} is, thus, unsubstantiated (especially when we
note that systematic uncertainties were neglected in the comparison
between the data of the two experiments).

\begin{figure}[ht]
\centering
\resizebox{0.45\textwidth}{!}{%
\includegraphics*{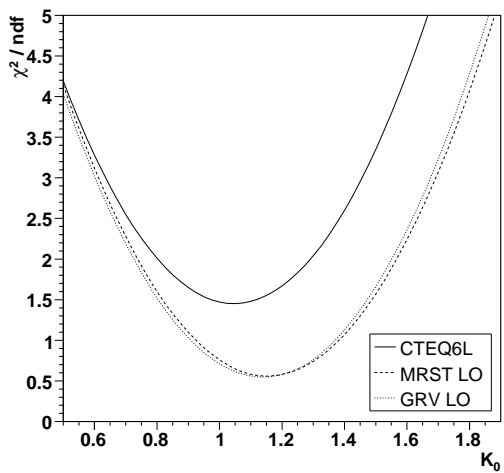}}
\caption{Quality of the fit to the charm production cross-section data
  shown in Fig.~\ref{fig:pdfVar} as a function of the $K_0$ factor
  used to scale the pp points.}
\label{fig:macdermott}
\end{figure}

We have revisited this issue using the charm production data currently
available, almost 20 years later.  Would the new measurements prefer a
$K_0$ value significantly higher than 1?  We fitted the charm
production cross-sections shown in Fig.~\ref{fig:pdfVar} after scaling
the pp values by a $K_0$ factor.  Figure~\ref{fig:macdermott} shows
how the fit quality, expressed in terms of $\chi^2$ per degrees of
freedom, changes when varying $K_0$.  We see that the best fit is
obtained with $K_0 \sim 1.1$, and that the fit quality significantly
degrades when using $K_0 \sim 1.5$.

\subsection{Beauty production cross-sections}

\begin{figure}[ht]
\centering
\resizebox{0.48\textwidth}{!}{%
\includegraphics*{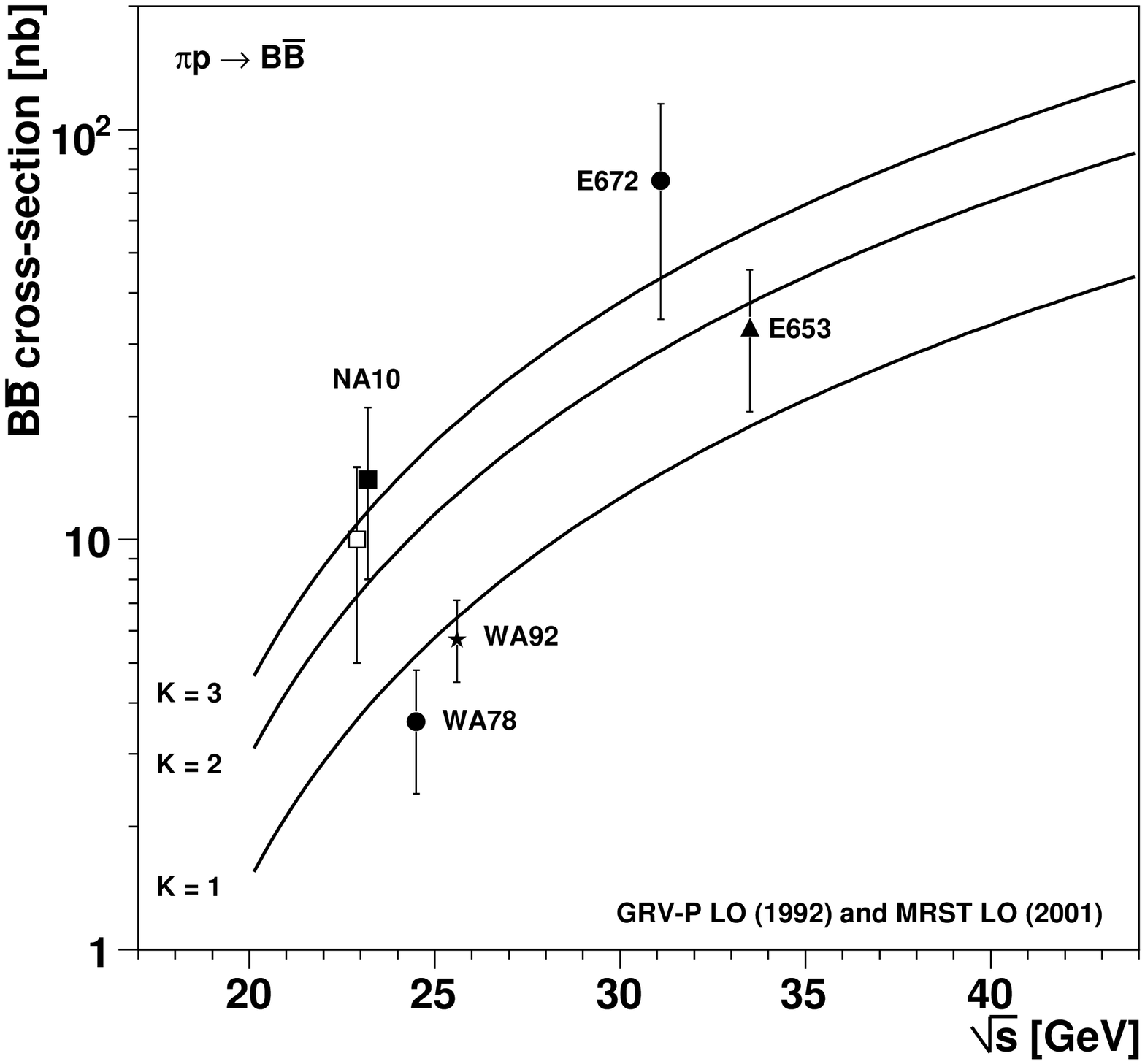}}
\resizebox{0.48\textwidth}{!}{%
\includegraphics*{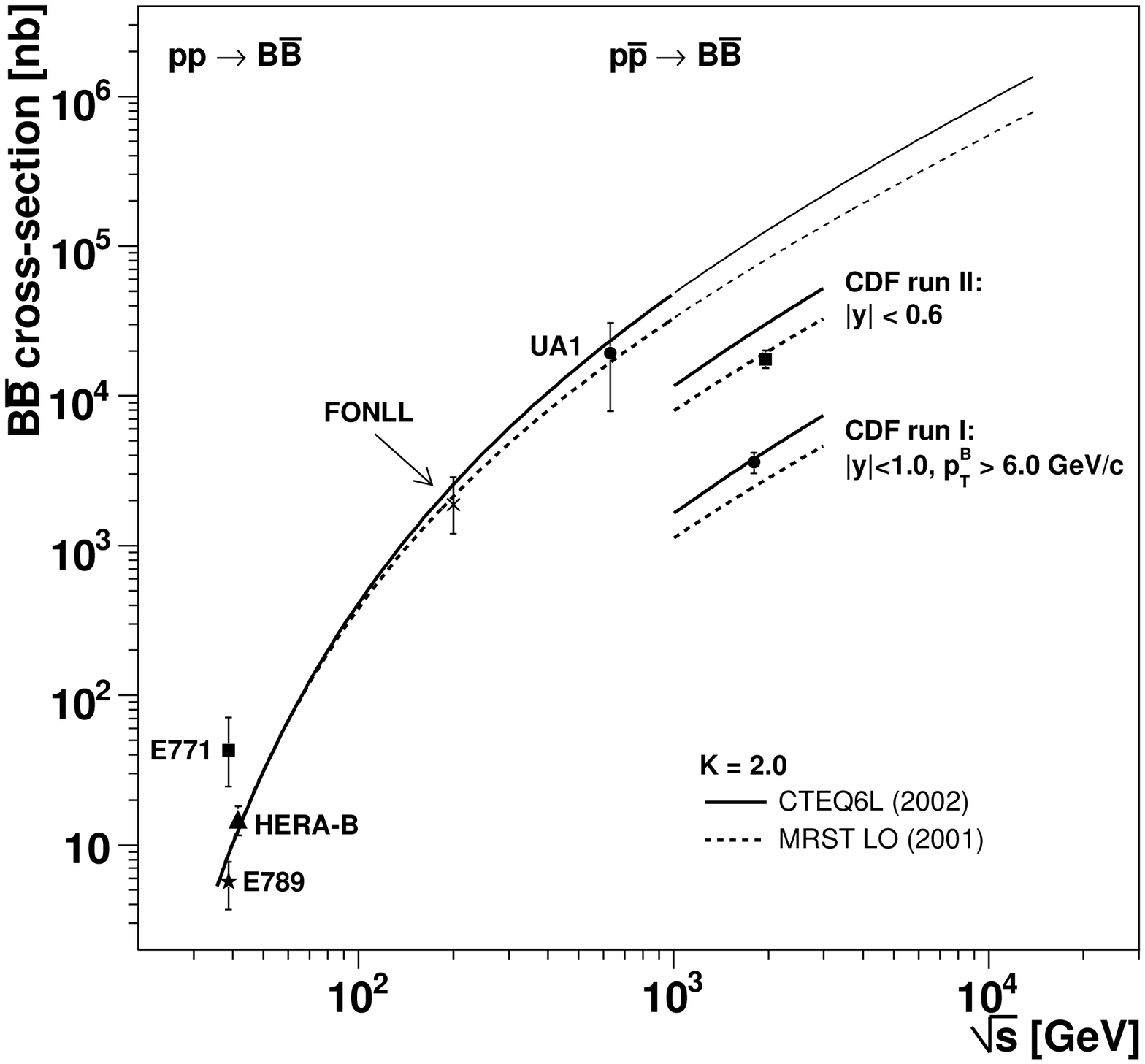}}
\caption{\BBbar\ production cross-sections in pion (left) and 
proton (right) collisions.}
\label{fig:beauty}
\end{figure}

As we have seen in Section~\ref{sec:dataBeauty}, not many experiments
have measured beauty production cross-sections, in pion or proton
induced collisions, and most of those measurements are derived under
model-dependent assumptions.  This might partially explain why some
data points, collected at essentially the same energy, differ by
factors of 5 (NA10 and WA78; E771 and E789).  The available
measurements are shown in Fig.~\ref{fig:beauty}, as a function of
$\sqrt{s}$, separated between pion (left) and proton (right) data.
Two points are shown for NA10, using open and closed squares, 
corresponding to different theoretical assumptions for the kinematical
distributions used to derive the full phase space value.

These measurements are compared to the results of the calculations we
have made with Pythia.  On the left panel we show the curves
calculated with GRV-P~LO (pion) and MRST~LO (proton) PDFs, scaled with
K-factors of 1, 2 and 3.  Given their significant spread, it is not
meaningful to use the data points to fit a specific K-factor value.
On the right panel we show the fixed-target proton data points,
together with \ppbar\ data from UA1 and CDF.  The two curves were
calculated with CTEQ6L and MRST~LO PDFs, and were (arbitrarily) scaled
up with a K-factor of 2.  We remark that the beauty production
cross-section at such high energies is dominated by gluon fusion and,
therefore, is identical for pp and \ppbar\ collisions.

It is clear that the fixed-target points, on their own, are not able
to give a meaningful normalisation of the calculations, and we must
profit as much as possible from the three measurements provided by the
\ppbar\ collider experiments: UA1 quotes the full phase space
cross-section at $\sqrt{s}=630$~GeV, extrapolated from $p_{\rm T}^{\rm
  b}>
6$~GeV/$c$ and $|y|<1.5$, while CDF gives cross-sections at $\sqrt{s}
= 1.8$ and 1.96~TeV, within their specific phase space window.

In Run~I, CDF measured a single B$^+$ meson cross-section of $3.6\pm 0.4\pm
0.4~\mu$b (dividing the sum of reconstructed B$^+$ and B$^-$ mesons by
2), in the kinematical window $|y|<1.0$ and $p_{\rm T}^{\rm B}>6.0$~GeV/$c$.
According to Pythia (version 6.326), this window covers 7.8\,\%
of the full phase space.  The corresponding calculated cross-section
is 1.9~$\mu$b if evaluated with the CTEQ6L PDF set, and 1.2~$\mu$b if
evaluated with the MRST~LO or GRV~LO PDF sets.  In Run~II, CDF
determined the single B \emph{hadron} (beauty only; not anti-beauty) 
production cross-section in $|y|<0.6$,
$17.6\pm 0.4^{+2.5}_{-2.3}~\mu$b, using displaced \jpsi's tagged to
come from beauty hadron decays.  In this kinematical window
($\sim$\,22\,\% of the full phase space), Pythia gives 16, 9 and
10~$\mu$b (corresponding to K-factors of 1.1, 1.9 and 1.7), with the
CTEQ6L, MRST~LO and GRV~LO PDF sets, respectively.  These two
measurements are also shown in Fig.~\ref{fig:beauty}, and should be
compared to the Pythia curves calculated for the corresponding phase
space windows.

Taking into account that the E771 and E789 data points (at the same
energy) differ by a significant amount, and that the collider values
have been measured in kinematical windows covering only a relatively
small fraction of full phase space, it is remarkable that the Pythia
curves, with a K-factor of 2, go through essentially all the data
points within around a factor 2, over \emph{four orders of magnitude}
in beauty production cross-section, between the fixed-target and the
CDF energies.  It is particularly remarkable that the comparison
between the two most reliable measurements, HERA-B and CDF, does not
indicate any increase of the K-factor with energy.

Using the Pythia curves we derive a beauty production cross-section in
pp collisions at RHIC energies, $\sqrt{s} = 200$~GeV, of around
2.5~$\mu$b.  This value is in good agreement with the prediction of a
QCD FONLL calculation, which goes beyond the (``fixed-order'') NLO
result by including the resummation of next-to-leading logarithms
(``NLL'').  See Ref.~\cite{CacciariPRL95} and references therein for
further details on the calculation and on its ``parameters'': the
heavy quark mass, the strong coupling, $\Lambda_{\rm QCD}$, the PDF
set, the factorisation and renormalisation scales, the fragmentation
functions, etc.  The result of this calculation, $\sigma({\rm
b}\bar{\rm b}) = 1.87^{+0.99}_{-0.67}$~$\mu$b, has been included in
Fig.~\ref{fig:beauty}, for comparison purposes.  The error bar
represents the uncertainty of the calculation, estimated by varying
certain input parameters within reasonable ranges.

\subsection{Beauty kinematical distributions}

\begin{figure}[ht!]
\centering
\resizebox{0.50\textwidth}{!}{
\includegraphics*{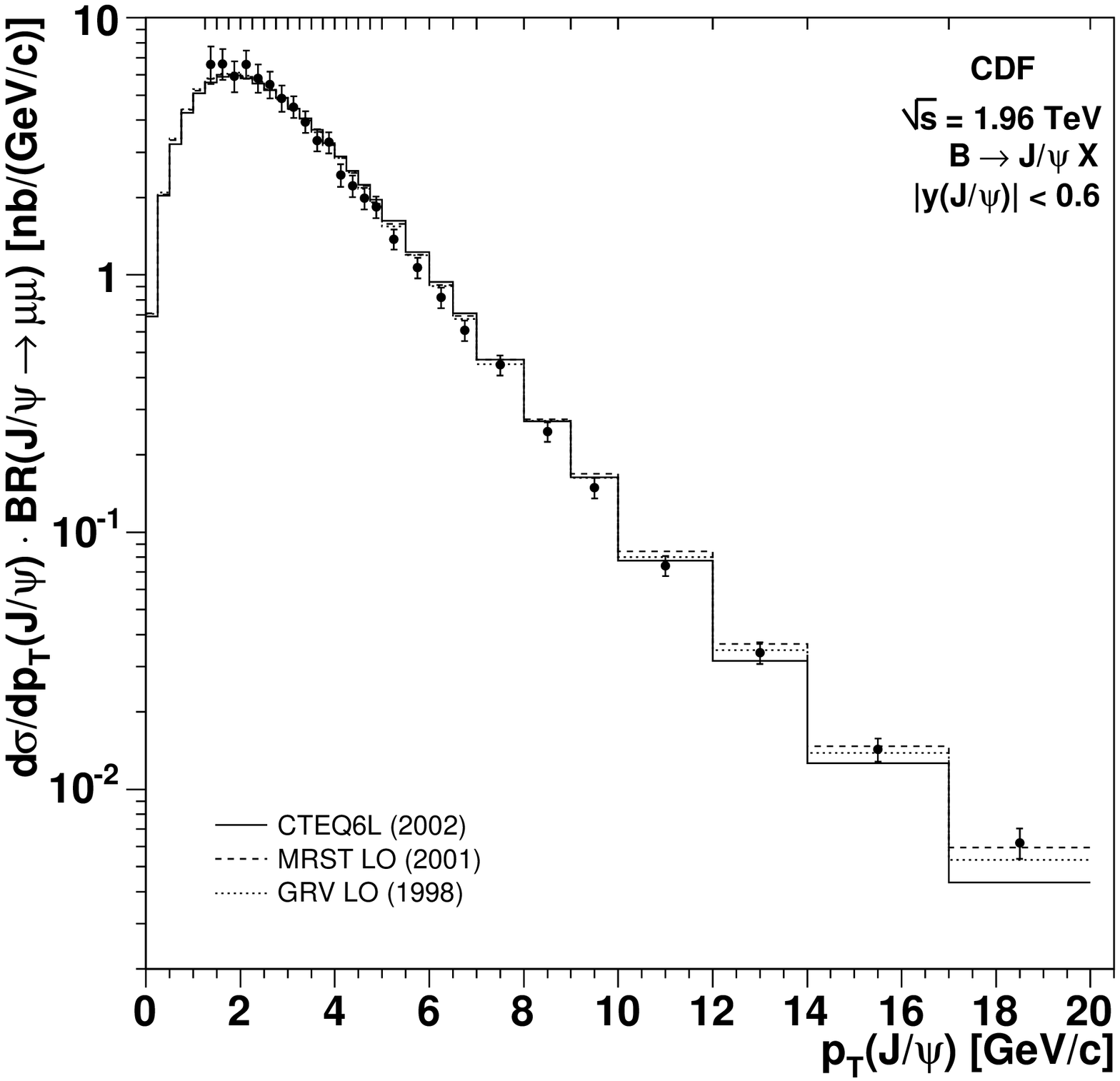}}
\resizebox{0.46\textwidth}{!}{
\includegraphics*{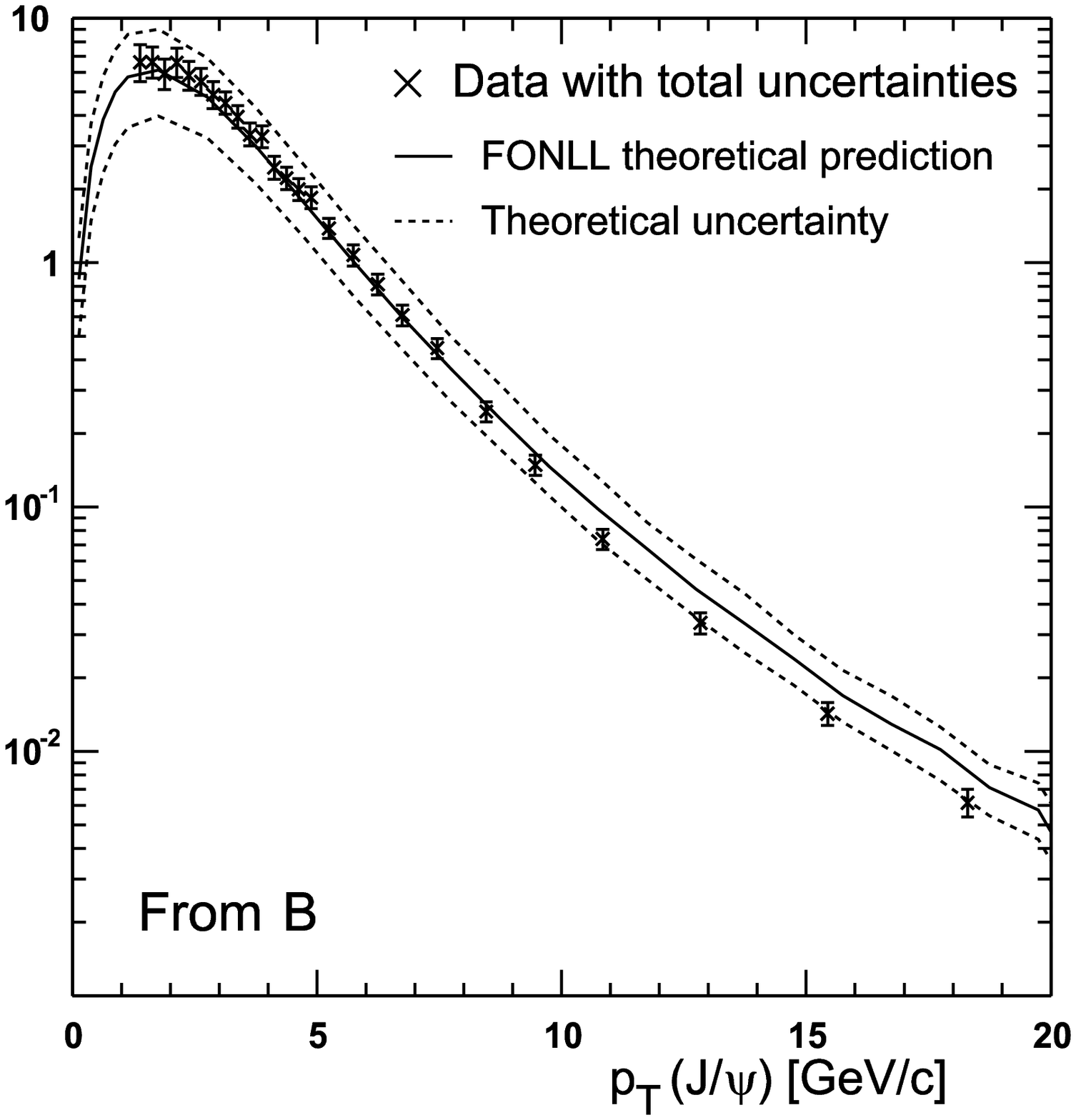}}
\caption{\pt\ distribution of \jpsi's from beauty decays, as measured
by CDF, within $|y|<0.6$.  The error bars include an overall 6.9\,\%
systematic uncertainty~\cite{cdf-beautyJpsi}.  The curves on the left
panel were calculated with Pythia, version 6.326; those on the
right panel are a FONLL calculation~\cite{cacciari3} (figure taken from
Ref.~\cite{cdfII}).}
\label{fig:kinCDF-beauty}
\end{figure}

Measurements of kinematical distributions of beauty mesons are rare.
UA1 published a \pt\ distribution on the basis of around 3000~events
(merging four independent analyses), for $p_{\rm T}^{\rm b} > 6$~GeV/$c$ and
$|y|<1.5$~\cite{ua1}.  From Run~I, CDF published a \pt\ distribution
for B$^+$ mesons with $p_{\rm T} > 6$~GeV/$c$ and $|y| < 1.0$, based
on 387 events~\cite{cdf}.  
Using around 40~pb$^{-1}$ of Run~II data, CDF tagged $\sim$\,40\,000
\jpsi\ mesons, within $|y({\rm J}/\psi)| < 0.6$, as being produced
away from the interaction vertex, presumably by beauty
decays~\cite{cdfII}.  The corresponding \pt\ distribution is shown in
Fig.~\ref{fig:kinCDF-beauty}, where it is compared with Pythia
calculations made with three PDF sets and with $\rm PARP(91)
=1$~GeV/$c$, on the left panel, and with a QCD FONLL
calculation~\cite{cacciari3}, on the right panel.

The agreement between the calculations and the measured data is quite
good, over three orders of magnitude and down to very low \pt\ values,
where theoretical uncertainties are particularly important.  The
curves calculated with Pythia were normalised using the K-factors
derived in the previous section.

\subsection{Beauty feed-down to \jpsi}

The study of the decay mode $\B \to {\rm J}/\psi\, X$ allowed several
experiments to determine the beauty production cross-section.  On the
other hand, for studies of \jpsi\ production this decay channel
constitutes a source of background which should be carefully
evaluated.  This feed-down source of \jpsi\ mesons is particularly
important in the context of the study of \jpsi\ suppression (or
enhancement) in
heavy-ion collisions at sufficiently high energies, as those at the
RHIC and LHC colliders.  Since beauty mesons should not be affected by
the medium formed in heavy-ion collisions, the fraction of \jpsi's
which come from beauty decays will not be suppressed.  Therefore, the
pattern of \jpsi\ suppression measured at RHIC or LHC energies cannot
be directly compared with predictions based on \jpsi\ melting by QGP
formation, for instance, without considering the beauty
``contamination''.  This observation emphasises the importance of
measuring the beauty yield in the RHIC and LHC heavy-ion experiments.

\begin{figure}[ht!]
\centering
\resizebox{0.48\textwidth}{!}{%
\includegraphics*{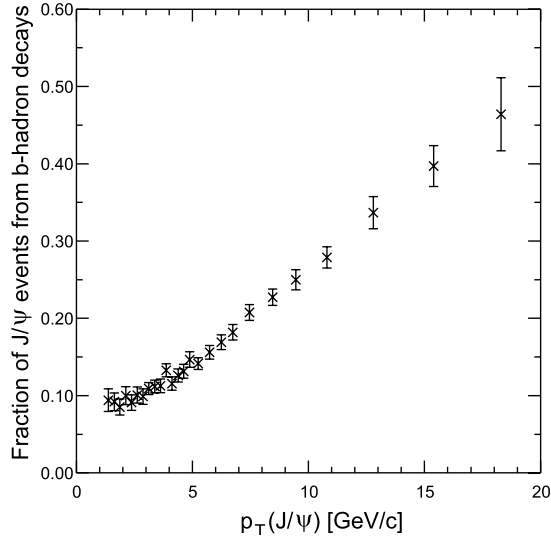}}
\caption{Fraction of \jpsi\ mesons from beauty decays as a function of
the \jpsi's transverse momentum, as measured by the CDF experiment for
$|y(\rm J/\psi)|<0.6$, at $\sqrt{s}=1.96$~TeV~\cite{cdfII}.  Error
bars include statistical and systematic uncertainties.}
\label{fig:BeautyToJpsi}
\end{figure}

We will now make a rough evaluation of the level of this problem.
Figure~\ref{fig:BeautyToJpsi} shows the fraction of \jpsi's coming
from the decay of beauty hadrons, as a function of the \jpsi's
transverse momentum, as measured by CDF in Run~II, within $|y|<0.6$.
We see that the fraction of \jpsi's from beauty decays rises steeply
with increasing \pt, from $\sim$\,10\,\% at $p_{\rm T} \lesssim
3$~GeV/$c$ to $\sim$\,50\,\% at $p_{\rm T} \sim 20$~GeV/$c$.  At
$\sqrt{s} = 630$~GeV, UA1 observed that $31\pm 2 \pm 12$\,\% of all
\jpsi's produced within $5 < p_{\rm T} \lesssim 25$~GeV/$c$ and $|y| <
2.0$ come from beauty decays~\cite{UA1-JPsi}.  This fraction becomes
negligible (less than 0.1\,\%) at the much lower energy of
HERA-B~\cite{heraB-beauty2}.


Since beauty production is expected to scale linearly with the mass
number of the colliding nuclei, while \jpsi\ production scales less
than linearly (as $A^{0.93}$ at SPS energies~\cite{NA50-alphapsi}), in
Au-Au or Pb-Pb collisions the relative fraction of \jpsi\ mesons
resulting from beauty decays will be higher than in pp collisions.  If
\emph{direct} \jpsi\ production is further suppressed in heavy-ion
collisions (NA50 measured a factor 2 of extra suppression in central
\mbox{Pb-Pb} collisions at the SPS~\cite{NA50-psisupp}), beauty
production could account for a significant fraction of the observed
\jpsi\ yield, especially at LHC energies and at high \pt\ values.
Studies of \jpsi\ suppression as a function of \pt\ are particularly
sensitive to this feed-down source, given its strong \pt\ dependence.


\subsection{Charm cross-section measurements without vertexing}

Besides the charm production cross-section measurements
mentioned in Section~\ref{sec:dataCharm}, mostly performed with
especially designed detectors and affected by relatively low
background levels, there are a few measurements made by other
experiments, in more difficult
conditions.  We will consider in this section three of these
``indirect measurements'', recently made by experiments working on the field of
``quark matter physics'': NA50 at the SPS; PHENIX and STAR at RHIC.
While these experiments are mostly devoted to the study of high-energy
nuclear collisions, they have also taken pp, p-A or d-Au data.  We
will only consider results from these more elementary collisions.

The NA38 and NA50 Collaborations studied dimuon production, of mass
above 1.5~GeV/$c^2$, in p-A, S-U and Pb-Pb collisions, at SPS
energies~\cite{NA50IMR}, to look for evidence of thermal dimuon
production from a quark-gluon plasma, presumably formed in heavy-ion
collisions.  The measured mass distribution, in the continuum
surrounding the \jpsi\ and \psip\ resonances, was compared to the
superposition of two expected sources: dimuons from the Drell-Yan
mechanism and muon pairs from simultaneous semi-muonic decays of pairs
of D mesons.  The proton-nucleus data, collected with a 450~GeV proton
beam and several nuclear targets (Al, Cu, Ag, W), could be rather well
described, both in terms of mass and \pt\ distributions, as the sum of
these two contributions, simulated with Pythia 5.7, with MRS~A PDFs
and $\rm PARP(91)=0.9$~GeV/$c$.  Pythia was also used to calculate the
``extrapolation factor'' needed to go from the elementary
(nucleon-nucleon) charm production cross-section used in the event
generation to the yield of muon pairs from D meson pair decays
detected in the phase space window probed by the experiment: $3<y_{\rm
lab}<4$ and $|\cos(\theta_{\rm CS})|<0.5$, where $y_{\rm lab}$ is the
dimuon rapidity in the laboratory frame and $\theta_{\rm CS}$ is the
Collins-Soper polar angle~\cite{Collins-Soper}.  
Using this calculated factor and a global
fit to the four p-A dimuon mass distributions, NA50 derived
$\sigma^{\rm pN}_{\rm c\bar{c}}=36.2\pm 9.1$~$\mu$b as the full phase
space charm production cross-section, per nucleon, which best
reproduces the dimuon data collected in p-A collisions at 450~GeV.
The derivation used a luminosity deduced from the number of \jpsi\
events (and a previously measured \jpsi\ cross-section), and assumed a
linear dependence of $\sigma_{\rm c\bar{c}}$ with the target mass
number.

\begin{figure}[ht]
\centering
\resizebox{0.48\textwidth}{!}{%
\includegraphics*{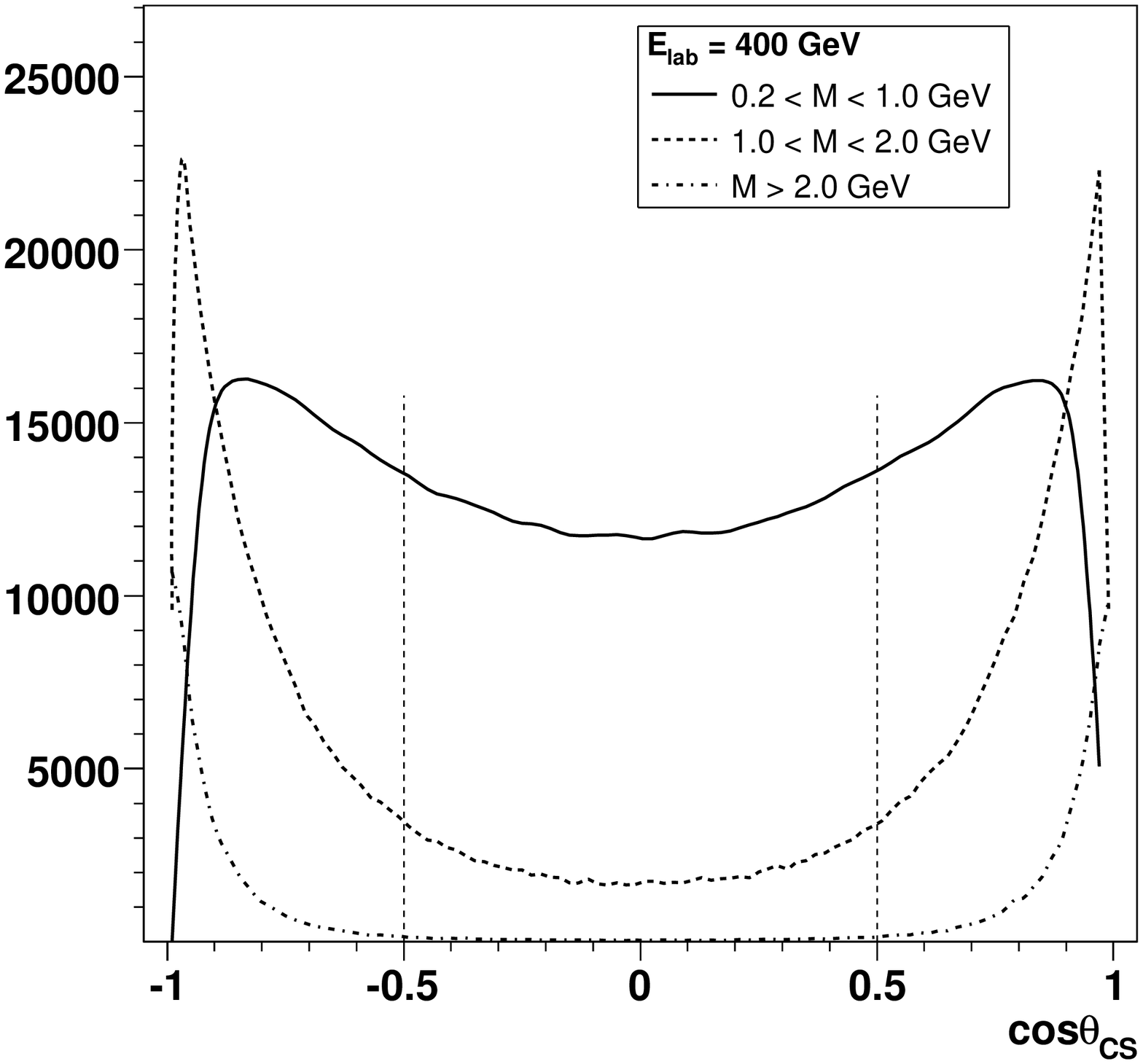}}
\resizebox{0.48\textwidth}{!}{%
\includegraphics*{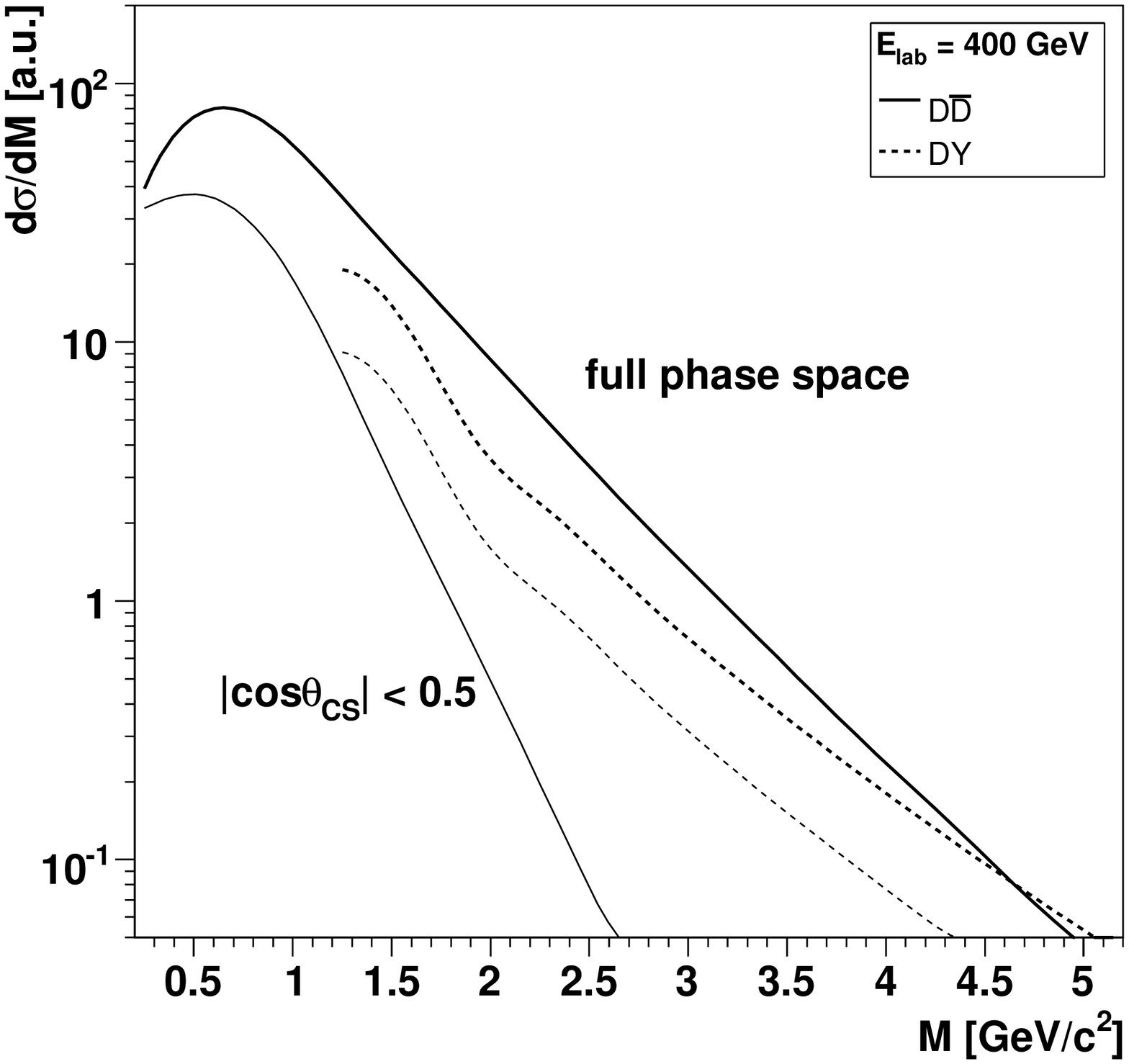}}
\caption{Left: The dimuon mass and $\cos(\theta_{\rm CS})$ variables
  are strongly correlated for muon pairs from charm decays.  Right:
  Dimuon mass distributions for Drell-Yan and charm decays, before
  (thick lines) and after (thin lines)
  applying the $|\cos(\theta_{\rm CS})|<0.5$ cut.}
\label{fig:costheta}
\end{figure}

We should be cautious when comparing this value with those directly
obtained from the study of hadronic decays of D mesons.
The ``charm yield'' is extracted from the dimuon mass spectra after
subtracting a very important ``combinatorial background'' of muon
pairs due to pion and kaon decays, and a Drell-Yan contribution which,
at such low dimuon masses (1.5--2.5~GeV/$c^2$), is rather uncertain.
Furthermore, the relative contribution of open charm decays and
Drell-Yan dimuons to the mass distribution is quite sensitive to the
kinematical cuts imposed on the events, given the strong correlation
between the dimuon mass and the $\cos(\theta_{\rm CS})$ kinematical
variables.  Figure~\ref{fig:costheta}-left shows that when we select
high mass muon pairs from \DDbar\ decays we are forcing them to have
$\cos(\theta_{\rm CS})$ close to $-1$ or $+1$.  Such events cannot be
measured by the NA38/50 spectrometer, which imposes the kinematical
cut $|\cos(\theta_{\rm CS})|<0.5$.  The influence of this cut on the
mass distributions of muon pairs resulting from charm decays and on
Drell-Yan dimuons is shown in Fig.~\ref{fig:costheta}-right, for the
case of pp collisions at $E_{\rm lab}=400$~GeV.  We see that the
Drell-Yan dimuon mass distribution is simply scaled down while the
reduction of the charm spectrum is much more pronounced for the higher
masses, leading to a significant change of the shape.  It is curious
to note, in particular, that the Drell-Yan contribution would
\emph{not} dominate the mass range above the \jpsi\ peak (a common
assumption at these collision energies) if the $|\cos(\theta_{\rm
CS})|<0.5$ cut would not be applied.

The PHENIX Collaboration measured inclusive single electron \pt\
spectra in pp collisions at $\sqrt{s} = 200$~GeV~\cite{phenix05},
using data collected in RHIC Run-2 (2001/2002), in the following phase
space window: $0.4 < p_{\rm T} < 5.0$~GeV/$c$, $|\eta| < 0.35$,
$\Delta \phi = \pi/2$. Charged particles were tracked in a drift
chamber and a pad chamber layer. Electrons were selected using the
information provided by an electromagnetic calorimeter and a ring
imaging \v{C}erenkov detector. The ratio of ``non-photonic'' electrons,
assumed to come from charm and beauty decays, to background (photonic)
sources is $\sim$\,0.4 for $p_{\rm T} < 1.5$~GeV/$c$ and $\gtrsim 1$
for higher $p_{\rm T}$ values.  The shape of the \pt\ spectrum
obtained after background subtraction is reasonably well described by
a superposition of charm and beauty contributions, as simulated with
Pythia, in the \pt\ range up to 1.5~GeV/$c$.  In the higher \pt\ range
the measured distribution exceeds the calculation, made with
Pythia~6.205, with $m_{\rm c} =1.25$~GeV$/c^2$, $\hat{s}$ as the $Q^2$
definition and the CTEQ5L set of PDFs, as described in
Ref.~\cite{phenix02}.  Since the total production cross-section is
dominated by the low \pt\ region, the high-\pt\ disagreement can be
neglected, and the normalised Pythia \emph{charm} curve was integrated
down to $p_{\rm T} = 0$.  The resulting mid-rapidity \ccbar\
production cross-section is $\dd \sigma_{\rm c\bar{c}}/ \dd y =
0.20\pm 0.03(\rm stat) \pm 0.11(\rm syst)$~mb and the estimated full
phase space value is $\sigma_{\rm c\bar{c}} = 0.92\pm 0.15\pm
0.54$~mb.

The STAR Collaboration extracted the \ccbar\ production cross-section
in d-Au collisions at $\sqrt{s} = 200$~GeV from data collected in
2003~\cite{star}.  Two independent measurements were made: using
inclusive electron data to probe charm semi-leptonic decays and using directly
reconstructed neutral D mesons, through the $\D \to \rm K \pi$
hadronic decay channel.  The yield of neutral D mesons with $|y| < 1$
and $p_{\rm T}$ in the range 0.1--3~GeV/$c$ was extracted from the
invariant mass spectrum of kaon-pion pairs, reconstructed in the TPC.
The lack of accurate vertexing information imposed the pairing of all
oppositely charged kaon and pion tracks of each event, resulting in a
signal-to-background ratio around 1/600.  After combinatorial
background subtraction, by event mixing, the $\D^0$ yield was
determined with an uncertainty estimated to be around 15\,\%.
Electrons and positrons were identified combining the $\dd E/\dd x$
measured in the TPC with the velocity information provided by a TOF
system (covering $\Delta \phi \simeq \pi/30$ and $-1 < \eta < 0$).
After subtracting the estimated contribution from photonic sources,
the electron \pt\ distribution in the range 1--4~GeV/$c$ was compared
to calculations of the yields expected from charm decays.  The $\D^0$
yield obtained from a combined fit to the $\D^0$ and electron data was
converted in a full phase space charm production cross-section, using
a pp inelastic cross-section of $\sigma_{\rm inel}^{\rm pp} = 42$~mb,
the average number of binary nucleon-nucleon collisions in d-Au
collisions (7.5), a factor of $4.7\pm 0.7$ to extrapolate from
mid-rapidity to full phase space, and a fraction of $\D^0$ mesons with
respect to all charmed hadrons of $0.54\pm 0.05$ (measured at LEP).
The result was $1.4\pm 0.2\pm 0.4$~mb per nucleon-nucleon interaction.
While this value is considerably higher than the PHENIX pp
cross-section, obtained at the same energy, the two results are
compatible with each other within (systematic) uncertainties.

\begin{figure}[htb]
\centering
\resizebox{0.48\textwidth}{!}{%
\includegraphics*{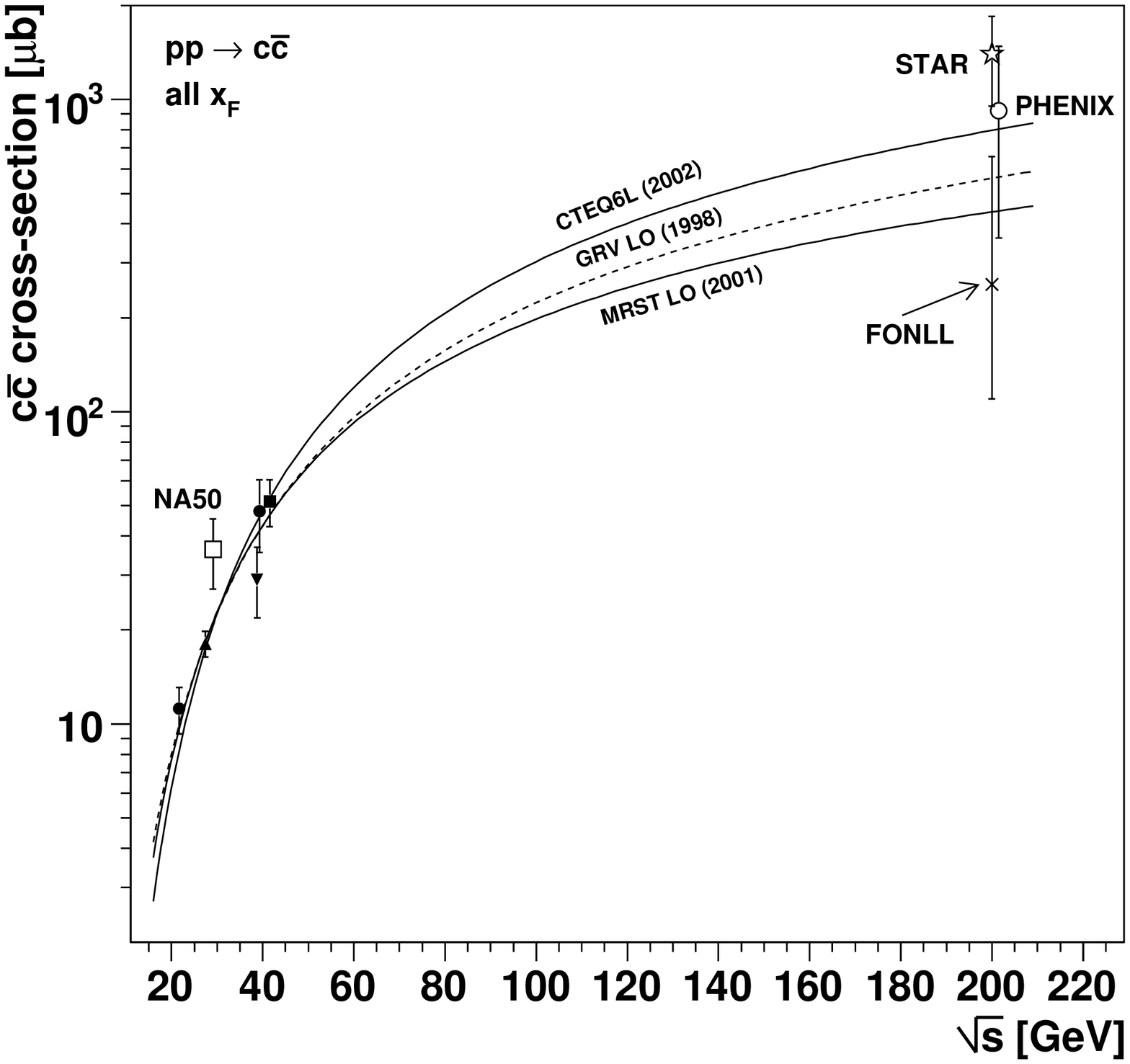}}
\resizebox{0.48\textwidth}{!}{%
\includegraphics*{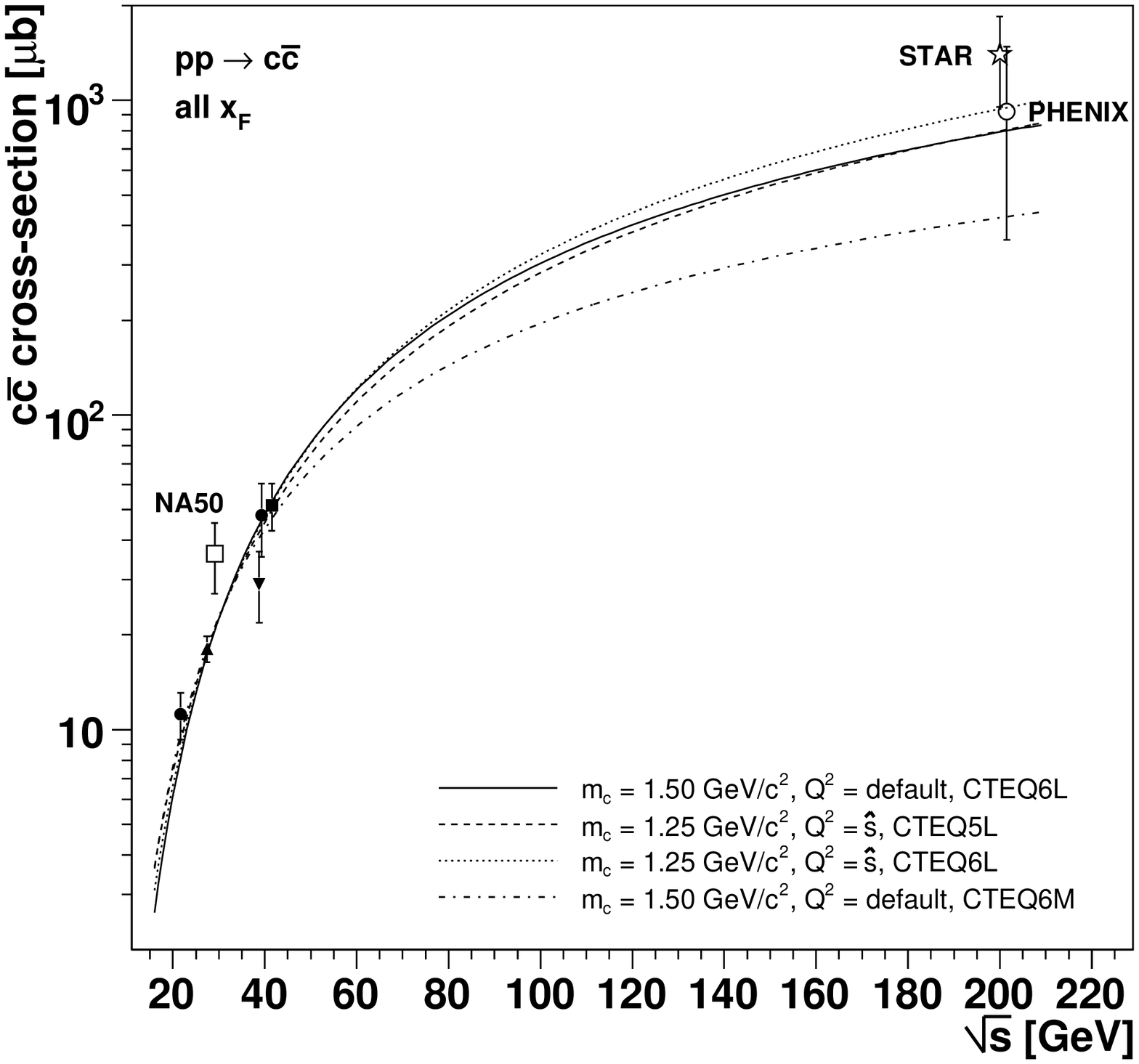}}
\caption{Left: Comparison of ``indirect'' $\sigma_{\rm c\bar{c}}$
measurements with Pythia calculations.  Right: Influence of the
$m_{\rm c}$, $Q^2$ and PDFs settings used by PHENIX on the calculated
\ccbar\ production cross-sections.}
\label{fig:indirMeas}
\end{figure}

In Fig.~\ref{fig:indirMeas}-left we compare the three
measurements mentioned above, made by NA50, PHENIX and STAR, to the
calculations we have made with Pythia and already presented in
Fig.~\ref{fig:pdfVar}.  The curves were normalised with the K-factors
determined from the fixed-target measurements, and reported in
Table~\ref{tab:ccbarPDF}.  Taking as reference the calculations made
with the CTEQ6L set of PDFs, the value derived from the NA50 p-A
dimuon data is a factor $2.1\pm0.5$ too high.  The PHENIX and STAR
values are closer to the calculations, differing by factors of
$1.15\pm0.7$ (PHENIX) and $1.7\pm0.6$ (STAR).  With respect to the
calculation made with MRST~LO PDFs, the NA50 value remains a factor~2
too high while the PHENIX and STAR points become too high by factors
$2.1\pm1.3$ (PHENIX) and $3.2\pm1.0$ (STAR).

We have also included in this figure, for comparison with the RHIC
data points and with Pythia's curves, the result of the FONLL
calculation~\cite{CacciariPRL95} for charm production at $\sqrt{s} =
200$~GeV: $\sigma({\rm c}\bar{\rm c}) = 256^{+400}_{-146}$~$\mu$b.  The
rather large uncertainty was evaluated by varying the input settings
of the calculation (see the text describing the FONLL point of
Fig.~\ref{fig:beauty} for further details).

On the right panel of Fig.~\ref{fig:indirMeas} we show four different
curves, all calculated with Pythia 6.326 but with different
settings.  The solid line is the curve obtained with our standard
settings and CTEQ6L PDFs, already presented on the left panel.  The
dashed line is obtained with the settings used by PHENIX: $m_{\rm c}
=1.25$~GeV$/c^2$, $\hat{s}$ as the $Q^2$ definition and the CTEQ5L set
of PDFs.  The dotted line uses ``intermediate'' settings: the same
mass and $Q^2$ definition as used by Phenix, but using the more recent
CTEQ6L PDFs.  This allows us to visualise the differences between the
settings in a more gradual way.  Finally, the dashed-dotted curve is
done with the same ``standard'' settings as the solid line, except
that we used CTEQ6M instead of CTEQ6L PDFs.  We recall once more that
the use of NLO PDFs, like CTEQ6M, is not appropriate in LO
calculations, such as those performed by Pythia.

It is a curious coincidence that the solid and dashed curves cross
each other at the energy of the RHIC data points, in spite of the fact
that only the fixed-target measurements contribute to their
normalisations.  The change in c quark mass, in $Q^2$ definition and
in PDF sets compensate each other, so that both the default settings,
with CTEQ6L, and the ``Phenix settings'' give a \ccbar\ cross-section
at $\sqrt{s}=200$~GeV of 800~$\mu$b.  The difference between the
dashed and dotted lines is due to changes between the CTEQ5L and
CTEQ6L PDFs, and probably originate in the harder gluon densities of
CTEQ6L: less gluons at low $x$ (see Fig.~\ref{fig:CompPdfs}) results
in somewhat lower cross-sections at high energies.

All the calculations give essentially the same values in the SPS
energy range, since they are ``forced'' by the existing measurements
to cross each other at around $\sqrt{s}=30$~GeV (see
Table~\ref{tab:ccbarPDF}).  In the context of the NA50 and NA60
experiments, however, it is worth recalling that at these energies
charm production is expected to be sensitive to nuclear effects on the
PDFs, according to the EKS~98 model (see Fig.~\ref{fig:nucl-ccbb}).
For $E_{\rm lab}=400$~GeV, the charm production cross-section per
nucleon is expected to be enhanced by $\sim$\,4\,\% in p-Be and by
$\sim$\,10\,\% in p-Pb collisions.  In the case of Pb-Pb collisions at
$E_{\rm lab}=158$~GeV per nucleon, the corresponding enhancement
factor should be around 12\,\%.

\section{Summary and conclusions}
\label{sec:summary}

In this report we reviewed some aspects of open charm and beauty
hadro-production from data collected in fixed-target and collider
experiments.  We considered measurements made with proton and pion
beams, at the CERN SPS, DESY and Fermilab, or in \ppbar\ colliders, at CERN
(UA1) and Fermilab (CDF).

We used the Monte Carlo event generator Pythia, version 6.326, as a theoretical model
to describe the reviewed data.  We have seen that Pythia provides a
reasonable description of the available charm data, in terms of energy
dependence of production cross-sections, D meson kinematical
distributions, and pair correlations, provided the intrinsic
transverse momentum of the colliding partons, $k_\mathrm{T}$, is
generated with a Gaussian of width PARP(91)\,$\sim$\,1~GeV$/c$.
However, we observed that the ratio between
charged and neutral D meson production cross-sections, as measured in
proton induced collisions, is around a factor of 2 higher than the
value expected from Pythia's calculations, with default settings.  The
comparison with data collected in other collision systems, and at many
different energies, shows that na\"{\i}ve spin counting arguments do not
seem to apply in the case of D meson production, presumably because of
the significant mass difference between the D and the D$^*$ states. This
indicates that $P_{\rm V}$ (parameter PARJ(13) in Pythia) should be set
to 0.6 (when generating charm events; for beauty the default value of
0.75 should be kept).  Except for PARP(91) and PARJ(13), we
used Pythia (6.326) with default settings.  Besides,
the calculations need to be up-scaled by an appropriate K-factor.  In
particular, using $m_{\rm c} =1.5$~GeV$/c^2$ and the CTEQ6L set of
PDFs, the best description of the measured charm cross-sections
requires a K-factor of 3.

Normalising the calculated charm production cross-sections to the
existing (fixed-target) data, we have deduced the values expected for
the RHIC energies, and compared them with the values extracted from
the measurements of PHENIX and STAR.  The issue of
nuclear effects in the parton densities was also addressed in some
detail.  At SPS energies, the EKS~98 parameterisation leads to a
$\sim$\,10\,\% higher \ccbar\ cross-section in p-Pb collisions, with
respect to a linear extrapolation from proton-proton collisions.  Such
a small effect cannot be confirmed by looking at existing data, given
the very large uncertainties of the measurements.  Maybe further
insight into these issues will result from the study of the
proton-nucleus data collected by the NA60 experiment in 2004, with
400~GeV protons incident on seven different nuclear targets
simultaneously placed on the beam line.

There are not many measurements of beauty production cross-sections
and most of those obtained with pion beams are significantly ``model
dependent'' (especially the oldest values).  In these conditions, it
is remarkable (even surprising) that the calculated energy dependence
of the beauty cross-section, for pp or \ppbar\ collisions, is able to
reproduce rather well the HERA-B and CDF measurements, which differ by
four orders of magnitude.  This results in a relatively accurate
prediction for the beauty production cross-section in the
``intermediate energies'' of the RHIC experiments, 2.5~$\mu$b.
Nevertheless, we must emphasise the relevance of a direct measurement
of the beauty production cross-section at RHIC, particularly important
to ensure a correct interpretation of the \jpsi\ measurements, in
heavy-ion collisions, as a function of collision centrality and of
\pt.

\section*{Acknowledgements}

We are particularly grateful to Torbjorn Sj\"ostrand, for his very
efficient help over the last few years and for a careful reading of
our manuscript, and to Andre David, for finding that $P_{\rm V}$
should be 0.6 rather than 0.75 in the case of charm production.
Several other colleagues helped us understanding some of the
measurements quoted in this report or gave us interesting feedback in
some other issues.  Among them, we would like to explicitly thank
R.~Averbeck, M.~Cacciari, S.~Conetti, A.~Devismes, A.~Drees,
E.~Gottschalk, T.~LeCompte, M.~Leitch, M.~Mangano, A.~Morsch,
M. zur~Nedden, L.~Ramello, C.~Salgado, J.~Schukraft, R.~Stefanski,
T.~Ullrich and P.~Weilhammer.  We are also very grateful to Gerry
Brown for his interest in our work; without his request and
encouragement, this review paper would not exist.

\section*{Note added in proof}

After this paper was completed, the HERA-B Collaboration finished
their open charm analysis~\cite{HERAB-final}.  The resulting
production cross-sections, per target nucleon and in the phase space
window $-0.15<x_{\rm F}<0.05$, are $10.7 \pm 1.2 \pm 1.4$~$\mu$b for
the charged D mesons and $26.3 \pm 2.4 \pm 2.6$~$\mu$b for the neutral
ones.  These values must be extrapolated to full phase space before
they can be compared to other measurements.  According to Pythia,
version 6.326, the HERA-B $x_{\rm F}$ window covers slightly different
fractions of full phase space for the leading and non-leading D
mesons, because they have different rapidity distributions: 49\,\% for
the $\D^+$ and $\D^0$; 58\,\% for the $\D^-$ and $\overline{\D^0}$.
These values were calculated with the CTEQ6L set of PDFs and would be
somewhat smaller if calculated with the MRST~LO (or GRV~LO) PDFs:
43\,\% and 55\,\%.  Averaging the extrapolation factors obtained with
these three PDF sets and taking into account the different particle
and antiparticle values, the total production cross-sections are $21.4
\pm 2.4 \pm 2.8$~$\mu$b and $52.6 \pm 4.8 \pm 5.2$~$\mu$b for the
charged and neutral D mesons, respectively.  These values replace the
preliminary ones given in Table~\ref{tab:ccCorr-p}.

\begin{figure}[hb]
\centering
\resizebox{0.48\textwidth}{!}{%
\includegraphics*{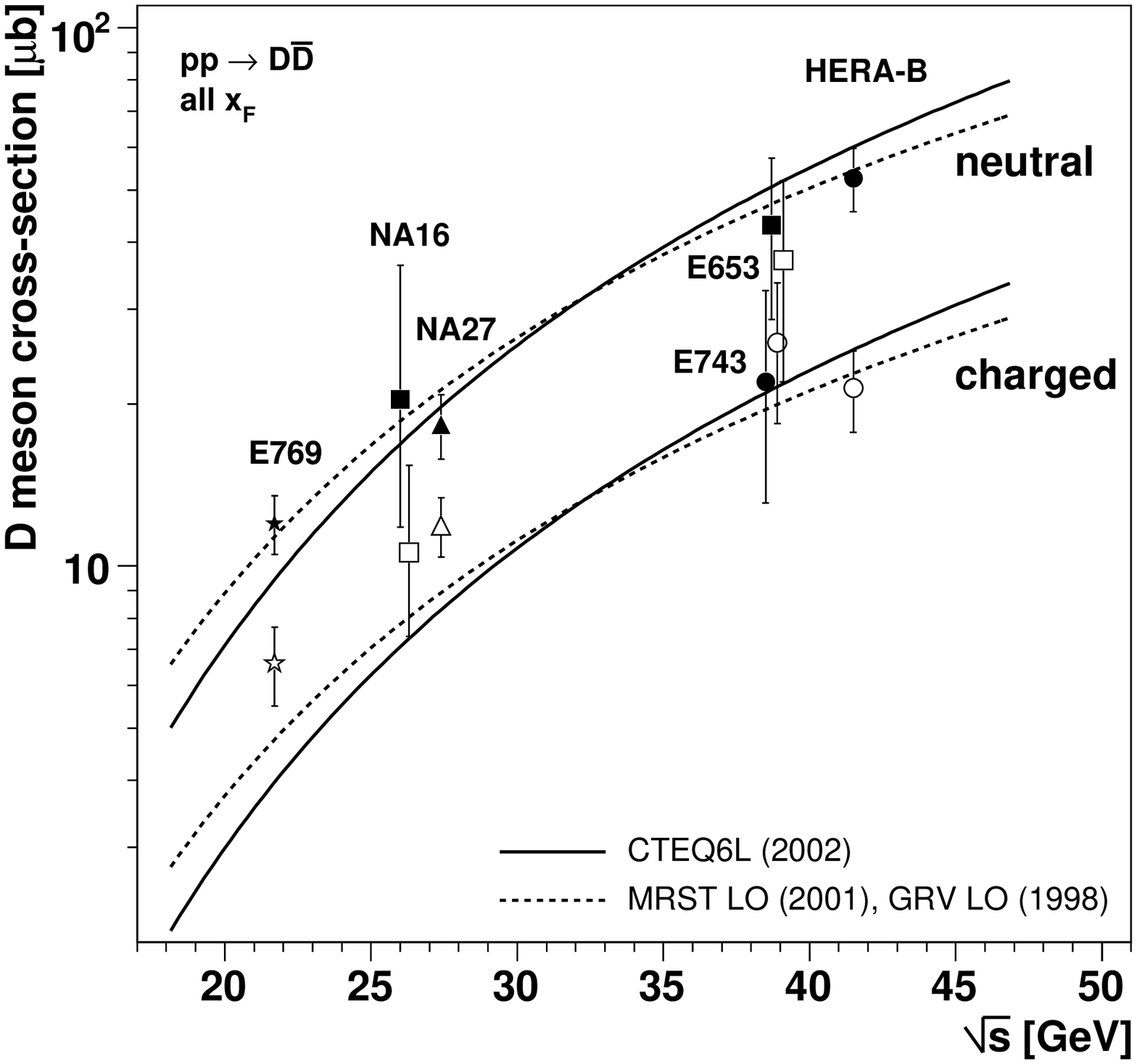}}
\resizebox{0.48\textwidth}{!}{%
\includegraphics*{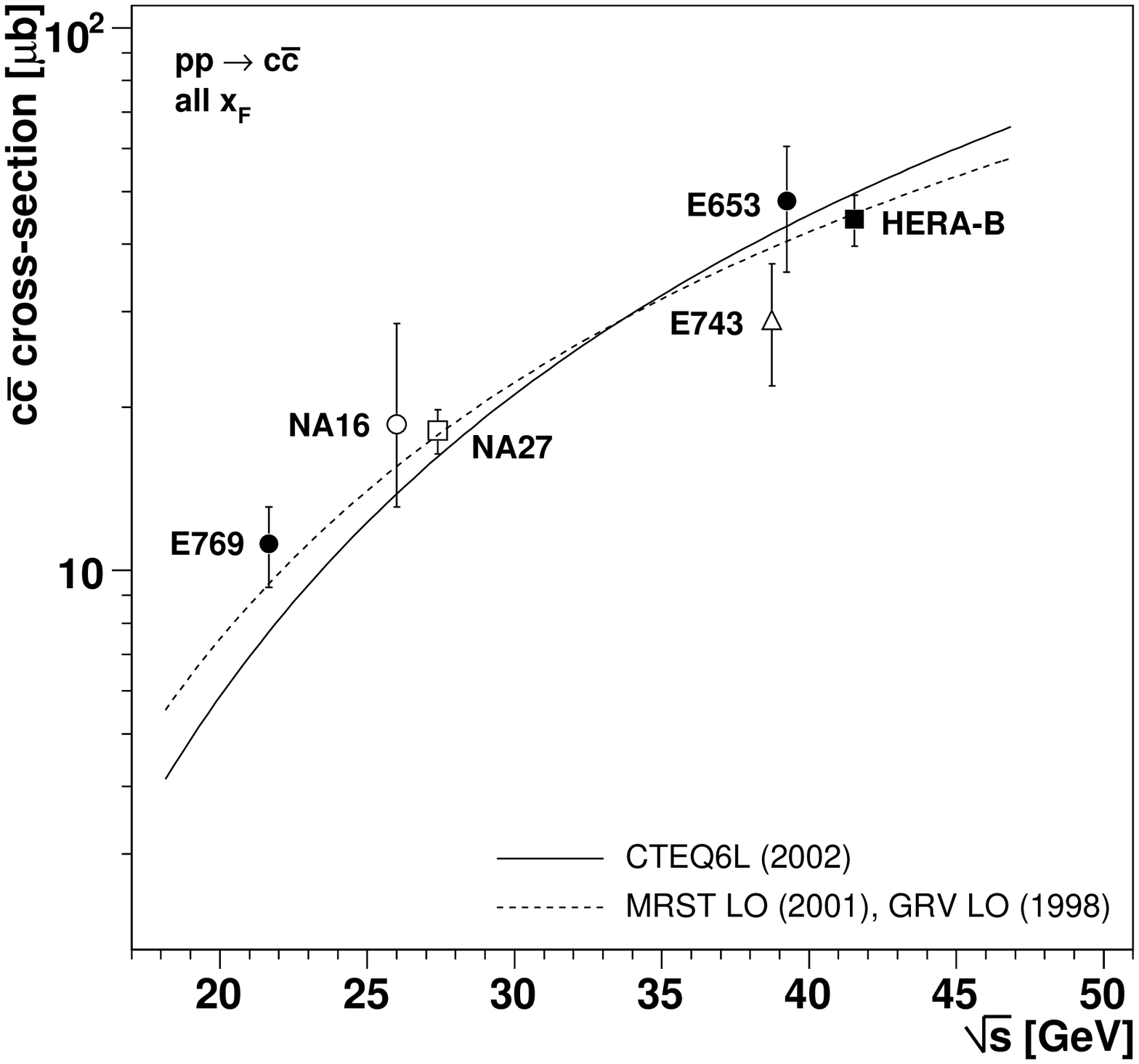}}
\caption{Neutral (closed symbols) and charged (open symbols) D meson
cross-sections (left) and corresponding total \ccbar\ cross-sections
(right), as a function of $\sqrt{s}$, compared to Pythia calculations
made with three different PDF sets.}
\label{fig:charm-herab}
\end{figure}

The left panel of Fig.~\ref{fig:charm-herab} shows the neutral and
charged D meson total production cross-sections, as a function of
$\sqrt{s}$, including the final HERA-B values, compared to curves
calculated with Pythia with $P_{\rm V}=0.6$ and three different LO PDF
sets (this figure supersedes the top panels of Fig.~\ref{fig:Pythia}).
The right panel of this figure updates the total \ccbar\
cross-sections, previously shown in Fig.~\ref{fig:pdfVar}-left,
replacing the preliminary HERA-B value (given in
Table~\ref{tab:ccbar}) by the new value: $44.4 \pm 3.2 \pm
3.5$~$\mu$b.  The K-factors resulting from the new fits to the data
points are 2.8, 3.6 and 4.4 for the CTEQ6L, MRST~LO and GRV~LO PDF
sets, respectively (these values supersede those given in
Table~\ref{tab:ccbarPDF}).

\begin{figure}[htb]
\centering
\resizebox{0.48\textwidth}{!}{%
\includegraphics*{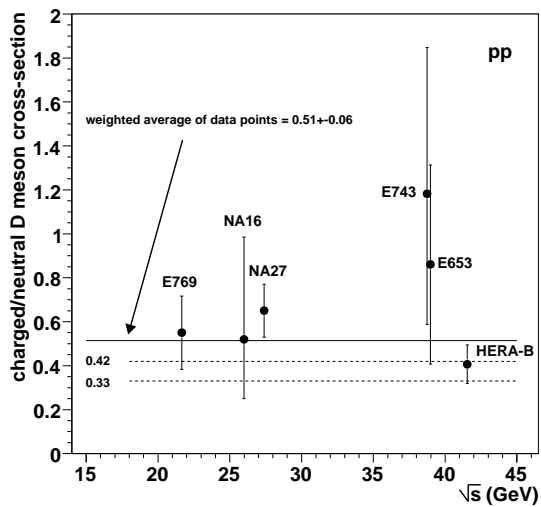}}
\caption{Ratio between the charged and neutral D meson cross-sections,
  for proton induced collisions, updated with the final HERA-B value.}
\label{fig:new-ratio}
\end{figure}

It is particularly remarkable that the final HERA-B value for the
ratio of charged to neutral D meson cross-sections is $0.40 \pm 0.06
\pm 0.04$, in very good agreement with the value expected when setting
$P_{\rm V}=0.6$.  Figure~\ref{fig:new-ratio} shows the updated version
of the left panel of Fig.~\ref{fig:charOverNeu}.

\end{document}